\documentclass[tikz,a4paper,12pt]{article}

\usepackage[hidelinks]{hyperref}
\usepackage{amsmath}
\usepackage{cleveref}
\usepackage{subfig}
\usepackage{amssymb}
\usepackage[utf8]{inputenc}
\usepackage[english]{babel}
\usepackage{cancel}
\usepackage{url}
\usepackage{mathtools}
\usepackage{tkz-euclide}
\usepackage[section]{placeins}
\usepackage{pgfgantt}
\usepackage[nottoc]{tocbibind}
\usepackage{chngcntr}
\usepackage{float}
\usepackage{verbatim}
\usepackage{graphicx,wrapfig}
\usepackage{geometry}

\floatstyle{plaintop}
\restylefloat{table}

\counterwithin{figure}{section}
\counterwithin{table}{section}

\usepackage{stackengine}
\def\delequal{\mathrel{\ensurestackMath{\stackon[1pt]{=}{\scriptstyle\Delta}}}}

\usetikzlibrary{calc}

\crefrangelabelformat{equation}{(#3#1#4--#5#2#6)}
\crefname{equation}{Eq.}{Eqs.}
\Crefname{equation}{Equation}{Equations}

\DeclarePairedDelimiter\abs{\lvert}{\rvert}%
\DeclarePairedDelimiter\norm{\lVert}{\rVert}%
\DeclareFontFamily{U}{wncy}{}
    \DeclareFontShape{U}{wncy}{m}{n}{<->wncyr10}{}
    \DeclareSymbolFont{mcy}{U}{wncy}{m}{n}
    \DeclareMathSymbol{\Sh}{\mathord}{mcy}{"58} 

\makeatletter
\let\oldabs\abs
\def\abs{\@ifstar{\oldabs}{\oldabs*}}
\let\oldnorm\norm
\def\norm{\@ifstar{\oldnorm}{\oldnorm*}}
\makeatother

\graphicspath{{pics/}}

\begin{document}
\label{Title}
\begin{titlepage}

%
%
%

\newcommand{\HRule}{\rule{\linewidth}{0.5mm}}
\center
\includegraphics[scale=0.15]{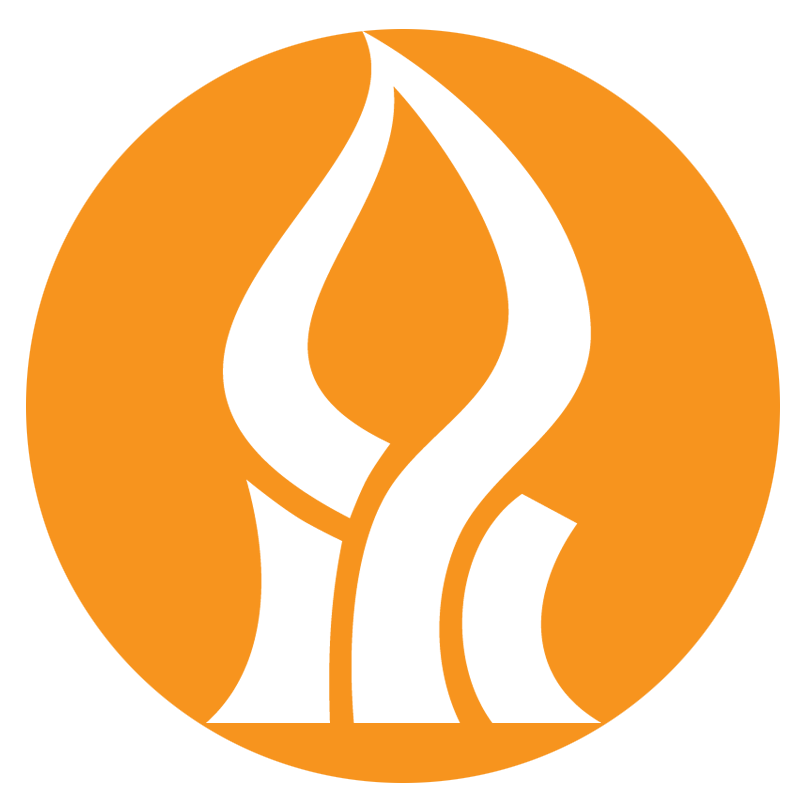} \vfill\vfill
\textsc{\LARGE Ben-Gurion University}\\[1.5cm]
\textsc{\Large Faculty of Engineering Sciences}\\[0.5cm] 
\textsc{\large Department of Mechanical Engineering}\\[0.5cm] 
\vfill
\textsc{\Large Final Project 18-40 - Final Report}\\[0.5cm] 
\HRule\\[0.4cm]
{\huge\bfseries SNIC bifurcation and its\vspace*{.5em} Application to MEMS}\\[0.4cm]
\HRule\\[1.5cm]
\begin{minipage}{0.4\textwidth}
\begin{flushleft}
\large
\textit{Author}\\
Shay \textsc{Kricheli}\\
302443692
\end{flushleft} 
\end{minipage}
~
\begin{minipage}{0.4\textwidth}
\begin{flushright}
\large
\textit{Supervisor}\\
Dr. Oriel \textsc{Shoshani}
\end{flushright}
\end{minipage}
\vfill\vfill\vfill
{\large June 14, 2018}
\end{titlepage}
\renewcommand{\thepage}{\roman{page}}
\newpage
\topskip0pt
\vspace*{\fill}
\noindent 
\parbox[b]{0.4\linewidth}{
\strut 
Signed by: \\[1cm]
\hrule
Shay Kricheli} 
\hspace{1cm}
\parbox[b]{0.4\linewidth}{
\strut 
\hrule
Oriel Shoshani, Ph.D.} 
\par\vspace{1cm} 
\vspace*{\fill}

\newpage
\section*{Abstract}

\textbf{This project focuses on a method to extract a frequency comb in mechanical means, for general interest and numerous practical applications in MEMS. The method of execution is the implementation of a beam that is exhibiting non-linear dynamics that is perturbed and analyzed for its transverse vibrations. The perturbation is an external harmonic driver with a chosen small amplitude and frequency (which is slightly detuned from the beam eigenfrequency), that when engaged with the unperturbed beam oscillations, causes it reach a state of "injection pulling" - an effect that occurs when one harmonic oscillator is coupled with a second one and causes it to oscillate in a frequency near its own. This causes the beam to reach SNIC bifurcation, rendering a frequency comb as desired. Theoretical analysis showed that the problem can be modelled using a non-linear equation of the beam, that translates to a form of the non-linear Duffing equation. While a solution to the dynamics function of the beam is hard to obtain in practice due to mathematical difficulties, a slow evolution model is suggested that is composed of functions of a amplitude and phase. Using several additional mathematical assumptions, the amplitude is seen to be related to the phase, while the phase equation solution is seen to be of the form of Adler's equation. These assumptions ultimately reduce the entire behaviour of the beam to a relatively simple solution to the Adler equation, which has a known analytical solution. Computerized numerical simulations are run on it to check the results and compare them to the theory and desired outcome. The results agreed with the theory and produce the expected frequency comb, showing the assumptions to be valid in extracting the comb.} \\
\newpage
\label{TOC}
\tableofcontents
\listoftables
\listoffigures
\newpage

\section*{Table of Symbols} \
\addcontentsline{toc}{section}{\protect\numberline{}Table of Symbols}%
\bgroup
\def\arraystretch{1}
\begin{table}[!htp]
\begin{center}
\resizebox{.31\paperheight}{!}{%
 \begin{tabular}{||c c c||}
 \hline
 Symbol & Units & Description\\ [0.5ex]
 \hline\hline
 $m$ & $[kg]$ & Mass\\
 \hline
 $\vec{r} / \vec{x} / \vec{y}$ & $[m]$ & Relocation or Position Vector\\
 \hline
 $\vec{v}$ & $\big[\frac{m}{sec}\big]$ & Velocity Vector\\
 \hline
 $\vec{a}$ & $\big[\frac{m}{sec^2}\big]$ & Acceleration Vector\\
 \hline
 $\vec{F}$ & $[N]$ & Force Vector\\
 \hline
 $T$ & $-$ & Temporal Function \\ 
 \hline
 $ \mathcal{T}$ & $[sec]$ & Period\\ 
 \hline
 $U$ & $[J]$ & Potential Energy\\
 \hline
 $E / V$ & $[Volt]$ & Voltage\\ 
 \hline
 $k$ & $[\frac{N}{m}]$ & Linear Stiffness Constant\\
 \hline
  $c$ & $[\frac{N s}{m}]$ & Damping Constant\\
 \hline
 $q$ & $[m]$ or $[rad]$ & Generalized Coordinate\\
 \hline
 $t$ & $[sec]$ & Time\\
 \hline
 $\tau$ & $-$ & Dimensionless Time\\
 \hline
 $\theta/\alpha$ & $[rad]$ & Angle\\
 \hline
  $\omega$ & $[\frac{rad}{sec}]$ & Angular Frequency\\
 \hline
 $\phi$ & $[rad]$ & Phase Shift\\
 \hline
 $\rho$ & $[\frac{kg}{m^3}]$ & Mass Density per Unit Volume\\
 \hline
 $A$ & $[m^2]$ & Cross Section Area\\
 \hline
 $E$ & $[Pa]$ & Young's Modulus\\
 \hline
 $l$ & $[m]$ & One Dimensional Length\\
 \hline
 $I$ & $[kg \ m^2]$ & Dynamical Moment Of Inertia\\
 \hline
 $u$ & $[m]$ & Beam Displacement\\
 \hline
 $Q$ & $-$ & Quality Factor\\
 \hline
  $\epsilon$ & $-$ & Infinitesimally Small Number\\
 \hline
 $\lambda$ & $-$ & Eigenvalue\\
 \hline
 $\zeta$ & $-$ & Damping Ratio\\
  \hline
 $\delta$ & $-$ & Duffing Non-Linearity Constant\\
 \hline
 $\psi$ & $-$ & Control Function\\
 \hline
 $d$ & $[m]$ & Beam Diameter\\
 \hline
 $i$ & $-$ & The Imaginary Number\\
  \hline
  $\Sh$ & $-$ & Shah Function\\
  \hline
  $\Phi$ & $[m]$ & Drive Amplitude\\
  \hline
  $\Gamma$ & $[m]$ & Control Amplitude\\
   \hline
  $X$ & $[m]$ & Displacement Function\\
  \hline
  $P$ & $-$ & Time Function\\
  \hline
\end{tabular} \
}
\caption{Table of Symbols}
\label{table:1}
\end{center} \
\end{table}\
\egroup

\clearpage
\renewcommand{\thepage}{\arabic{page}}
\setcounter{page}{1}
\section{Introduction}

\textit{Injection pulling} and \textit{locking} phenomenons occur in specific conditions when two general oscillators interact, in a way that can make them  synchronize and oscillate simultaneously to some extant. The idea of harnessing that physical phenomenon in our advantage was covered in great detail and has numerous applications in physics, electrical and mechanical engineering \cite{shoshani1} \cite{mansuri}. One interesting application of this interaction is to render a \textit{frequency comb} - a series of discrete, frequency lines in the mathematical frequency domain centered around a \textit{carrier frequency}. Perhaps the most practical application of such a frequency comb is transition from one frequency range to another, e.g., the carrier frequency can be at the extremely high frequency  (EHF) band, while the side band of the frequency comb can approach the radio frequency (RF) band. In this project we discuss the manner in which this phenomenon occurs mechanically, when a mechanical oscillator is excited in such a way that a saddle-node on invariant circle (SNIC) bifurcation occurs and a frequency comb is produced. The implementation of this could relate to many fields, and in this project the relation of the subject to \textit{MEMS - Microelectromechanical Systems} will be analyzed, hoping to provide some insight into the manner in which a fairly simple mechanical model can generate an output that other, more complex systems produced. The main setback in electrical systems is that they usually involve numerous components that each carry a degree of \textit{noise} in the corresponding output signal they produce, which is often unwanted and usually disrupts the desired outcome and its applications. In that context, a mechanical model can be favourable in that it produces a reduced amount of noise and can be more precise in the results it will provide. The goal of this project is to establish the theoretical background to the problem and to discuss its condition and assumption of validity --- using models that were previously examined. We will be using a model of vibrations of a non-linear double-clamped beam \cite{lifshitz} that will be perturbed with a harmonic signal and then closed with a feedback loop in order to stimulate the behaviour near a SNIC bifurcation that will hopefully produce a frequency comb as desired. We will not use models that include complex constraints such as parametric resonance reaction model \cite{lei li} or a model involving stochastic model, but try to provide a model as simple as possible. Afterwards, the mathematical model will be put to the test with thorough experimentation using computerized numerical analysis. Then the results will be presented, discussed and finally - conclusions will arise towards the completion of the project.

\normalsize


\newpage

\section{Theoretical Background}

In this report, there will be numerous mentions of basic and maybe a little more advanced notions in mathematics and physics, and this section is dedicated to introduce the general concept of the report and to lay out some main ideas that are presumed to be introductory.

\subsection{The Harmonic Oscillator}

When modeling a system, one usually makes certain assumptions that relate the motion in question to some familiar dynamical phenomena having a specific behavior similar to the modeled system. One very useful and frequently used mechanical entity is the \textit {harmonic oscillator.}\ This dynamical system behaves in such a way, that when displaced from its equilibrium position - a restoring force is exerted upon it, that is a function of the displacement from the aforementioned equilibrium position. 


\subsubsection{Restoring Spring Force}
 
In many practical and theoretical discussions - the force is taken to be \textit{linearly proportional}\ to the displacement and the proportionality constant is stated to be the  \textit {effective stiffness} of the system. In many cases, the oscillator in itself is a mechanical spring, and in that case that stiffness is the \textit {spring's stiffness} -  a mechanical property of the spring depending on the material in which it was made of, geometrical factors of the spring and more. In a formulated linear equation of motion of a particle subjected to a restoring force, the coefficient of the linear term with respect to the position coordinate - is said to be that exact effective stiffness of the system. Mathematically
\begin{equation}
       \vec{F}_{spring}=-k\vec{x}
\end{equation} 
 Where $\vec{F}_{spring}$ is the force applied by the spring, $k$ is the effective stiffness and $\vec{x}$ is the position vector of the object. This stiffness factor is the constant given in the first derivative of the equation of motion - which the first element in the \textit{Taylor Series} expansion of the motion function (under the assumption that the motion is due to a physical potential and therefore sufficiently \textit{smooth} and \textit{differentiable}). This project deals with a non-linear spring, and therefore implements a different approach, one that introduces the first step taken into non-linear dynamics analysis - the use of the Duffing equation. This equation incorporates the next element in the Taylor Series expansion of the motion function, which (under the assumption that the motion's potential is symmetric, thus cancelling all even powers constituents) is the 3rd degree polynomial constituent.
\subsubsection{Frequency of Oscillation}
 
A mechanical entity that is oscillating has a \textit {frequency} of oscillation, that is the scalar measure of a rate at which the entity oscillates. In harmonic oscillators - oscillators that exhibit periodic, repeating motion, this measure is reasoned by an analogy of being angular and relates the motion to an angular displacement per unit time in an imaginary circle. When the oscillator is initially perturbed and then left \textit {autonomous}, meaning the system is not excited by an external force for the motion time, that steady-state frequency of oscillation is then called the \textit {natural frequency} of the oscillator. When dealing with a \textit{linear} system - the coefficient of the linear term in the dynamics equation of motion is said to be the square of that natural frequency
\begin{equation}
    \frac{k}{m} \delequal \omega_n^2
\end{equation}
where $m$ is the effective mass of the system. The discussion in this project is of a non-linear system that exhibits non-linear dynamics and therefore requires a higher degree of analysis, as the Duffing equation offers. The expression for the natural frequency in the Duffing case is somewhat more complex.
\subsection{The Duffing Equation}
The Duffing equation - named after Georg Wilhelm Christian Caspar Duffing (1861 - 1944) is non-linear second-order differential equation used to model certain damped and driven oscillators. It is the first step from a linear system to a non linear one in that it includes a cubic term that can be thought of as an additional term considered from the \textit{Taylor Series} of the Sine function. The forced form of the equation is given by
\begin{equation} \label{duffingeq}
       \ddot{x} + 2 \zeta \omega _n \dot{x} + \omega _n ^2 x + \delta x^3 = A\cos{(\omega_d t)}
\end{equation} 
where $x$ is the unknown displacement of the oscillator and $\zeta$, $\omega _n$, $A$, $\omega_d$ and $\delta$ are given constants, that each represent a physical characteristic of the system in question:
\begin{itemize}
\item{\makebox{$\zeta$} - represents the amount of damping in the system}
\item{\makebox{$\omega _n$} - represents the natural frequency of oscillation of the system, which is the square of the linear effective stiffness}
\item{\makebox{$\delta$} - represents the amount of non-linearity in the restoring force}
\item{\makebox{$A$} - is the amplitude of the periodic driving force}
\item{\makebox{$\omega_d$} - is the angular frequency of the periodic driving force}
\end{itemize}
Equation \ref{duffingeq} describes the motion of a damped oscillator with a more complex \textit{potential} than in simple harmonic motion. The Duffing equation in its various forms is used to describe many nonlinear systems. Although most physical systems cannot be described accurately in this way for a wide range of operating conditions, such as frequency and amplitude of excitation, in many cases it is possible to use this equation as an approximate description so that their behaviour can be studied qualitatively. In some situations, quantitative analysis can be conducted for small amplitudes of excitation. In many cases, it is the first step in moving from a linear to a nonlinear system. The Duffing equation is an example of a relatively simple dynamical system that exhibits \textit{chaotic} behavior \cite{duffing}.

\subsubsection{Duffing Equation Normalization}

With use of \textit{dimensional analysis}, the duffing equation can be reduced to a more simple form, that will be defined by a single parameter - the damping. To do so, we'll define a normalized time scale
\begin{equation}
    \tau \delequal \omega _n t
\end{equation}
where $t$ is the regular-scale time. Let us now define two additional non-dimensional parameters
\begin{equation}
    y \delequal \frac{x}{\sigma} \ ; \sigma \delequal \frac{\omega _n}{\sqrt{\delta}}
\end{equation}
Using these two and the \textit{Chain Rule} from Calculus, we can express the time derivatives of $x$ as follows
\begin{equation}
\begin{gathered}
    \dot{x} = \frac{dx}{dt} = \frac{d(\sigma y)}{dt} = \sigma \ \frac{dy}{dt} = \sigma \ \frac{dy}{d\tau} \frac{d\tau}{dt} = \omega _n \sigma \frac{dy}{d\tau}\\
    \ddot{x} = \frac{d^2x}{dt^2} = \frac{d\dot{x}}{dt} = \frac{d(\omega _n \sigma \frac{dy}{d\tau})}{dt} = \omega _n \sigma \frac{d(\frac{dy}{d\tau})}{dt} = \omega _n \sigma \frac{d(\frac{dy}{d\tau})}{d\tau} \frac{d\tau}{dt} = \omega _n^2 \sigma \frac{d^2y}{d\tau^2} 
    \end{gathered}
\end{equation}
Plugging into equation \ref{duffingeq}
\begin{equation}
    \omega _n^2 \sigma \frac{d^2y}{d\tau^2} + 2 \zeta \omega _n^2 \sigma \frac{dy}{d\tau} + \omega _n ^2 \sigma y + \delta \sigma^3 y^3 = A\cos{(\omega_d t)}
\end{equation}
Dividing by $\omega _n^2 \sigma$
\begin{equation}
    \frac{d^2y}{d\tau^2} + 2 \zeta \frac{dy}{d\tau} + y + \frac{\delta \sigma^3}{\omega _n^2 \sigma} y^3 = \frac{A}{\omega _n^2 \sigma} \cos{(\omega_d t)}
\end{equation}
Using the definition of $\sigma$ we get the final form
\begin{equation} \label{duff_norm}
    \frac{d^2y}{d\tau^2} + 2 \zeta \frac{dy}{d\tau} + y + y^3 = \frac{A}{\omega _n^2 \sigma} \cos{(\omega_d t)}
\end{equation}

 \subsection{Phase Space Analysis}
 
\begin{figure}[h]
\centering
\includegraphics[scale=0.6]{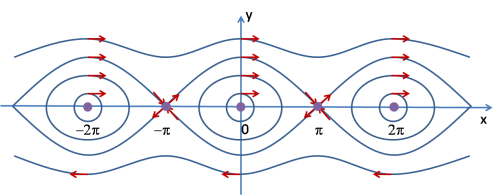}
\caption[Phase Space Portrait of a Pendulum]{Phase Space Portrait of a Pendulum\protect\footnotemark} \label{phaseportrait}
\end{figure}

 The use of Phase Space is widely accepted in dynamical systems analysis and is in many cases a more convenient approach while dealing with a non-linear problem. In classical mechanics, the phase space is the space of all possible states of a physical system - being positions of all particles located at generalized coordinates $q$, and maybe also their velocities or momenta, for one needs both the position and momentum of system in order for it to be \textit{deterministic} - as in such that it is possible to determine its future behavior. Mathematically, the configuration space of $n$ degrees of freedom might be defined by a manifold $M$ in   $\mathbb{R}^n$ while $n$ can even be infinite. Each possible state corresponds to one unique point in the phase space and in that manner it is possible to examine systems with high degrees of freedom. The \textit{phase space portrait} is a graphical representation of the phase space, and is many cases servers as a useful solution method in non-liner dynamics. In the simplest of cases, the phase portrait is $2$-dimensional as see in figure \ref{phaseportrait}.
 
A typical analysis in the phase plane starts with first defining the state variables - a set of generalized coordinates that define our system. Then, we'd like to express the state variables' time derivative as a function of their current state - namely construct a set of differential equations for each one of them, in the form
\begin{equation}
    \dot{\vec{x(t)}} = \vec{f}[\vec{x (t)}]
\end{equation}
When the analysis is made on a linear system, these equations can be written in a simple matrix form and the whole analysis is made much easier to cope with
\begin{equation}
    \dot{\vec{x(t)}} = A(t) \vec{x(t)} + B(t) \vec{u(t)}
\end{equation}
where $A(t)$ and $B(t)$ are appropriate matrices and $u(t)$ is the input of the system. Alas, in non-linear dynamics we have to use other, more complex methods.
\footnotetext{Modern Physics Lectures, BGU Physics \url{http://physweb.bgu.ac.il/COURSES/ModernPhysCohen/LectureNotes/topic18.html}}

\subsubsection{Phase Portrait}
A phase portrait is a geometric representation of the trajectories of a dynamical system in the phase plane. Each set of initial conditions is represented by a different curve, or point. Phase portraits are an invaluable tool in studying dynamical systems. They consist of a plot of typical trajectories in the state space. This reveals information such as whether an attractor, a repellor or limit cycle is present for the chosen parameter value. A phase portrait graph of a dynamical system depicts the system's trajectories (with arrows) and stable steady states (with dots) and unstable steady states (with circles) in a state space. The axes are of state variables - $x$ and $\dot{x}$ \cite{strogatz}. Phase portraits are highly useful in non-linear dynamics due to their depiction of \textit{nullclines} - lines wherein one variable does not change over time (as in $\dot{x_1}= 0$ or $\dot{x_2}= 0$ for example) and of \textit{critical points} - intersections of nullclines where neither variable changes over time (as in $\dot{x_1} = \dot{x_2}= 0$ for example).

\subsection{Injection Pulling and Locking}
 
\begin{figure}[h] 
\centering
\includegraphics[scale=0.5]{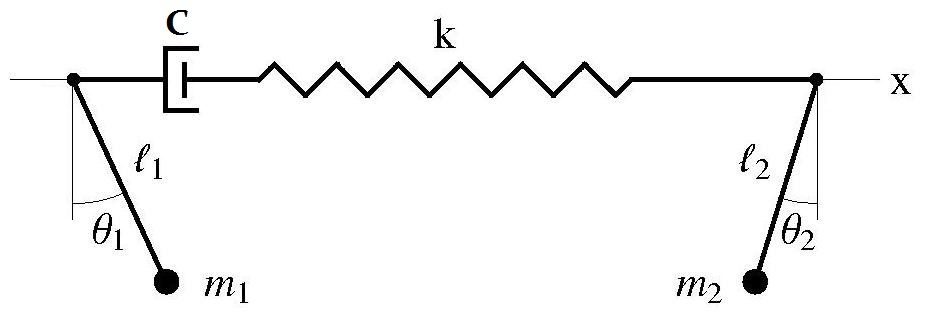}
\caption[Two pendulums in injection pulling]{Two pendulums in injection pulling \protect\footnotemark}
\label{two_pendulums}
\end{figure}

 \textit {Injection Pulling} is a phenomenon occurring when a harmonic oscillator is externally disturbed by a second oscillator that is oscillating at a nearby frequency (illustrated in figure \ref{two_pendulums}). The two oscillators then become \textit {coupled}, meaning they affect each other in their oscillation frequencies. If the coupling is strong enough then the frequencies can get near enough, which will cause the second oscillator to capture the first oscillator and making it to have essentially identical frequency as the second. When that two oscillators move at the same frequency - then this is called \textit {Injection Locking}. When the second oscillator merely disturbs the first but does not capture it entirely, the effect is called injection pulling. These phenomenons were studied by Adler \cite{adler}, Razavi \cite{razavi}, Kurokawa \cite{kurokawa} each from a slightly different perspective, and many others, while one of the first mentions of them on paper were by the mid 17th century by the Dutch scientist Christiaan Huygens. He noticed that the pendulums of two clocks on the wall moved in unison if the clocks were hung close to each other \cite{ramirez}. He deducted that the coupling of the mechanical vibrations through the wall drove the clocks into synchronization \cite{dilao} \cite{tiebout}.

\footnotetext{On the problem of synchronization of identical dynamical systems: The Huygens’s clocks, Rui Dil\~ao}
Injection locking has been used in many beneficial ways in the design of early television sets and oscilloscopes, allowing the equipment to be synchronized to external signals at a relatively low cost \cite{tiebout}. Injection locking has also been used in high performance frequency doubling circuits. However, injection locking and pulling, when unintended, can degrade the performance of phase locked loops and RF integrated circuits. Synchronization properties of periodic self-sustained oscillators are based on the existence of a special variable, called the \textit {phase}, usually noted with the Greek letter $\phi$. Mathematically, self-sustained oscillations correspond to a stable \textit {limit cycle} in the \textit {state space} of an autonomous continuous-time dynamical system. The phase $\phi$ can be introduced as the variable parametrizing the motion along this cycle. \cite{CMOS}
\subsection{The Adler Equation}
 Robert Adler (1913 - 2007) was an Austrian-born American inventor with a Ph.D. in Physics from the University of Vienna. Adler was best known as the co-inventor of the television remote control. What was less known about him is that he also did early work on certain highly nonlinear oscillating circuits, which led eventually to the Kuramoto model of synchronization, which is a generalization of the equation Adler proposed, named after him. \cite{adler2}. The equation is of the form
\begin{equation} \label{adlerstatement}
    \dfrac{d\phi}{dt} = B[1 - K \sin{(\phi})]
\end{equation}
 The Adler equation models the rate of phase rotation of a oscillator at a given instant from the phase (noted $\phi$) and \textit{amplitude} relations between the excited oscillator and the external signal, while $A$ and $B$ are constants. Mathematically, this manifests as a differential equation for the oscillator phase ($\phi$) as a function of time. The derivation of the equation and a thorough explanation of its parameters as presented by Adler, is shown in appendix \ref{adler_derivation}.
\subsection{Bifurcation Theory}
In most cases when analyzing complex system of non-linear behaviour, the use of \textit{bifurcation} analysis is a powerful tool for understanding the system's characteristics and reaction to a change in one of its affection parameters. A bifurcation occurs when a small smooth change made to the parameter values (the bifurcation parameters) of a system causes a sudden 'qualitative' or topological change in its behaviour. Bifurcations is a mathematical study of change that surfaces in continuous systems (described by differential equations - ordinary and partial) and discreet systems (described by maps). A local bifurcation occurs when a parameter change causes the stability of an equilibrium (or fixed point) to change. In continuous systems, this corresponds to the real part of an eigenvalue of an equilibrium passing through zero. These changes are witnessed in the phase portrait, where the topological changes in the phase portrait of the system can be confined to arbitrarily small neighbourhoods of the bifurcating fixed points by moving the bifurcation parameter close to the bifurcation point (hence 'local') \cite{bifurcation}. Mathematically, we can consider the continuous dynamical system described by the vector ODE
\begin{equation}
    \dot{\vec{x}} = \vec{f}(\vec{x},\lambda)
\end{equation}
A local bifurcation occurs at point $(x_0 , \lambda_0 )$ if the \textit{Jacobian matrix} of the function $f$ has an eigenvalue with zero real part. If the eigenvalue is equal to zero, the bifurcation is a steady state bifurcation, but if the eigenvalue is non-zero but purely imaginary, this is a Hopf bifurcation. Examples of local bifurcations include

\begin{itemize}
\item{\makebox{} Saddle-node (fold) bifurcation}
\item{\makebox{} Transcritical bifurcation}
\item{\makebox{} Pitchfork bifurcation}
\item{\makebox{} Period-doubling (flip) bifurcation}
\item{\makebox{} Hopf bifurcation}
\item{\makebox{} Neimark–Sacker (secondary Hopf) bifurcation}
\end{itemize}

\subsubsection{SNIC Bifurcation}

\begin{figure}[h]
\centering
\includegraphics[scale=0.6]{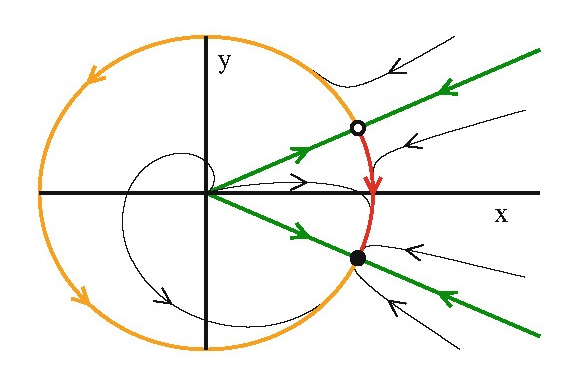}
\caption[Illustration of a SNIC bifurcation]{Illustration of a SNIC bifurcation\protect\footnotemark}
\label{snic_ill}
\end{figure}

The Saddle-Node on an Invariant Circle (SNIC) bifurcation is one of the basic scenarios for creation of a periodic orbit in smooth continuous-time dynamical systems. Intuitively, one can say it is a bifurcation that causes the system to reach an infinite limit cycle in the phase plane and oscillate freely around it. Mathematically, a SNIC bifurcation for a vector field that lies in $ \mathbb{C} ^n$ (where $n \ge 2$) and represented by $\dot{\vec{x}} = \vec{f}(\vec{x})$ on a manifold $M$ is an elementary saddle-node - an equilibrium with a simple eigenvalue 0 with no other eigenvalues on the imaginary axis with a trajectory that is asymptotic to the saddle-node in both directions of time - resulting in movement of the states of the system along an invariant circle in $ \mathbb{C} ^n$ \cite{SNIC}. Such a bifurcation results in an appearance of a limit cycle of an infinite period, as illustrated in figure \ref{snic_ill}.
\footnotetext{C. Gros, Complex and Adaptive Dynamical Systems, Chapter 2 - Bifurcations and Chaos in Dynamical Systems}

\begin{figure}[h]
\centering
\includegraphics[scale=0.5]{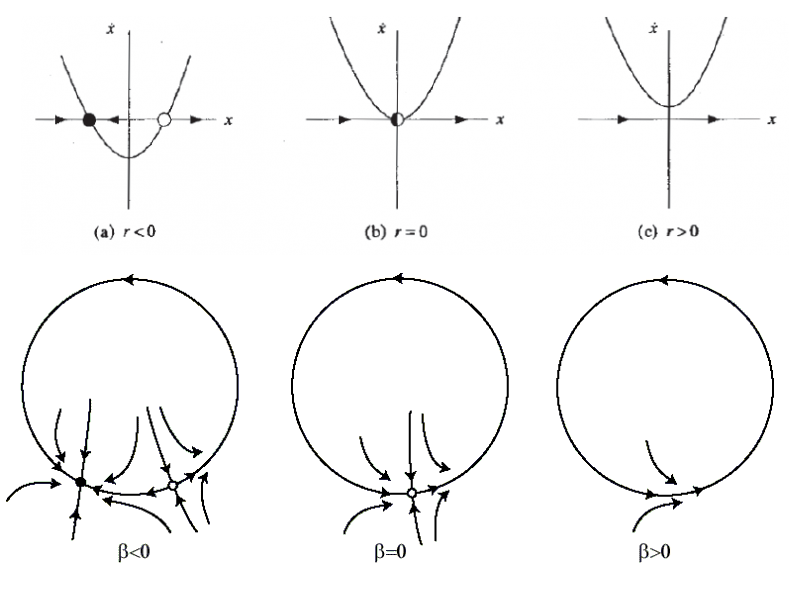}
\caption[SNIC Bifurcation States]{SNIC Bifurcation States\protect\footnotemark}
\label{snic_states}
\end{figure}

One has to note that it is possible to calculate the time period of the limit cycle using Adler's equation \cite{SNIC2}. Calculating the period of the limit cycle of this bifurcation for $\abs{K} > 1$ (the case in which there is no synchronization)
\begin{equation}
    \mathcal{T} = \int_0^\mathcal{T} dt = \int_0^{2\pi} \frac{dt}{d\phi} d\phi = \int_0^{2\pi} \frac{d\phi}{\dot{\phi}} = \frac{1}{B} \int_0^{2\pi} \frac{d\phi}{1 - K \sin{(\phi)}} = \frac{1}{B} \frac{2\pi}{\sqrt{K^2 - 1}}
\end{equation}
One can see that
\begin{equation}
    \lim_{\abs{K}\to 1} \mathcal{T} \rightarrow \infty
\end{equation}
Figure \ref{snic_states} shows the three states of the SNIC bifurcation - the left is when the system has two points of equilibrium- one stable (black) and one not stable (white), the middle is when the two point intersect and the bifurcation occurs and the right is when the system doesn't have an equilibrium point and then it oscillates freely. SNIC bifurcation can occur only in non-linear systems and therefore we wish to build a non-linear model.
\footnotetext{(A Bit of) Biological Neural Networks – Part I, Spiking Neurons, Jack Terwilliger}

\subsection{Frequency Comb Modulation}
This section will lay out the relevant details regarding the modulation of the Frequency Comb. A Frequency Comb is a set of equally spaced discrete frequencies that are centered around a carrier frequency and obtained from a vibrating physical system. In this project we investigate theoretical aspects of frequency comb generation from a closed-loop MEMS beam, which is exposed to an external periodic perturbation. When the detunning between the carrier frequency and the perturbation frequency is sufficient, a SNIC bifurcation occurs and the desired frequency comb is obtained. To simplify the notion, we can discuss about the Dirac comb first.

\subsubsection{The Dirac Comb}

\begin{figure}[h]
\centering
\includegraphics[scale=0.5]{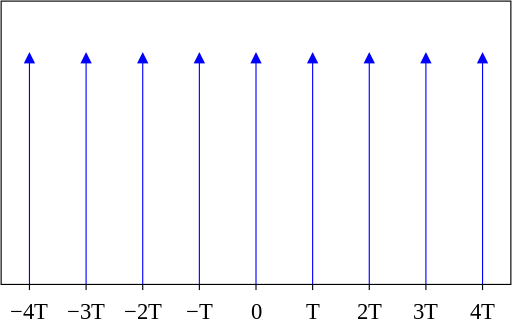}
\caption[Illustration of a Dirac Comb]{Illustration of a Dirac Comb\protect\footnotemark}
\label{Dirac_comb}
\end{figure}

The Dirac Comb is strictly a mathematical notion and is a periodic tempered distribution constructed from a set of Dirac delta functions. The generalized Dirac Delta function can be defined in the following manner: Let $\delta_{\tau}^{\epsilon}$ denote the following function
\begin{equation}
    \delta_{\xi}^{\epsilon} (t) \delequal \begin{cases} 
      0 & ; \ t < \xi \\
     \frac{1}{\epsilon} & ; \ \xi < t < \xi + \epsilon \\
      0 & ; \ t > \xi + \epsilon
   \end{cases}
\end{equation}
Now the Dirac Delta function is (loosely) defined as
\begin{equation}
    \delta_{\xi} (t) \delequal \lim_{\epsilon \to 0} \delta_{\xi}^{\epsilon} (t) = \begin{cases} 
      + \infty & ; \ t = \xi \\
      0 & ; \ t \neq \xi
   \end{cases}
\end{equation}
where we say "loosely" due to the fact that $\infty$ is of course - not a real value. The function has two special property of the form
\begin{equation}
    \int^{\infty}_{-\infty} \delta_{\xi} (t) dt = 1
\end{equation}
\begin{equation} \label{dirac2}
    \int^{\infty}_{-\infty} f(t) \delta_{\xi} (t) dt = f(\xi)
\end{equation}
for all continuous compactly supported functions $f(t)$. The Dirac Comb then is a sum of the form
\begin{equation}
    \Sh _\mathcal{T} (t) \delequal \sum_{k=-\infty}^{\infty} \delta_{0} (t - k\mathcal{T})
\end{equation}
Where $\Sh _\mathcal{T} (t)$ - the Shaw function denotes the math function defining the Dirac Comb (illustrated in figure \ref{Dirac_comb}) and $\mathcal{T} \in \mathbb{R}$ is some period. This function as written is defined in the time domain, and when thinking of the analogous notion in the frequency domain - we the get a frequency comb.
\footnotetext{Dirac comb, Wikipedia}
\subsubsection{Frequency Comb definition}
Mathematically, the frequency domain representation of a \textit{perfect frequency comb} (a frequency comb that is defined $\forall \omega \in \mathbb{R}$) is a series of delta functions spaced according to the following formula
\begin{equation}
    f_n \delequal f_0 + n f_r
\end{equation}
where $n \in \mathbb{N}$, $f_r$ is the \textit{comb spacing} (equal to the modulation frequency), and $f_{0}$ is the \textit{carrier offset frequency}, such that $f_0 \leq f_r$. The carrier offset frequency or \textit{center frequency} is the nominal frequency of an analog frequency modulation that is formed around it. \cite{frequency_comb}
\begin{figure}[h]
\centering
\includegraphics[scale=1]{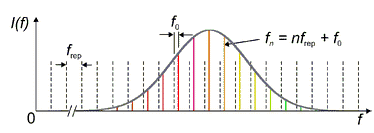}
\caption[Schematics of a frequency comb---the carrier frequency is at the center of the Lorentzian (orange) and there is a set of discrete frequency lines around it, i.e., a frequency comb]{Schematics of a frequency comb---the carrier frequency is at the center of the Lorentzian (orange) and there is a set of discrete frequency lines around it, i.e., a frequency comb\protect\footnotemark}
\label{frequency_comb}
\end{figure}
\footnotetext{Spectroscopic applications of femtosecond optical frequency combs, Helen S. Margolis}
\subsubsection{Mechanisms for Frequency Comb Generation}
As mentioned before, frequency combs are generally generated using optical and electrical means. The most popular way of generating a frequency comb is with a mode-locked laser. Such lasers produce a series of optical pulses separated in time. The spectrum of such a pulse train approximates a series of Dirac delta functions separated by the repetition rate of the laser. Others techniques involve using electro-optic modulation of a continuous-wave laser, using four-wave mixing and forming low-frequency combs using electronic components.


\section{Research Problem Formulation}

\begin{figure}[h]
\centering
\includegraphics[scale=0.3]{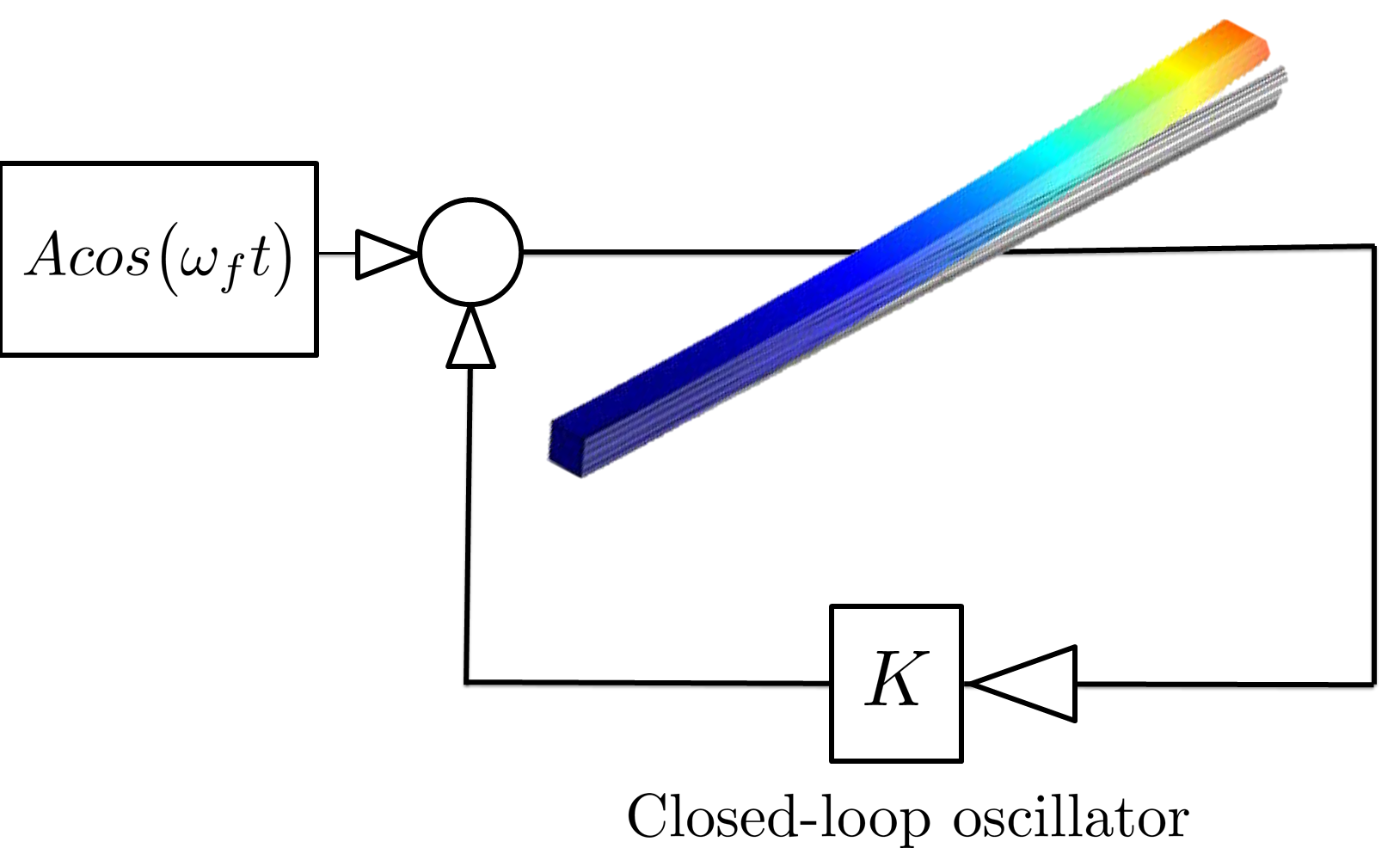}
\caption[Schematics of the MEMS based device]{Schematics of the MEMS based device}
\label{beam}
\end{figure}

This study's problem is centered around a perturbed beam that is excited externally and is controlled with a feedback loop. In this section the full extant of the problem - with excitation, damping and control - will be presented with a known model for a non-linear beam from solid mechanics. The effect of the control feedback loop and external excitation will be taken into consideration. An illustration of the system in the form of a MEMS - based device is presented in figure \ref{beam}.

\subsection{Beam Equation Derivation}

The beam in question is taken to be with boundary conditions of clamped-clamped. We will show that this beam's dynamics equation will reduce to the Duffing equation \cite{shoshani2}.
Using a model from Lifshitz and Cross' work (\cite{lifshitz}), the controlled, driven and damped transverse vibrations in one spatial axis for a slender beam with uniform mass density per unit volume $\rho$, length $l$, cross section $A$, area moment of inertia in the perpendicular direction, relative to the neutral axis $I$ and Young's modulus $E$ are described by the following partial differential equation 
\begin{equation} \label{beameq}
    \begin{gathered}
    \rho A \frac{\partial^2 u(x,t)}{\partial t^2} +2 \zeta \frac{\partial u(x,t)}{\partial t} - \tau[u(x,t)] \frac{\partial^2 u(x,t)}{\partial x^2} + EI \frac{\partial^4 u(x,t)}{\partial x^4} \\
    = \big[ A\cos{(\omega_d t)} + \psi [u(x,t)] \big] \delta(x-x_0)
\end{gathered}
\end{equation}
where $u(x,t)$ is a function that describes the transverse displacement of the beam at a point $x$ and time $t$, ,$\tau$ is a functional of $u$ that accounts for tension due to the midline stretching, $\psi$ is the control function - a saturated closed loop feedback input that is dependant on $u(x,t)$ (which we will introduce later on) and $A\cos{(\omega_d t)}$ represents the forcing element - both act on the beam at the location $0 \leq x_0 \leq l$. $A$ is the amplitude of the driving external signal such that $A \ll 1$, $\omega_d$ is the drive angular frequency, which is close to the natural frequency ($\omega _d = 1 - \Delta \omega$; $\Delta \omega \ll 1$), the damping ratio $\zeta$ is inserted artificially to model the actual physical behaviour of the beam.
\subsubsection{Boundary Conditions}
To solve this partial differential equation, one has to define appropriate boundary conditions. In our case, the beam is taken to be clamped from both sides. A clamped-clamped bean has boundary condition of
\begin{equation} \label{boundary}
    u (0,t) = u (l,t) = \frac{\partial u}{\partial x} (0,t) = \frac{\partial u}{\partial x} (l,t) = 0\\
\end{equation}
as the displacement of the beam at the end points is zero, as well as the angle - which is the derivative with respect to the spatial coordinate. 
\subsubsection{Axial Tension non-Linearity}
One way to model the axial tension $\tau$ is in following form 
\begin{equation}
    \tau[u(x,t)] = \frac{EA}{2l}\int_0^l\bigg(\frac{\partial u(x,t)}{\partial x}\bigg)^2 dx
\end{equation}
As one can see, plugging this into eq. \ref{beameq} yields a non-linear term, as we want to produce a Duffing equation
\begin{equation}
    \begin{gathered}
    \rho A \frac{\partial^2 u(x,t)}{\partial t^2} +2 \zeta \frac{\partial u(x,t)}{\partial t} - \frac{EA}{2l}\int_0^l\bigg[\frac{\partial u(x,t)}{\partial x}\bigg]^2 dx \cdot \frac{\partial^2 u(x,t)}{\partial x^2} + EI \frac{\partial^4 u(x,t)}{\partial x^4} \\
    = \big[ A\cos{(\omega_d t)} + \psi [u(x,t)] \big] \delta(x-x_0)
\end{gathered}
\end{equation}
We wind up with this integro-differential equation that looks pretty rough but can be solved using some assumptions. 
\subsubsection{Existence and Uniqueness}
With the boundary conditions set and appropriate initial conditions our problem will be a \textit{Cauchy problem} - a partial differential equation that satisfies certain conditions that are given on a surface in the domain - in our case - $\mathbb{R}^3$. The Duffing equation has several initial conditions that can be applicable and moreover it exhibits chaos in certain condition so a small change in the initial conditions can result in different behaviour entirely. Thus we don't exactly know which ones are valid in our case, but we nonetheless know that for the appropriate ones, the existence and uniqueness theorem holds. The \textit{Cauchy–Kowalevski} theorem states that local existence and uniqueness for analytic partial differential equations associated with Cauchy initial value problems holds for a system of $m$ differential equations in $n$ dimensions when the coefficients are analytic functions. Therefore we can say that we'll get a solution and it will be unique.

\subsection{Solution to the Beam PDE}

Partial differential equations can be rather complicated to solve, but in our case the problem can be manipulated using the fairly familiar technique of variable separation.

\subsubsection{Variable Separation}
The general solution to the beam equation can be achieved by we using variable separation (also known as the Fourier method) and taking $u(x,t)$ to be the sum of the multiplications of a spatial function $X(x)$ and a temporal function $P(t)$ for the infinite DOF's as follows
\begin{equation}
    u(x,t) = \sum ^{\infty}_{n=1}X_n(x)P_n(t)
\end{equation}
Furthermore, we now use a \textit{Single Mode Assumption}. To consider the vibrations of an individual eigen-mode, under the assumption that it does not interact with other modes, we take $u_n(x,t) = X_n(x)P_n(t)$ for one of the beam's eigenmodes $n$. Plugging $u_n(x,t)$ into equation \ref{beameq} we can solve the equation for the spatial and temporal functions. The entire mathematical derivation is shown in appendix \ref{solving_PDE}.


\section{Mathematical Analysis and Hypothesis}

In this section, the mathematics of the subject will be shown in detail to provide a firm grasp on the assumptions and calculations made in this project. First we will discuss the general equation we would want to solve and then introduce several mathematical and physical assumptions to simplify our expression to improve its analytic computability. We will present parallel numerical analysis throughout.

\subsection{The Spatial Equation Analysis}

\begin{figure}[h]
\centering
\includegraphics[scale=0.35]{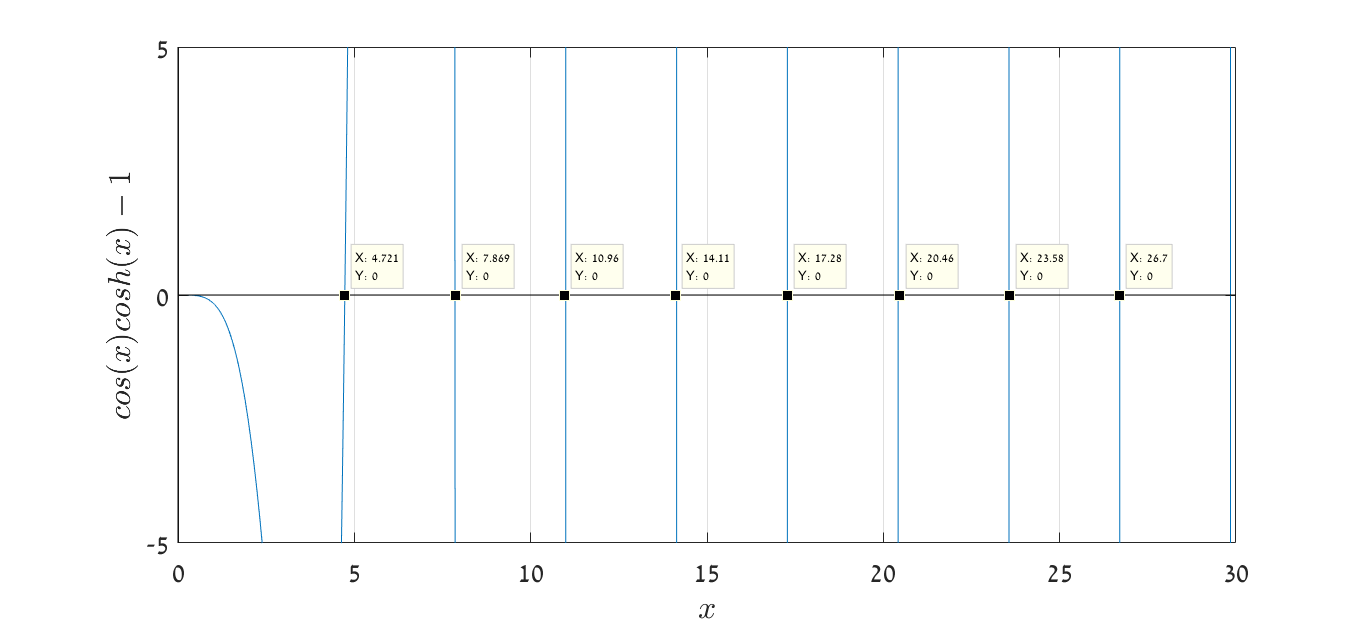}
\caption{Graph of $\cos{(\alpha_n l)} \cosh{(\alpha_n l)} = 1$} \label{coscosh}
\end{figure}

  As mentioned in the PDE derivation - to solve for the spatial function $X_n(x)$, one takes the homogenous equation and set $\tau = 0$ to yields a linear equation that produces a solution for the eigenfunctions $X_n(x)$ that can be expressed in the form
\begin{equation}
\begin{split}
    X_n(x) = \frac{1}{a_n} \big[(\sin{(\alpha _n l)} - \sinh{(\alpha_n l)})(\cos{(\alpha_n x)} - \cosh{(\alpha_n x)}) -\\
    -(\cos{(\alpha_n l)} - \cosh{(\alpha_n l)})(\sin{(\alpha_n x)} - \sinh{(\alpha_n x)})\big]
    \end{split}
\end{equation}
where we use the convention that the local maximum of the $n$'th eigenmode $X_n(x)$ that is nearest to the center of the beam is scaled to 1. $a_n$ is the value of the function in the square brackets at its local maximum that is closest to $x = 0.5l$.
Note that as the beam has an infinite number of DOF's, it has an infinite number of eigenfunctions induced by a corresponding infinite number of eigenmodes - $\alpha_n l$ that are the solutions to the transcendental equation given in equation \ref{coscosh_eq} in the derivation, which we'll show here to clarify
\begin{equation}
    \cos{(\alpha_n l)} \cosh{(\alpha_n l)} = 1
\end{equation}
This equation can be solved using a graphical solution as shown in figure \ref{coscosh}. Note that this figure does not illustrate a physical entity but a purely mathematical solution to equation \ref{coscosh_eq}. The code for this simulation is given in appendix \ref{char_eq_code}.

One can see that the first solutions in the infinite set of eigenvalues are approximately
\begin{equation}
    \left\{\alpha_n l\right\} \approx \left\{4.721, 7.869, 10.96, 14.11, 17.28, 20.46 23.58, 26.7...\right\}
\end{equation}
and the remaining ones tend towards odd-integer multiples of $\frac{\pi}{2}$, as $n$ increases.
We are interested in the first mode of vibration for the beam, so we can calculate $X_1(x)$
\begin{equation}
\begin{split}
    X_1(x) = \frac{1}{a_1} \big[(\sin{(\alpha _1 l)} - \sinh{(\alpha_1 l)})(\cos{(\alpha_1 x)} - \cosh{(\alpha_1 x)}) -\\
    -(\cos{(\alpha_1 l)} - \cosh{(\alpha_1 l)})(\sin{(\alpha_1 x)} - \sinh{(\alpha_1 x)})\big]
    \end{split}
\end{equation}
Calculating $a_1$
\begin{equation}
\begin{split}
    a_1 = \bigg[\sin{(\alpha _1 l)} - \sinh{(\alpha_1 l)}\bigg] \bigg[\cos{ \bigg(\alpha_1 \frac{l}{2} \bigg)} - \cosh{ \bigg(\alpha_1 \frac{l}{2} \bigg)} \bigg] -\\
    -\bigg[ \cos{(\alpha_1 l)} - \cosh{(\alpha_1 l)} \bigg] \bigg[ \sin{\bigg(\alpha_1 \frac{l}{2}\bigg)} - \sinh{ \bigg(\alpha_1 \frac{l}{2} \bigg)} \bigg] = 93.357
    \end{split}
\end{equation}
Plugging $a_1$ and the rest of the coefficients we get
\begin{equation}
\begin{split}
    X_1(x) = \frac{1}{93.357} \big[57.758 \cdot(\sin{(\alpha_1 x)} - \sinh{(\alpha_1 x)}) -\\
    -58.787 \cdot (\cos{(\alpha_1 x)} - \cosh{(\alpha_1 x)}) \big] =\\
    = 0.618 \cdot \big[ \sin{(\alpha_1 x)} - \sinh{(\alpha_1 x)} \big] -0.629 \cdot \big[ \cos{(\alpha_1 x)} - \cosh{(\alpha_1 x)} \big]
    \end{split}
\end{equation}
We can also note that
\begin{equation}
    \alpha_1 = \frac{4.75}{l}
\end{equation}
So the equation can be written as
\begin{equation}
\begin{split}
    X_1(x) = 0.618 \cdot \bigg[ \sin{\bigg(4.75 \frac{x}{l}\bigg)} - \sinh{\bigg(4.75 \frac{x}{l}\bigg)} \bigg] -\\
    -0.629 \cdot \bigg[ \cos{\bigg(4.75 \frac{x}{l}\bigg)} - \cosh{\bigg(4.75 \frac{x}{l}\bigg)} \bigg]
    \end{split}
\end{equation}
Moreover, because $x \in [0,l]$, we can take $s = 4.75 \frac{x}{l}$ and then
\begin{equation} \label{final_spatial}
    X_1(s) = 0.618 \cdot \bigg[ \sin{(s)} - \sinh{(s)} \bigg] -0.629 \cdot \bigg[ \cos{(s)} - \cosh{(s)} \bigg]
\end{equation}
where $s \in [0,4.75]$.
\subsection{The Temporal Dynamics Equation Analysis}
As we saw earlier, the previous chapter left us with the beam vibrations equation that we solved for the spatial coordinate, so we now have to solve the non-linear, non-homogenous equation to get a corresponding function of the beam's dynamics in time. 
\subsubsection{Reduction of the Temporal Equation to the Duffing Equation}
Taking equation \ref{rearraned_eq}, multiplying all of the equation by $X_n(x)$, integrating over the whole length of the beam with respect to the position ($\int_0^ldx$) and doing some integration by parts yields the time-dependant solution for the $n$'th mode's dynamics of the beam
\begin{equation}
\begin{gathered}
    \ddot{P_n(t)} + \frac{2 \zeta}{\rho A} \dot{P_n(t)} + \frac{EI\int_0^l [X_n''(x)]^2 dx}{\rho A \int_0^l X^2_n(x) dx} P_n(t) + \frac{E \big[\int_0^l [X_n'(x)]^2 dx\big]^2}{2 \rho l \int_0^l X^2_n(x) dx} P^3_n(t) =\\
    = \int_0^l \bigg[ A\cos{(\omega_d t)} + \psi (u_n (x,t)) \bigg] X_n(x)  \delta(x-x_0) dx 
    \end{gathered}
\end{equation}
We can see that the integral in the RHS is non-dependant on the forcing and control. Moreover, using the property of the Dirac Delta function mentioned in equation \ref{dirac2}, we can write the equation in the final from
\begin{equation} \label{duffbeam}
\begin{gathered}
    \ddot{P_n(t)} + \frac{2 \zeta}{\rho A} \dot{P_n(t)} + \frac{EI\int_0^l [X_n''(x)]^2 dx}{\rho A \int_0^l X^2_n(x) dx} P_n(t) + \frac{E \big[\int_0^l [X_n'(x)]^2 dx\big]^2}{2 \rho l \int_0^l X^2_n(x) dx} P^3_n(t) =\\
    = \bigg[ A\cos{(\omega_d t)} + \psi ((u_n (x,t)) \bigg] X_n(x_0)
    \end{gathered}
\end{equation}

Equation \ref{duffbeam} is of the form of the Duffing equation, assuming the constants are known. Now we'd like to reduce this equation to a normalized form as shown in equation \ref{duff_norm}. To do so, let us equate coefficients in the following manner
\begin{equation} 
    \omega _n ^2 = \frac{EI\int_0^l [X_n''(x)]^2 dx}{\rho A \int_0^l X^2_n(x) dx} \rightarrow \boxed{\omega _n \delequal \sqrt{\frac{EI\int_0^l [X_n''(x)]^2 dx}{\rho A \int_0^l X^2_n(x) dx}}}
\end{equation}
\begin{equation} 
    2 \Tilde{\zeta} \omega _n = \frac{2 \zeta}{\rho A} \rightarrow \boxed{\Tilde{\zeta} \delequal \frac{\zeta}{\rho A \omega _n} = \frac{\zeta}{\sqrt{\rho A E I}} \sqrt{\frac{\int_0^l X^2_n(x) dx}{\int_0^l [X_n''(x)]^2 dx}}}\\
\end{equation}
\begin{equation} 
    \boxed{\delta \delequal \frac{E \big[\int_0^l [X_n'(x)]^2 dx\big]^2}{2 \rho l \int_0^l X^2_n(x) dx}}
\end{equation}
Moreover, let us define
\begin{equation}
    T (t) \delequal \frac{P(t)}{\xi} \ ; \xi \delequal \frac{\omega _n}{\sqrt{\delta}} \ ; \tau \delequal t \omega _n
\end{equation}
Using the normalization technique shown, the equation can be written in the form
\begin{equation}
    \frac{d^2T}{d\tau^2} + 2 \Tilde{\zeta} \frac{dT}{d\tau} + T + T^3 = \frac{1}{\omega_n ^2 \xi} \bigg[ A\cos{\bigg(\frac{\omega_d}{\omega_n} \tau \bigg)} + \psi ((u_n (x,t)) \bigg] X_n(x_0)
\end{equation}
Now, we can take the function $X_n(x)$ for the first mode and actually calculate all of the constants mentioned (with use of the beam's physical parameters) but we can alternatively determine the constants of damping, driving and control independently and thus forego brute-force calculation. To do so we'll define
\begin{equation}
    \frac{A X_n(x_0)}{\omega_n^2 \xi} \delequal \Phi \ ; \frac{X_n(x_0)}{\omega_n^2 \xi} \delequal \Omega
\end{equation}
So the final dynamics equation is
\begin{equation} \label{final_duff_norm}
    \frac{d^2T}{d\tau^2} + 2 \Tilde{\zeta} \frac{dT}{d\tau} + T + T^3 = \Phi \cos{\bigg(\frac{\omega_d}{\omega_n} \tau \bigg)} + \Omega \psi (u_n(x,t))
\end{equation}
\subsubsection{Determining the Control Function}
The control function now has to be determined for it to maintain the beam in a limit cycle as detailed in the theoretical background. Let us consider the function for the $n$'th mode
\begin{equation}
    \psi (u_n) = \frac{\Gamma}{\Omega } \frac{\dot{u_n}}{\abs{\dot{u_n}}} = \frac{\Gamma}{\Omega} \frac{\dot{u_n}}{\abs{\dot{u_n}}} = \frac{\Gamma}{\Omega} \frac{\dot{P(t)} X(x)}{\abs{\dot{P(t)} X(x)}} = \frac{\Gamma}{\Omega } \frac{\dot{P(t)}}{\abs{\dot{P(t)}}} = \frac{\Gamma}{\Omega } \frac{\frac{dT}{d\tau}}{\abs{\frac{dT}{d\tau}}} 
\end{equation}
as we assumed that for the first mode, the beam does not bend in the negative direction of the $x$ axis.
We will show that the following function effectively cancels the damping effect and maintains the beam in constant vibration \cite{seshia}. To do so, we'll leave out the driving and look at the dynamics equation with the control only
\begin{equation}
    \frac{d^2T}{d\tau^2} + 2 \Tilde{\zeta} \frac{dT}{d\tau} + T + T^3 = \Gamma \frac{\frac{dT}{d\tau}}{\abs{\frac{dT}{d\tau}}} 
\end{equation}
We'd like to calculate the \textit{work} done by the system, meaning by definition to evaluate
\begin{equation}
    W = \int_0^\mathcal{T} \vec{F} \cdot \vec{dx} =  \int_0^\mathcal{T} \vec{F} \cdot \frac{\vec{dx}}{dt} \ dt = \int_0^\mathcal{T} \vec{F} \cdot \dot{\vec{x}} \ dt
\end{equation}
where $\vec{F}$ is the external force acting upon the system, $\vec{x}$ is the position vector (which is $\vec{T}$ in our case) and $x$ is the coordinate in which the work is done and $T$ is the total time interval. We assume the force is in the $x$ direction and we'd like to express it in terms of $T(t)$. We can look at the system as a simple dynamical system (one with a second derivative and liner terms) that the damping, control and duffing function acting upon it as external forces and then
\begin{equation}
    \frac{d^2T}{d\tau^2} + T= \Gamma \frac{\frac{dT}{d\tau}}{\abs{\frac{dT}{d\tau}}} - 2 \Tilde{\zeta} \frac{dT}{d\tau} -T^3 = F \bigg( T,\frac{dT}{d\tau} \bigg)
\end{equation}
Let us now assume the solution for $T(\tau)$ in the steady-state is a simple sinusoidal with a constant amplitude of the form $T(\tau) = a_{ss} cos(\tau)$ where $a_{ss}$ is the steady-state amplitude, and define the work in terms of $T(\tau)$
\begin{equation}
\begin{gathered}
    W = \int_0^\mathcal{T} F \bigg( T,\frac{dT}{d\tau} \bigg) \dot{T} \ dt = \int_0^\mathcal{T} F \bigg( T,\frac{dT}{d\tau} \bigg) \frac{dT}{dt} \ dt = \\
    = \int_0^\mathcal{T} F \bigg( T,\frac{dT}{d\tau} \bigg) \frac{dT}{d \tau} \frac{d \tau}{dt} \ dt = \int_0^{2 \pi} F \bigg( T,\frac{dT}{d\tau} \bigg) \frac{dT}{d \tau} \ d \tau
    \end{gathered}
\end{equation}
Plugging $F$ and $T$ we get
\begin{equation}
\begin{gathered}
    W = \int_0^{2 \pi} \bigg[ \Gamma \frac{\frac{dT}{d\tau}}{\abs{\frac{dT}{d\tau}}} - 2 \Tilde{\zeta} \frac{dT}{d\tau} -T^3 \bigg] \big[-a_{ss} sin(\tau) \big] \ d \tau = \\
    = -a_{ss} \int_0^{2 \pi} \bigg[ \Gamma \frac{-a_{ss} sin(\tau)}{\abs{-a_{ss} sin(\tau)}} + 2 \Tilde{\zeta} a_{ss} sin(\tau) - a_{ss}^3 cos^3(\tau) \bigg] sin(\tau) \ d \tau =\\
    = -a_{ss} \int_0^{2 \pi} \bigg[ -\Gamma \frac{sin(\tau)}{\abs{sin(\tau)}} + 2 \Tilde{\zeta} a_{ss} sin(\tau) - a_{ss}^3 cos^3(\tau) \bigg] sin(\tau) \ d \tau =\\
    a_{ss}  \Gamma \int_0^{2 \pi} \frac{sin^2(\tau)}{\abs{sin(\tau)}} \ d \tau -2a_{ss}^2 \Tilde{\zeta} \int_0^{2 \pi} sin^2(\tau) \ d \tau + a_{ss}^4 \int_0^{2 \pi} cos^3(\tau) sin(\tau) \ d \tau
    \end{gathered}
\end{equation}
Let us note that the third term is an integral on a multiplication of an odd function \big[$sin(\tau)$ \big] and an even one \big[$cos^3(\tau)$ \big] which results in an odd function. The integral is over a complete interval of these trigonometric functions, and is therefore $0$. Now let us evaluate the first term
\begin{equation}
\begin{gathered}
    a_{ss}  \Gamma \int_0^{2 \pi} \frac{sin^2(\tau)}{\abs{sin(\tau)}} \ d \tau = a_{ss}  \Gamma \int_0^{\pi} sin(\tau) \ d \tau - a_{ss}  \Gamma \int_\pi^{2 \pi} sin(\tau) \ d \tau =\\
    = 2 a_{ss}  \Gamma + 2 a_{ss}  \Gamma = 4 a_{ss}  \Gamma
    \end{gathered} 
\end{equation}
Now the second term is
\begin{equation}
\begin{gathered}
    -2a_{ss}^2 \Tilde{\zeta} \int_0^{2 \pi} sin^2(\tau) \ d \tau = -2a_{ss}^2 \Tilde{\zeta} \int_0^{2 \pi} \frac{1 - cos(2 \tau)}{2} \ d \tau = -2a_{ss}^2 \pi \Tilde{\zeta}
    \end{gathered} 
\end{equation}
Now in order for the damping to fade, we need to total work of the system to be zero, meaning that the whole term has to equal to $0$
\begin{equation} \label{ss_amp}
\begin{gathered}
    4 a_{ss}  \Gamma - 2a_{ss}^2 \pi \Tilde{\zeta} = 0 \rightarrow \boxed{a_{ss} = \frac{2 \Gamma}{\pi \Tilde{\zeta} }}
    \end{gathered} 
\end{equation}
when $B$ is the steady state amplitude that will result in our model. So the final equation with forcing will be
\begin{equation} \label{final_duff_norm_with_force}
    \frac{d^2T}{d\tau^2} + 2 \Tilde{\zeta} \frac{dT}{d\tau} + T + T^3 = \Phi \cos{\bigg(\frac{\omega_d}{\omega_n} \tau \bigg)} + \Gamma \frac{\frac{dT}{d\tau}}{\abs{\frac{dT}{d\tau}}} 
\end{equation}
\subsection{Reduction of the Duffing Equation to the Slow Evolution Model}
Though we can solve the Duffing equation with numerical methods, the main problem with the final equation shown in the previous section is that it does not admit an exact analytic solution. In the special, more simple scenario in which the beam is undamped and undriven - an exact solution can be obtained using \textit{Jacobi's elliptic functions}. However, these type of functions, defined using the complex plane, require a high degree of knowledge and are not easily dealt with. Moreover, we have both damping and driving in our equation. So alongside a numerical simulation we can perform on the equation, a different solution to the problem is suggested which is the Slow Evolution model - of the form $T(\tau) = a(\tau) \cos{[\frac{\omega _d}{\omega _n} \tau + \theta(\tau)]}$ - one that describes the time-dependant function as a multiplication of an amplitude $a(\tau)$ with a periodic component $\cos{[\frac{\omega _d}{\omega _n} \tau + \theta(\tau)]}$ that is composed of the frequency ratio $\frac{\omega _d}{\omega _n}$ and a function of the phase $\theta(\tau)$ we call the \textit{biased oscillator phase}. This phase function is dependent a solution to Adler's equation and it describes the phase pull with respect to time. The functions that describe the phase and amplitude are found with that is called the \textit{slow evolution equations}.  So, under that assumption, in order for us to actually propose a solution to eq. \ref{duffbeam}, we first have to find $a(\tau)$ and $\theta(\tau)$. These two functions can be found after solving two coupled differential equations that are established due to several assumptions made in Shoshani's article on the matter \cite{shoshani1}
\begin{equation} \label{amp}
    \dot{a} = \frac{2 \Gamma}{\pi} - \Tilde{\zeta} a - \Phi \sin{(\theta)}
\end{equation}
\begin{equation} \label{theta}
    \dot{\theta} = \Delta\omega + \frac{3 \omega _n}{8\omega_d} a^2 - \frac{\Phi}{a} \cos{(\theta)}
\end{equation}
where the dot operator denotes a derivative with respect to $\tau$, $\Delta \omega = 1 - \frac{\omega_n}{\omega _n}$ and $\Delta \omega \ll 1$. We can see that in the absence of the external signal ($\Phi = 0$), the closed-loop oscillator has a stable limit-cycle with an amplitude of $a_{ss} = \frac{2 \Gamma}{\pi \Tilde{\zeta} }$, which describes the condition for which the loop dynamics sustains the isolated oscillator - as mentioned earlier. Note that the resonator considered in this work is assumed to operate in an amplitude range for which a weakly nonlinear model holds, and while the resonator in the experimental device has a hardening nonlinearity (the Duffing parameter is larger than $0$), the analysis presented is also applicable for a resonator with a softening nonlinearity.
\subsection{Solution Layout}
The main idea in solving the equations in the case where an external forcing exists is relying on the assumption that the external signal is small in comparison with the closed-loop gain . Thus, we can assume that the perturbed amplitude is given by
\begin{equation} \label{amp3}
    a(\tau) = a_{ss} + \epsilon r(\tau) + O (\epsilon^2) = \frac{2 \Gamma}{\pi \Tilde{\zeta} } + \epsilon r(t) + O (\epsilon^2)
\end{equation}
where $\epsilon \ll 1$. Plugging into equation \ref{amp} we get the following governing equation for the amplitude perturbation away from $a_{ss}$
\begin{equation}
    \dot{r} = -\Tilde{\zeta} r - \frac{\Phi}{\epsilon} \sin(\theta)
\end{equation}
One can see that in contrast to the phase dynamics, dissipation is explicitly present in the amplitude dynamics. Hence, perturbations of the amplitude decay relatively rapidly. This point can also be seen from the free-running oscillator equations. Consequently, one can assume that the amplitude is following the phase,
meaning $\dot{r} \approx 0$. Using this result, one gets
\begin{equation}
    r(\tau) = - \frac{\Phi}{\epsilon \Tilde{\zeta}} \sin(\theta)
\end{equation}
Plugging into eq. \ref{amp3} we get the perturbed amplitude
\begin{equation} \label{amp4}
    a(\tau) = \frac{2 \Gamma}{\pi \Tilde{\zeta} } - \frac{\Phi}{\Tilde{\zeta}} \sin(\theta) + O (\epsilon^2)
\end{equation}
From here on out, we define the biased oscillator phase as $\phi = \theta + \theta_0$ where $\tan{(\theta_0)} = \frac{1}{3} \frac{\omega _d}{\omega _n} \frac{\Tilde{\zeta} ^3 \pi ^2}{\Gamma ^2 }$, which represents the phase difference between the external signal and the oscillator, as elaborated before. Plugging eq. \ref{amp4} into eq. \ref{theta}, neglecting terms of $O(\epsilon^2)$ and retaining terms up to $O(\epsilon)$ yields an approximated Adler equation for $\phi$ given by
\begin{equation}
    \dot{\phi} \approx \Delta \omega ' - B \sin{\phi}
\end{equation} 
where
\begin{itemize}
\item{\makebox{$\Delta \omega ' = \Delta \omega + \frac{3}{2} \frac{\omega _n}{\omega _d} \frac{\Gamma ^2}{\pi ^2 \Tilde{\zeta}}$} is the frequency offset}
\item{\makebox{$B = \frac{\Phi}{a_{ss}} \bigg[1 + \Big( 3 \frac{\omega _n}{\omega _d} \frac{\Gamma ^2}{\Tilde{\zeta} ^3 \pi ^2} \Big)^2 \bigg] ^ {\frac{1}{2}} = \Phi \frac{\Tilde{\zeta} \pi}{2 \Gamma} \sqrt{1 + cot^2(\theta_0)} = \frac{\Phi}{sin(\theta_0)} \frac{\Tilde{\zeta} \pi}{2 \Gamma}$} \\ is the mean phase modulation amplitude}
\end{itemize}
So after much work this problem reduces to solving the Adler equation.
\subsection{Adler Equation's Potential Function}
The point of interest in near the bifurcation, that occurs when $\abs{K} = \abs{\frac{B}{\Delta \omega '}} = 1$, from the direction where $\abs{\frac{B}{\Delta \omega '}} > 1$, where there's no phase locking and harmonic beats occur. So we look for points that live up to
\begin{equation} \label{lim}
    \lim_{\abs{\frac{B}{\Delta \omega '}} \to 1^+} \phi(t)
\end{equation}
Those points have a specific characteristic that make the beam enter SNIC bifurcation and produce the desired frequency comb. Adler's equation can be seen as a dynamics equation where the mass term was neglected. Let us rewrite it in the form
\begin{equation} \label{duffre}
    \epsilon \ddot{\phi} + \dot{\phi} = \Delta \omega'- B\sin{\phi}
\end{equation}
where $\epsilon \ll \frac{1}{B}$. One can see that the RHS corresponds to the term of the sum of the forces in Newton's Second Law. Let us remember that the potential of a system is defined as
\begin{equation}
    U = - \int \sum_{i=1}^{n}\vec{F_i} \cdot \vec{dr} 
\end{equation}
Using this on the RHS in eq. \ref{duffre}
\begin{equation}
U(\phi) = - \int f(\phi) d\phi = - \int [\Delta \omega ' - B\sin{\phi}] d\phi = -\Delta \omega ' \phi - B\cos{\phi}
\end{equation}
\begin{figure}[!h] 
\centering
\includegraphics[scale=0.7]{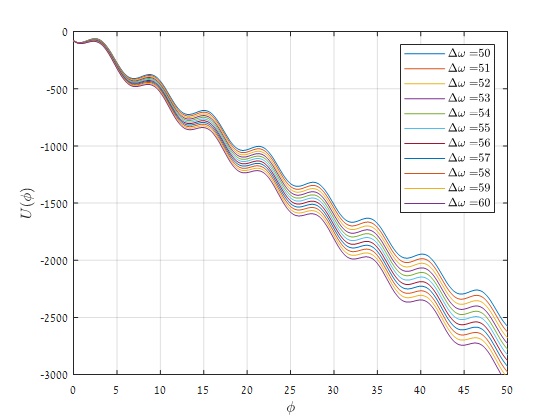}
\caption{Illustration of washboard potential for different values of $K$ and $B = 80$} \label{wash}
\end{figure}
This result is what is knows as a \textit{"washboard potential"} for it is a function that has a harmonic term (usually due to gravity in potential function) and an additional linear term, both being negative - so as the constant $\Delta \omega '$ increases, the graph of the function rotates clockwise and therefore resembles an inclined washboard, while the coefficient of the cosine function - $B$ determines the amplitude of the "wave" of the function.  Figure \ref{wash} illustrates the behaviour of the washboard potential function
As one can see, these two parameters define $K$. The analogy to a washboard rests in that as the slope of the washboard potential is not very steep, a particle flowing through it can be trapped in one of the local minima in the washboard potential. 
In our discussion this is analogous to a state of injection locking - as in one oscillator can lock into another's frequency and therefore be "trapped" by it. As the slope increases (meaning it becomes more steep) - the "flow" down the potential curve becomes more evident, which means physically that a there's not enough coupling to trap the oscillator and it can oscillate in its frequency freely. We look for a bifurcation from values of $K$ that obey equation \ref{lim} - we wish to look for smaller and smaller values of $K$ from values larger than 1 - that correspond to greater values of $B$ and more importantly - smaller values of $\Delta \omega '$ - meaning smaller values of the two oscillators frequency difference. 
\begin{figure}[!h] 
\centering
\includegraphics[scale=0.35]{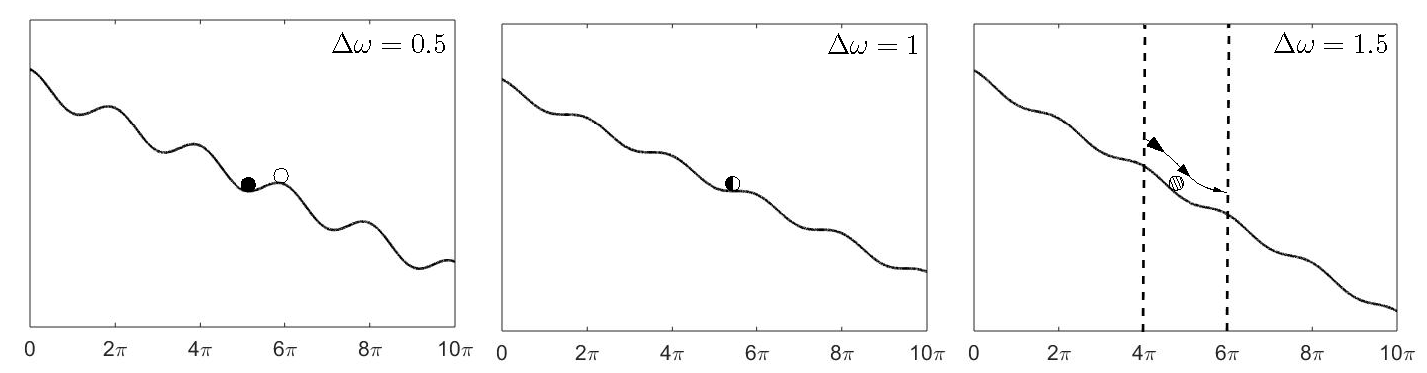}
\caption{Illustration of the Three Washboard Points} \label{wash2}
\end{figure}
An additional illustration of the washboard is given in figure \ref{wash2} - where the three points of bifurcation appear clearly - the first where $\Delta \omega ' < 1$ and there are two equilibrium values, one stable (the black one) and one not (the white one), the point where both values coincide and create one quasi-stable value that incites the bifurcation, and a third where there's no equilibrium value, the coupling decreases and the oscillator oscillates freely. The code for the washboard potential simulation is given in appendix \ref{washboard_code}.

\subsection{Solving Adler's Equation}
The equation is an ordinary differential non-linear equation that can be solved numerically, though an analytical solution is surprisingly possible. The layout for the solution is added to appendix no. 3. The final solution for the phase shift $\phi$ is of the form
\begin{equation}
    \phi(t) = 2\arctan{\Bigg[\frac{1+\sqrt{K^2-1}\tan{\big[\frac{B\sqrt{K^2-1}}{2}}(t-t_0)\big]}{K} \Bigg]}
\end{equation}
When $t_0$ is the integration constant. This equation in fact, describes the phase rotation of the oscillator at a given instant. Simplifying a third assumption made by Adler, the condition for synchronization can be expressed mathematically as $\abs{K} < 1$.
\section{Simulations}
In this section, we perform numeric simulations on the equations detailed in earlier sections. There are many equations that were mentioned and we can perform numerous analyses on each one of them, but we'd like to ultimately concentrate on the desired result - being the graphical solution producing the frequency comb. So we'd like to simulate several equations we derived in the previous chapters. These simulations will be done gradually to ultimately lay the foundations for the extraction of the frequency comb. All of the simulations were done using the MATLAB\textsuperscript \textregistered  \ numerical computing environment and its add-on interface Simulink \textsuperscript \textregistered. All of the codes were added as appendices.
\subsection{The Beam's Spatial Function}
A mentioned earlier, the beam's displacement is described by a multiplication of a temporal function of time and and another spatial function of the spatial coordinate $x$. The function of $x$ was fully determined for the first mode and was shown in equation \ref{final_spatial}, where a change of variables described the function as of a variable $s$. A simulation of the function was conduced and the results are shown in figure \ref{X(s)}. The code is added in appendix \ref{spatial_code}.
\begin{figure}[h]
\centering
\includegraphics[scale=0.37]{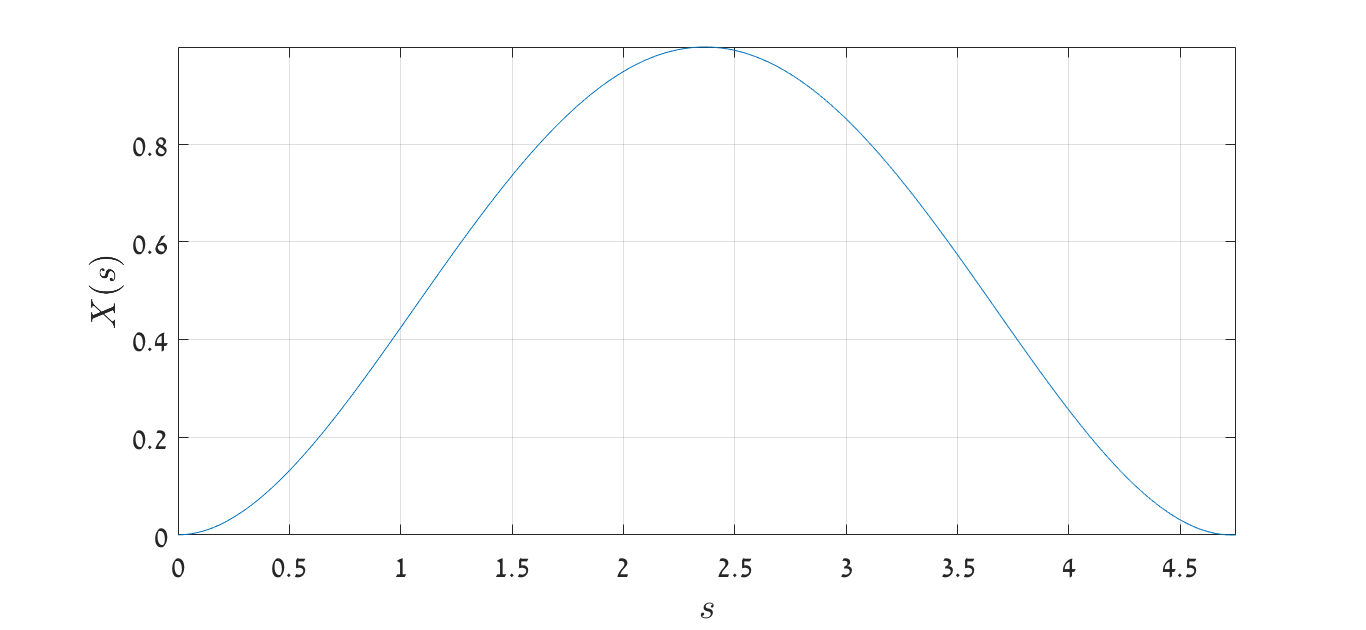}
\caption[The Spatial Function X(s)]{The Spatial Function X(s)}
\label{X(s)}
\end{figure}
\subsection{The Beam's Temporal Function}
\begin{figure}[h]
\centering
\includegraphics[scale=0.5]{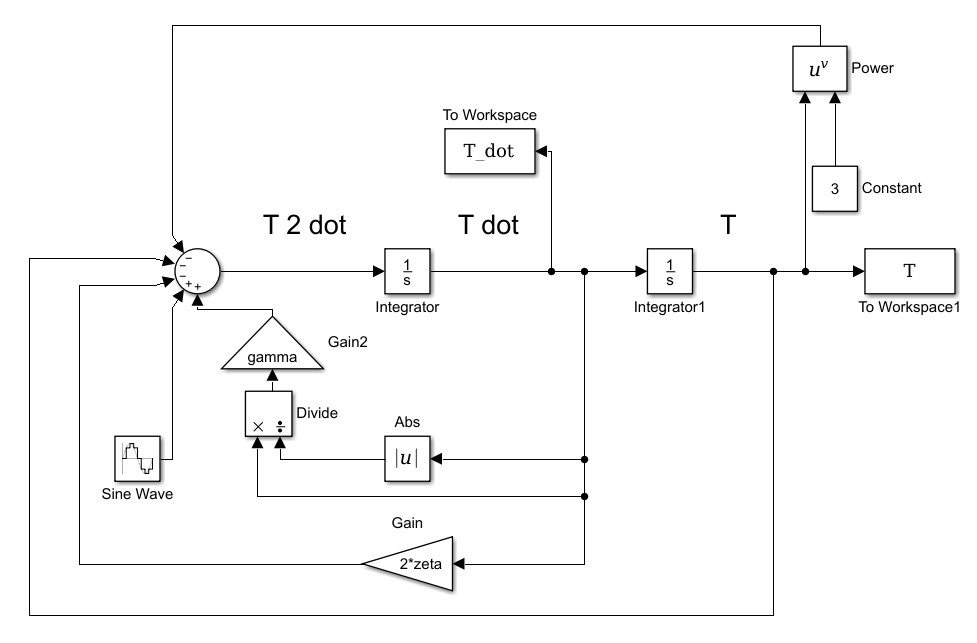}
\caption[Simulink Model of the Duffing Equation]{Simulink Model of the Duffing Equation}
\label{simulink_model}
\end{figure}

\begin{figure}[h]
\centering
\includegraphics[scale=0.36]{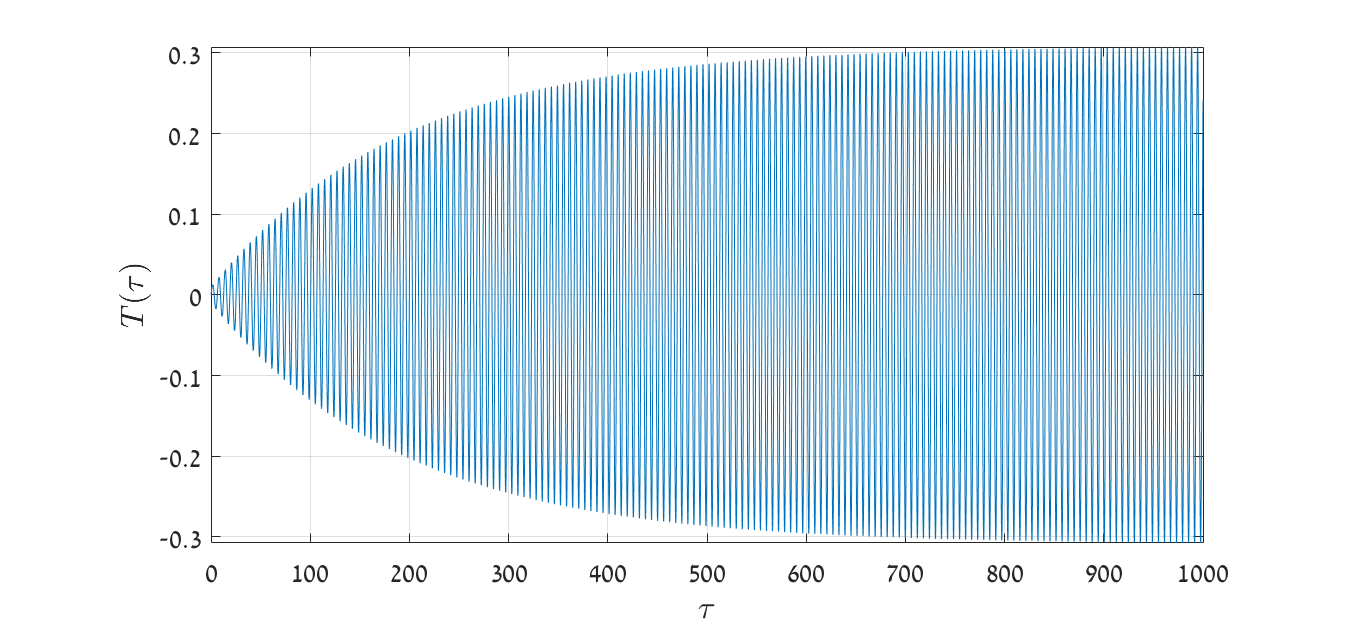}
\caption[Simulation for $T(\tau)$ Steady State Amplitude]{Simulation for $T(\tau)$ Steady State Amplitude}
\label{a_ss_sim}
\end{figure}

The temporal dynamics of the beam was derived up to a final normalized equation for the dimensionless time $\tau$ shown in equation \ref{final_duff_norm_with_force}. This equation can be represented by a Simulink model as illustrated in figure \ref{simulink_model}.
Note that the "dot" notation denotes a derivative with respect to the normalized time $\tau$. This model can actually solve the non-linear Duffing equation numerically and produce the function $T$ as a function of the normalized time $\tau$. This simulation purpose is to verify that the control function as mentioned does restrain the system to the steady-state amplitude and causes it to vibrate steadily after sufficient amount of time. To visualize this, a low quality factor was chosen to ensure large damping and quick convergence to the steady-state. Furthermore, as explained in the hypothesis, this model was built on the assumption that the dominant term in the Duffing equation is the linear term, thus the non-linear terms in the equation are in charge of applying the wanted tweaks to make the system behave as wanted. Due to this assumption, we can say that the dynamics of the actual beam with respect to the dimensional time will be similar to what the dimensionless model depicts, up to a multiplication by some constant. Mathematically, this assumption leads to the conclusion that the following relation must hold
\begin{equation}
    T^3(\tau) \ll T(\tau)
\end{equation}
Due to this conclusion, we must have that the steady-state is smaller than $1$. Therefore the control effort was chosen as $\Gamma = \frac{1}{4Q}$. By the relation given in equation \ref{ss_amp} the steady-state amplitude has to be
\begin{equation}
    a_{ss} = \frac{2 \Gamma}{\pi \Tilde{\zeta} } \underset{\Tilde{\zeta}= \frac{1}{2Q}}{=} \frac{4 \Gamma Q}{\pi} \underset{\Gamma = \frac{1}{4Q}}{=} \frac{1}{\pi} \approx 0.318
\end{equation}
The simulation code is given in appendix \ref{time_code_ss} and the results are shown in figure \ref{a_ss_sim}.
\subsection{The Beam's Dynamics Phase Portrait}
The model of the temporal function for the beam performed two integration operations on the second derivative of the time function, so we got the function itself and its derivative with respect to $\tau$. Note that the beam was given an initial perturbation it's velocity with a value of $0.01$ and the initial position was set to zero. With these two function in time, we can look at the phase portrait of the system, hoping that it will display the limit cycle we wish for. The phase portrait of the system is displayed in figure \ref{phase_portrait} and the code is displayed in appendix \ref{phase_code}.
\newpage
\begin{figure}[!h]
\centering
\includegraphics[scale=0.37]{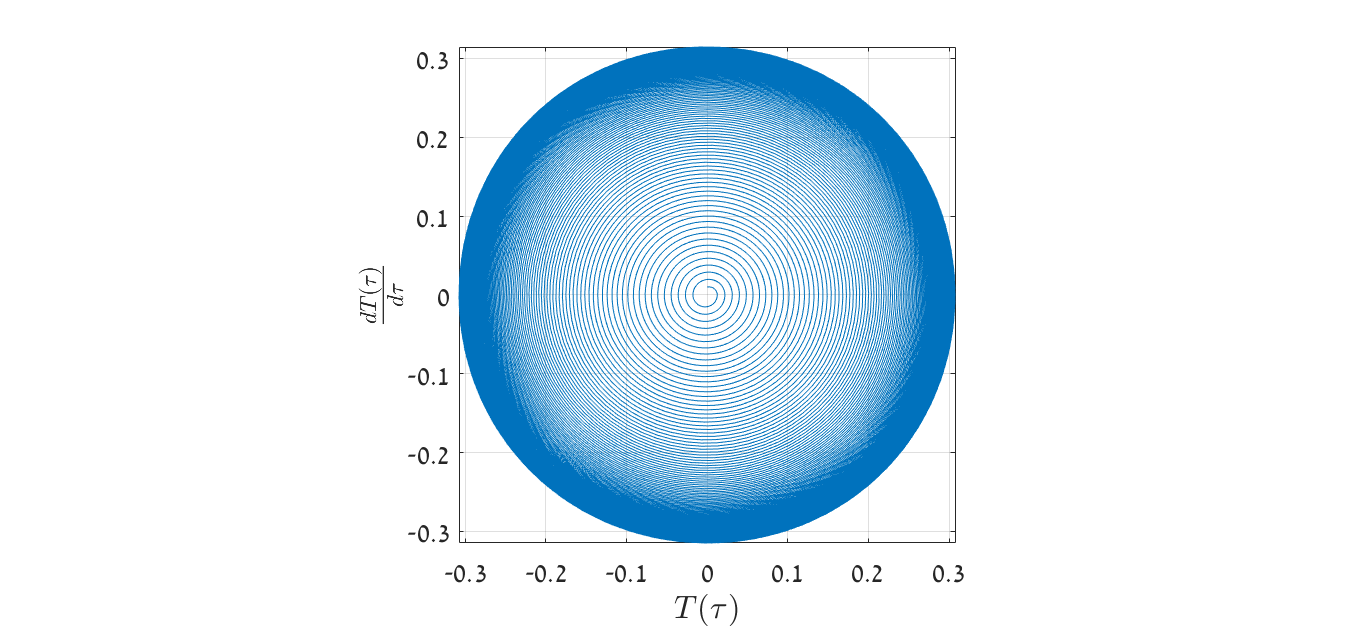}
\caption[Phase Portrait of the System]{Phase Portrait of the System}
\label{phase_portrait}
\end{figure}
\subsection{The Beam's Complete Vibration}
After obtaining both the spatial and temporal functions of the vibration of the beam, these functions can then be multiplied to result in the complete function of vibrations for the beam. Since it is a function of two variables, simulation is possible by extraction of an animation of the beam. This is done by iteratively plotting frames of the beam's location on the spatial coordinate and running through the simulation time. This animation is shown in the presentation of this project while a sample is shown in figure \ref{complete_vibartion} and the code is given in appendix \ref{complete_sim_code}.
\begin{figure}[!h]
\centering
\includegraphics[scale=0.41]{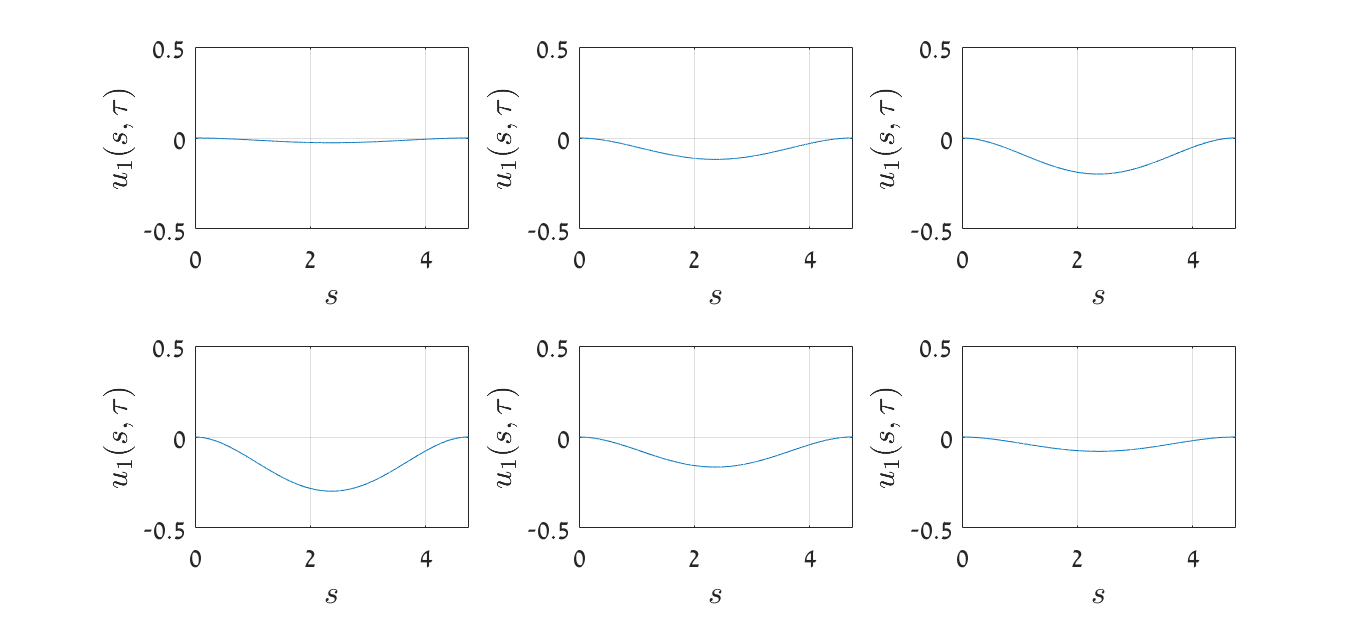}
\caption[Complete Vibration of the Beam]{A Sample of Half of a Complete Vibration of the Beam}
\label{complete_vibartion}
\end{figure}
\subsection{The Frequency Comb}
Generation of the frequency comb can be achieved by solving Adler's equation to get the phase as a function of time and then calculate the solution's \textit{Fourier Transform}, that transforms the function to the complex plane, in order to obtain the gain of the phase as a function of frequency. Then, in the range of values keeping up with the condition equation \ref{lim} defines, the comb will hopefully appear. We will perform numerical simulations in order to try and obtain actual solutions to the model of the beam (with forcing and control) that was presented and then check their correlation to the results obtained in this section, by actually deriving the equation for the vibration of the beam. Using the solution to Adler's equation, one can experiment using actual analytical analysis. In the context of the assumption made earlier, several simulations were made in order to witness the desired behaviour graphically. A Matlab function was written for this purpose, which is added to appendix no. \ref{adler_code}. The function allows to run multiple simulations for different values of $K$ and $B$. In this report, simulation results for 3 values of $B$ - 3,4 and 5 (which were chosen so as we assumed that the $B$ parameter was relatively small and in the range of a few units) were taken. The $K$ parameter was checked in $6$ different sets of values, presented in table no. \ref{Table of values for K}. Each iteration produced a graph for the phase $\phi (t)$ as a function of time, where the time series was taken in a interval of $100$ seconds, divided into increments of $0.1$ seconds. Finally, each iteration produced a graph for the absolute value of the \textit{Discrete Time Fast Fourier Transform} of the phase signal - as a function of frequency. This method utilizes an algorithm to compute the Fourier Transform for discrete samples in a fast way. The table of values and the first iteration are presented in the next page.

\bgroup
\def\arraystretch{1}
\begin{table}[!ht]
\footnotesize
\begin{center}
 \begin{tabular}{||c c||} 
 \hline
 Iteration & Values for $K$\\ [0.5ex] 
 \hline\hline
 $1$ & $\left\{10,9,8,7,6,5,4,3,2\right\}$\\ 
 \hline
  $2$ & $\left\{2,1.9,1.8,1.7,1.6,1.5,1.4,1.3,1.2\right\}$\\ 
 \hline
   $3$ & $\left\{1.1,1.09,1.08,1.07,1.06,1.05,1.04,1.03,1.02\right\}$\\ 
 \hline
   $4$ & $\left\{1.01,1.009,1.008,1.007,1.006,1.005,1.004,1.003,1.002\right\}$\\ 
 \hline
   $5$ & $\left\{1.001,1.0009,1.0008,1.0007,1.0006,1.0005,1.0004,1.0003,1.0002\right\}$\\ 
    \hline
   $6$ & $\left\{1.0001,1.00009,1.00008,1.00007,1.00006,1.00005,1.00004,1.00003,1.00002\right\}$\\
 \hline
\end{tabular} \
\caption{Table of values for $K$}

\label{Table of values for K}
\end{center} \
\end{table}\
\egroup

\newpage
\begin{center}

\subsubsection{Iterations for B = 3}
\hfill
\begin{figure} [!h]
    \centering
    \subfloat[Time Domain]{{\includegraphics[scale=0.20]{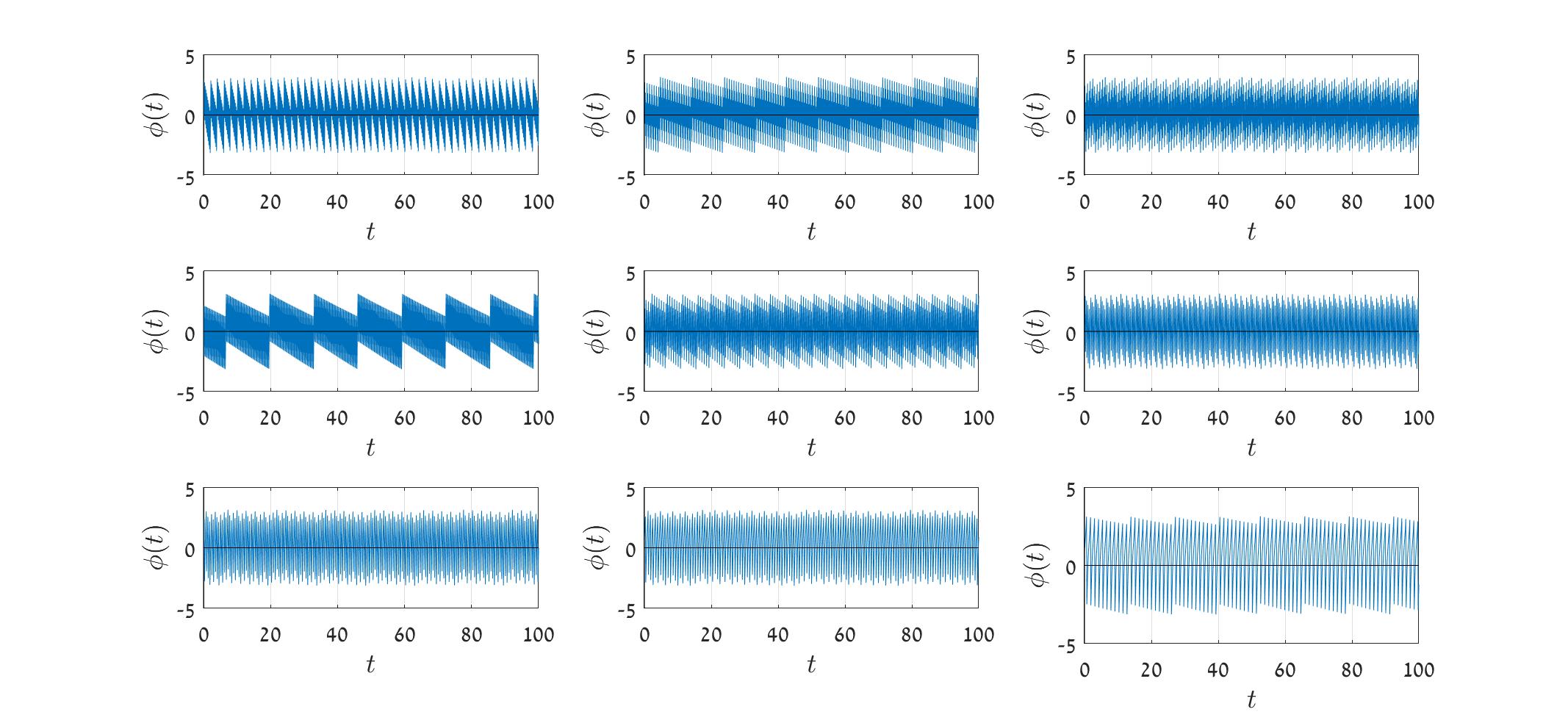}}}
    \qquad
    \subfloat[Frequency Domain]{{\includegraphics[scale=0.20]{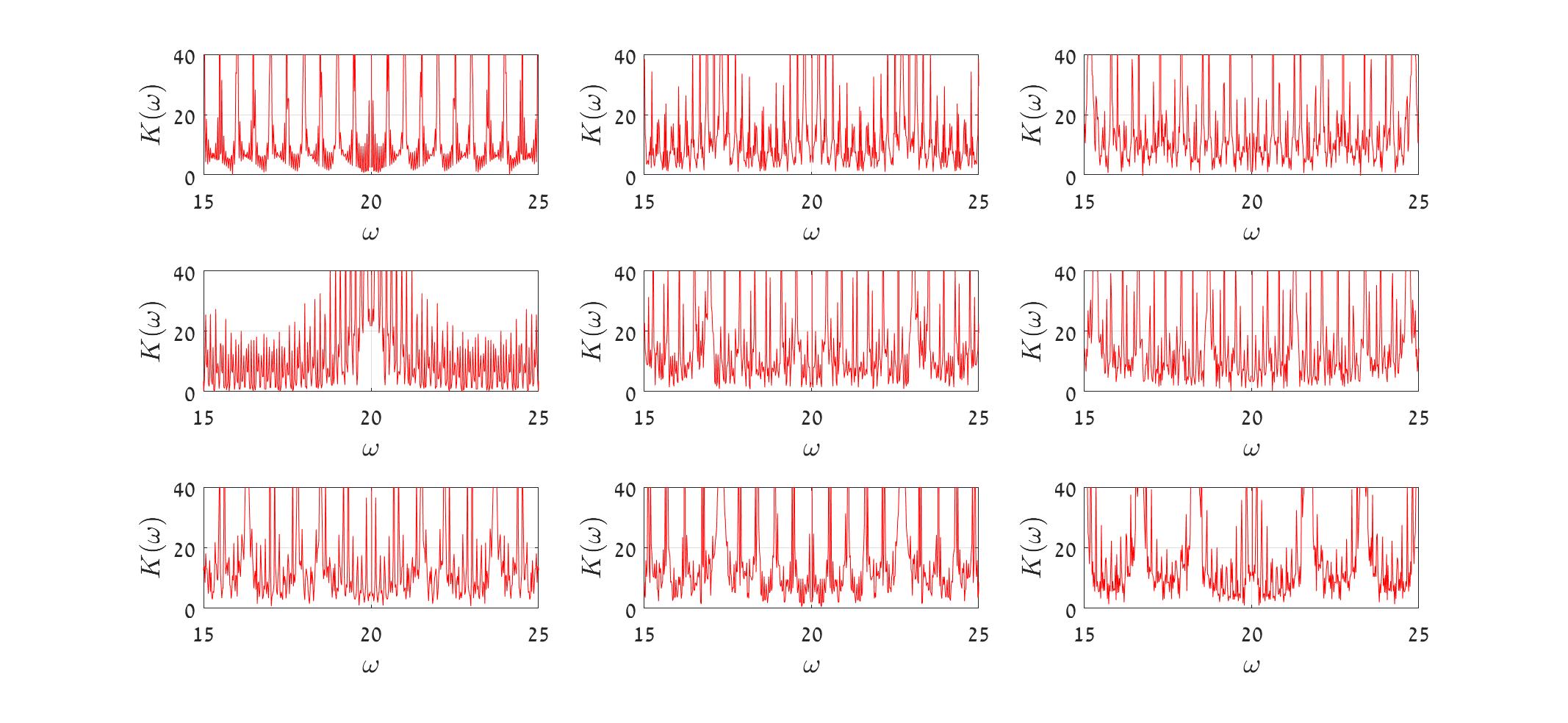}}}
    \caption{Iteration 1 for B = 3}
\end{figure}
\newpage
\hfill
\begin{figure} [!h]
    \centering
    \subfloat[Time Domain]{{\includegraphics[scale=0.20]{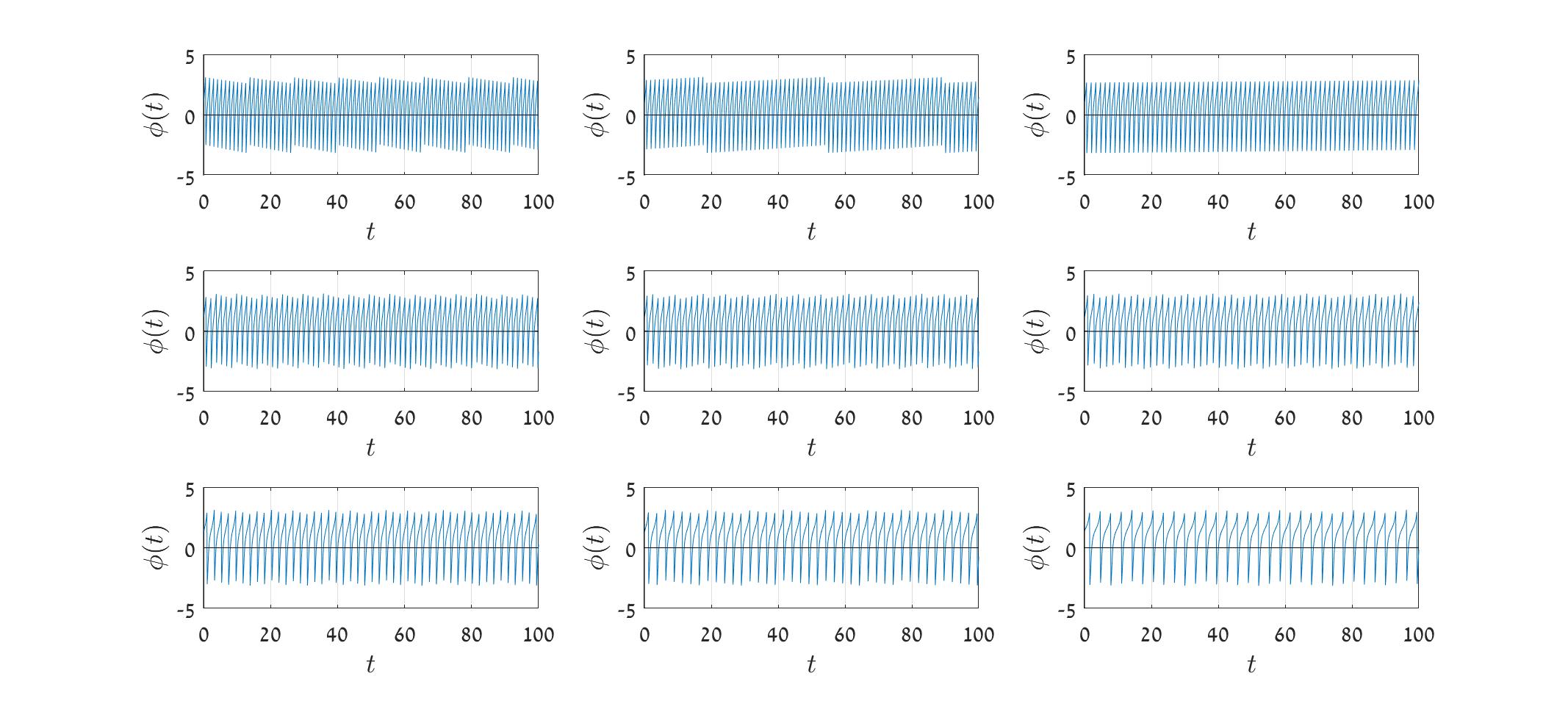}}}
    \qquad
    \subfloat[Frequency Domain]{{\includegraphics[scale=0.20]{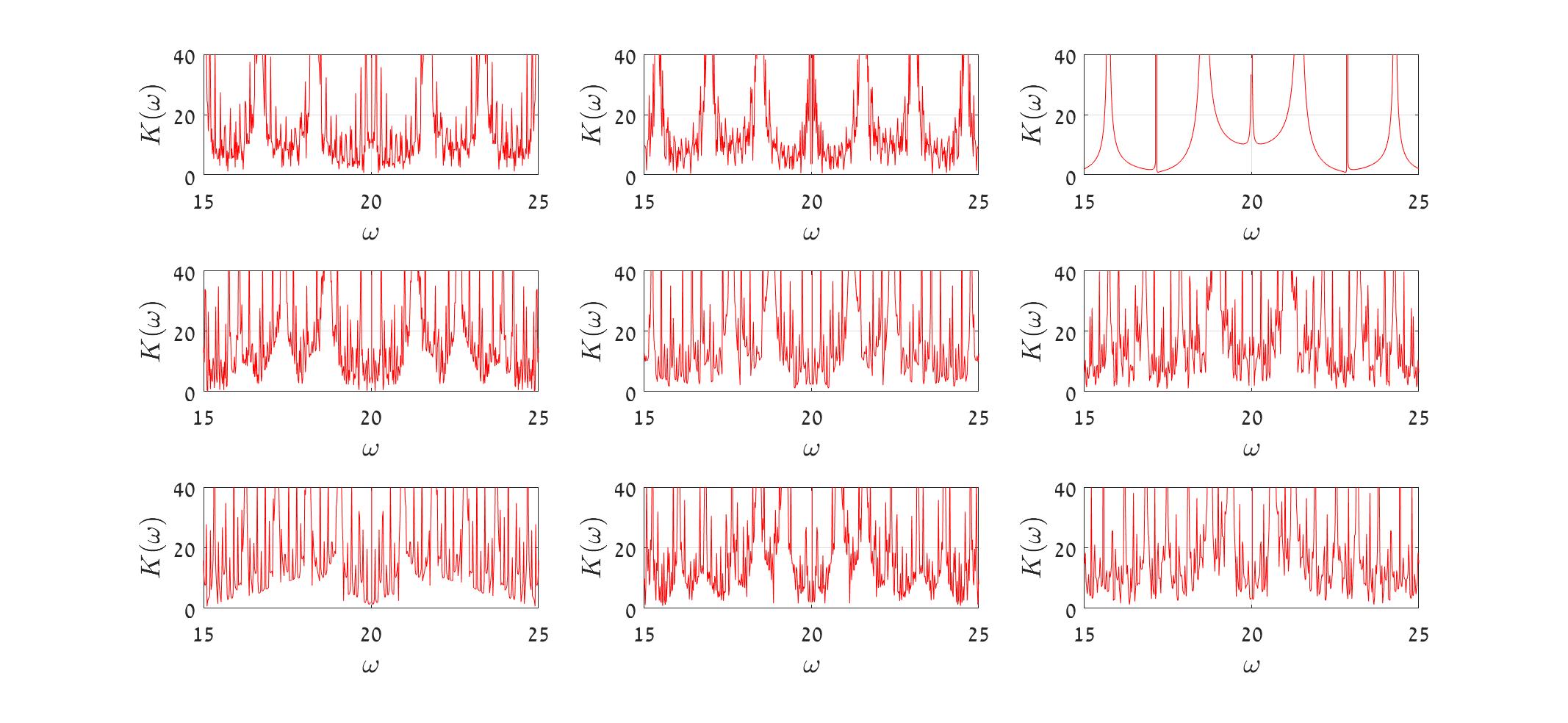}}}
    \caption{Iteration 2 for B = 3}
\end{figure}
\newpage
\hfill
\begin{figure} [!h]
    \centering
    \subfloat[Time Domain]{{\includegraphics[scale=0.20]{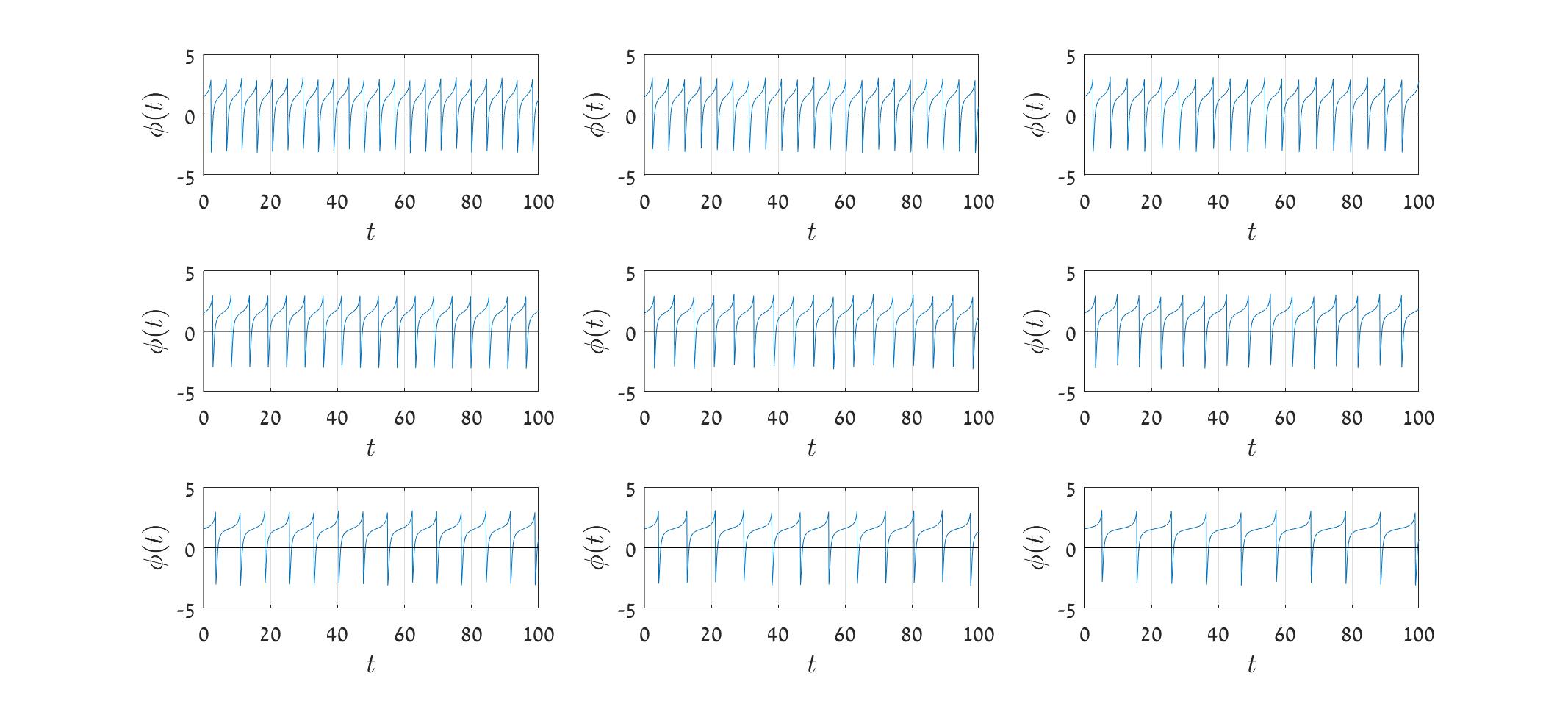}}}
    \qquad
    \subfloat[Frequency Domain]{{\includegraphics[scale=0.20]{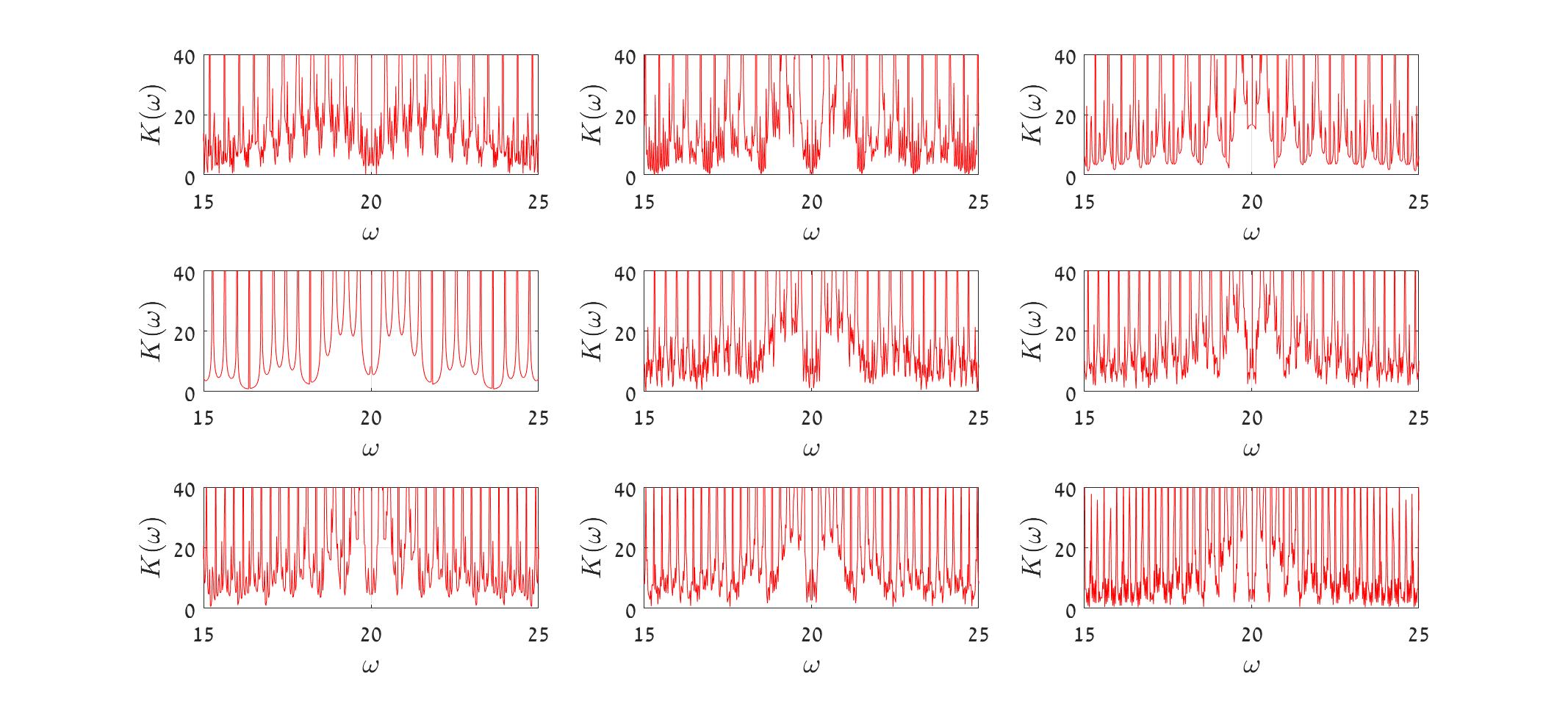}}}
    \caption{Iteration 3 for B = 3}
\end{figure}
\newpage
\hfill
\begin{figure} [!h]
    \centering
    \subfloat[Time Domain]{{\includegraphics[scale=0.20]{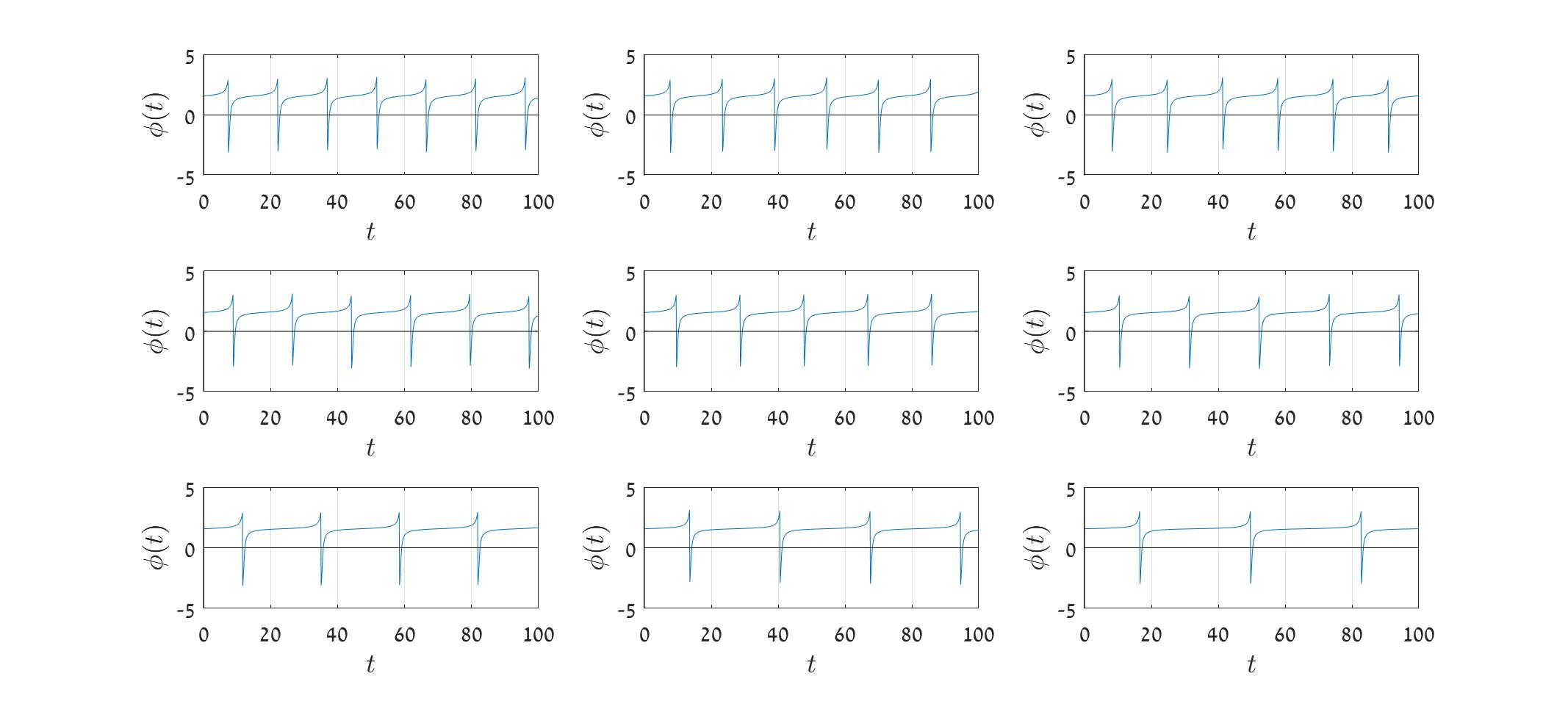}}}
    \qquad
    \subfloat[Frequency Domain]{{\includegraphics[scale=0.20]{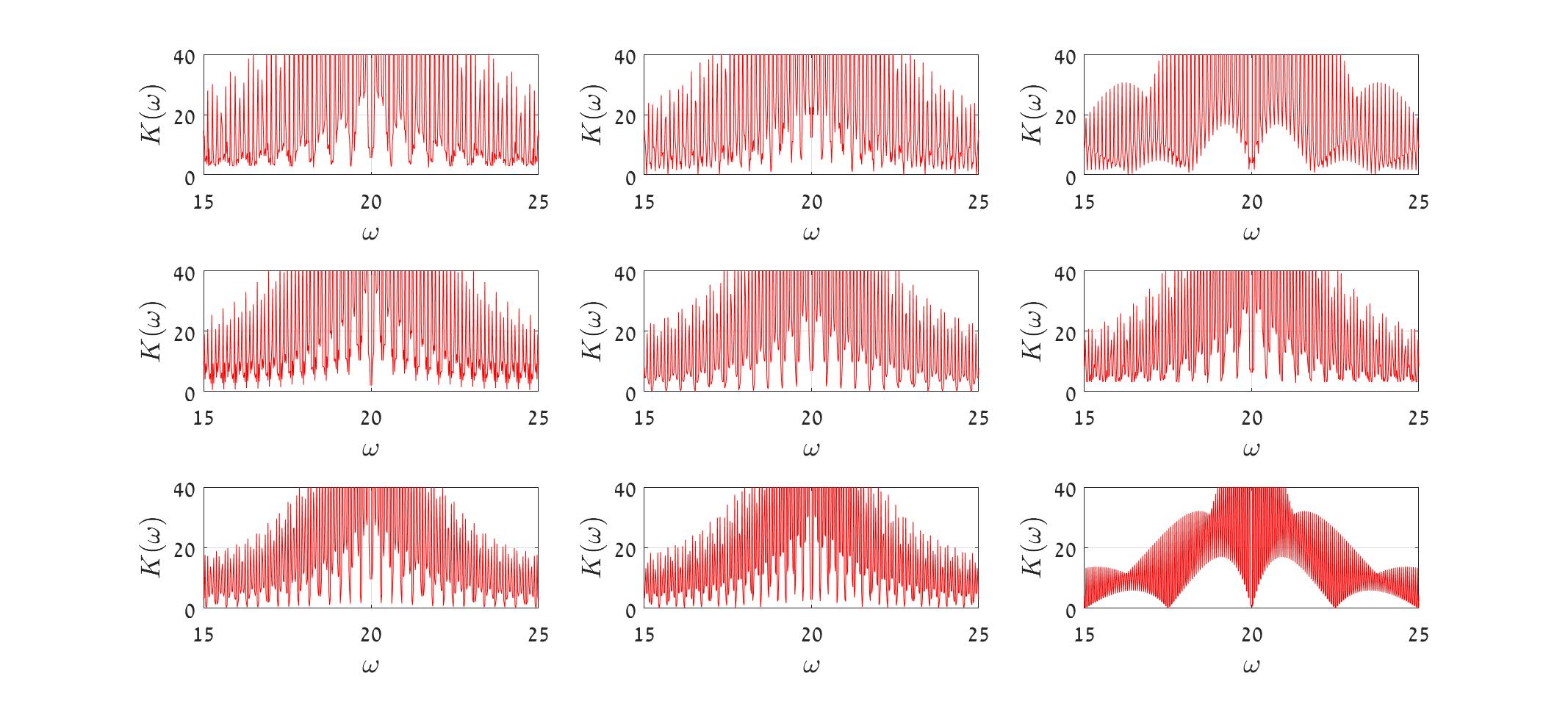}}}
    \caption{Iteration 4 for B = 3}
\end{figure}
\newpage
\hfill
\begin{figure} [!h]
    \centering
    \subfloat[Time Domain]{{\includegraphics[scale=0.20]{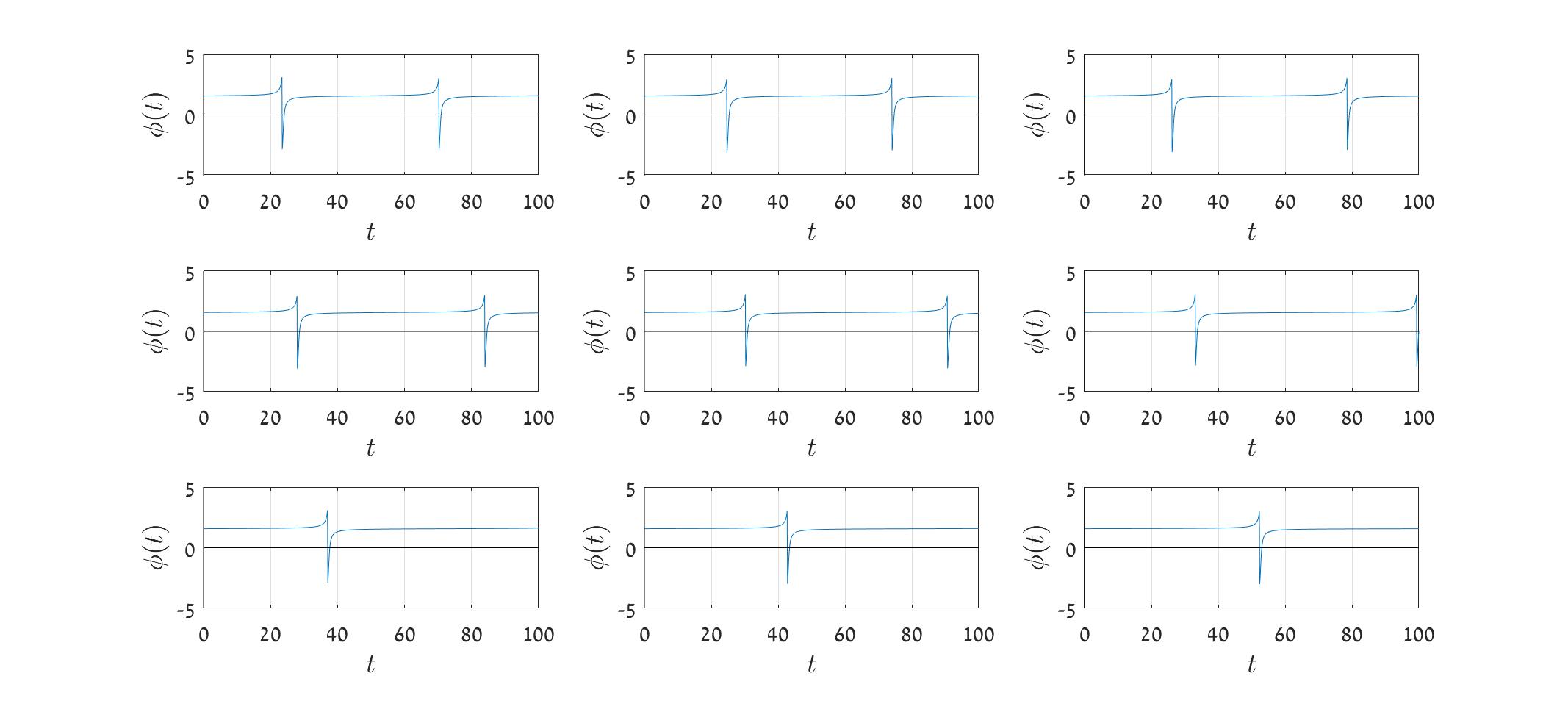}}}
    \qquad
    \subfloat[Frequency Domain]{{\includegraphics[scale=0.20]{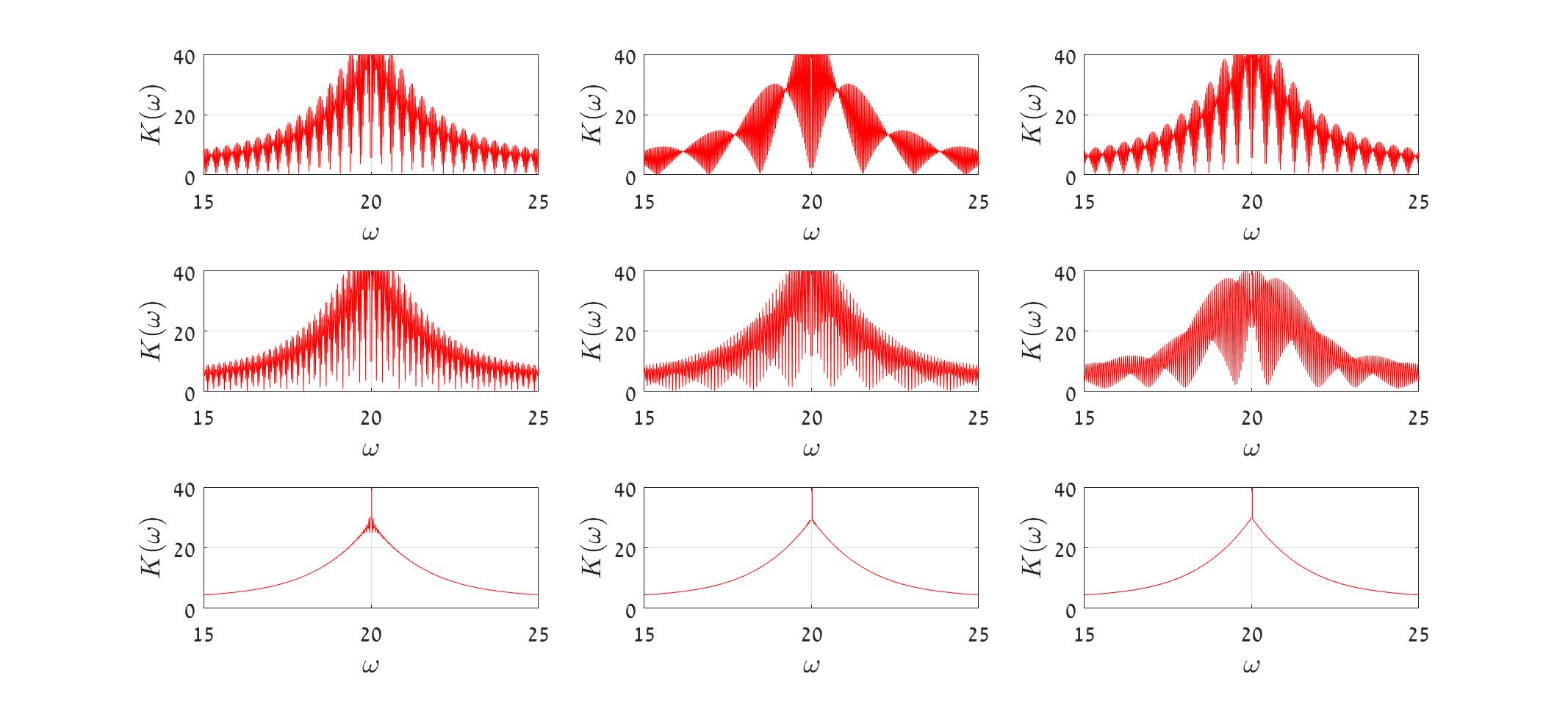}}}
    \caption{Iteration 5 for B = 3}
\end{figure}
\newpage
\hfill
\begin{figure} [!h]
    \centering
    \subfloat[Time Domain]{{\includegraphics[scale=0.20]{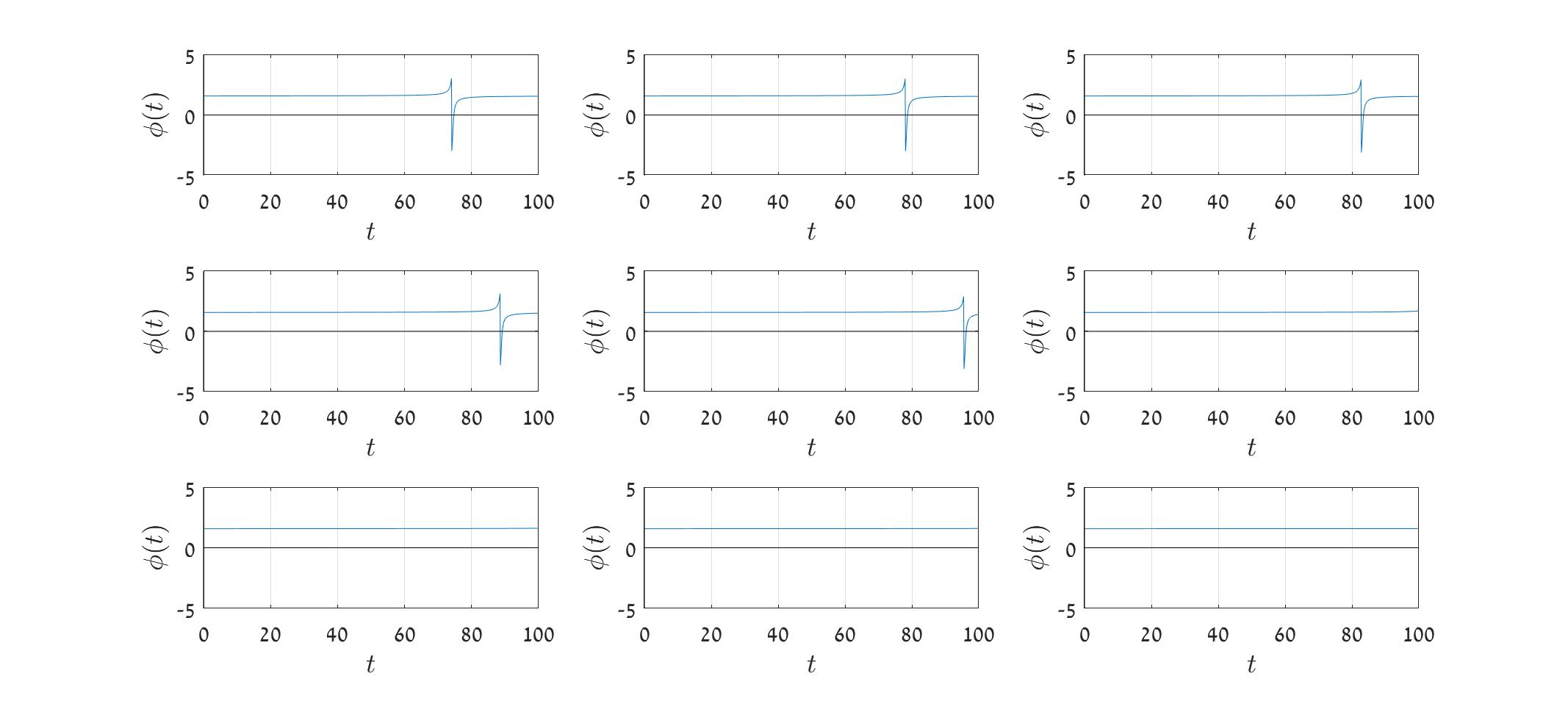}}}
    \qquad
    \subfloat[Frequency Domain]{{\includegraphics[scale=0.20]{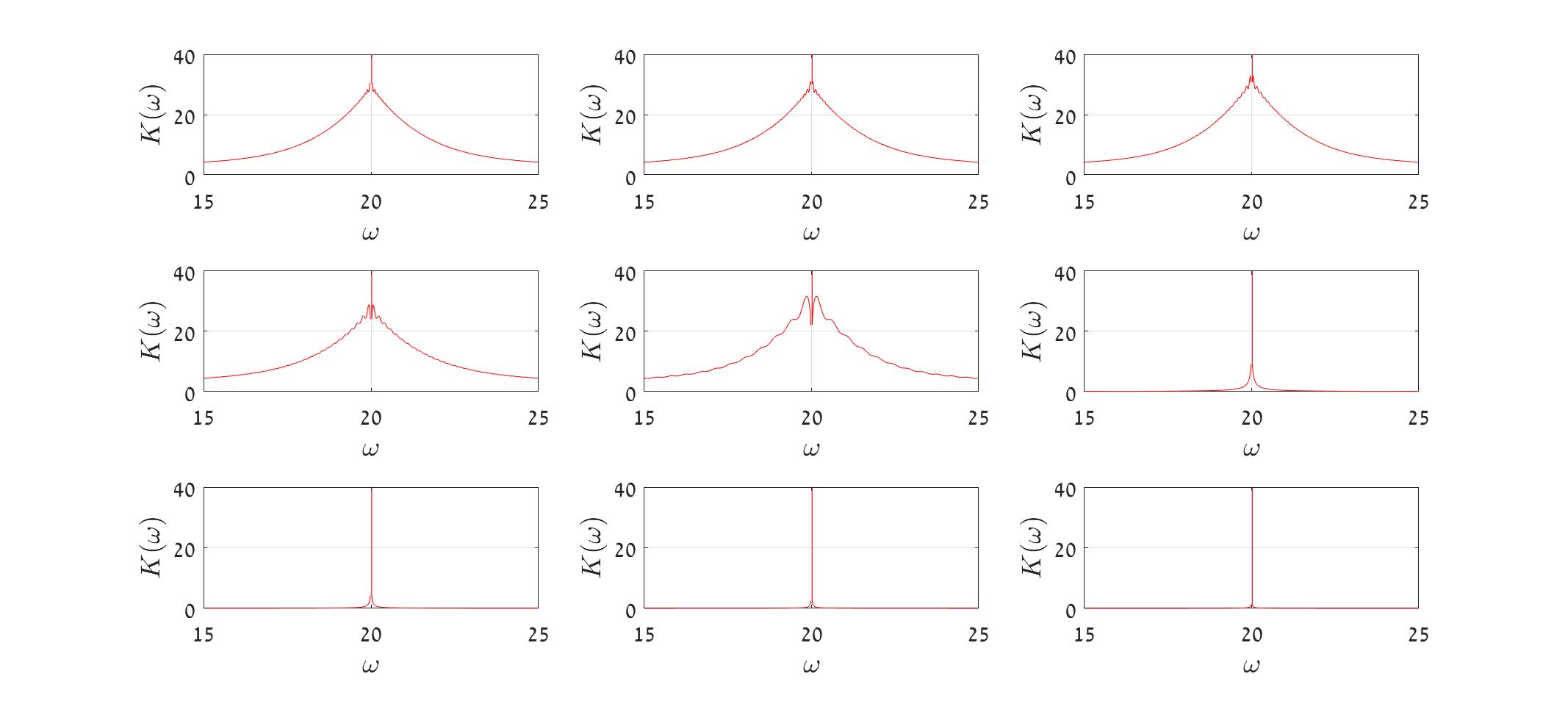}}}
    \caption{Iteration 6 for B = 3}
\end{figure}

\newpage
\subsubsection{Iterations for B = 4}
\hfill
\begin{figure} [!h]
    \centering
    \subfloat[Time Domain]{{\includegraphics[scale=0.20]{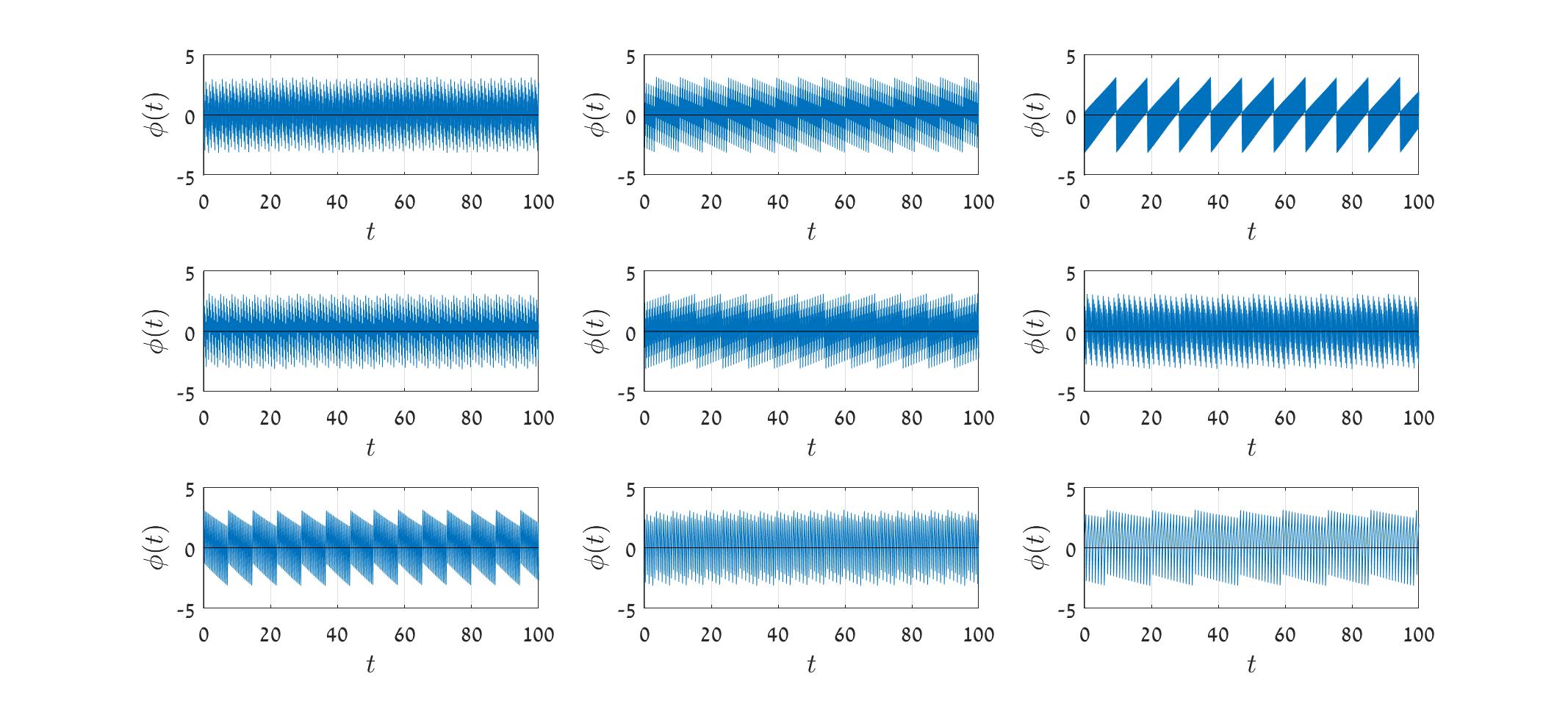}}}
    \qquad
    \subfloat[Frequency Domain]{{\includegraphics[scale=0.20]{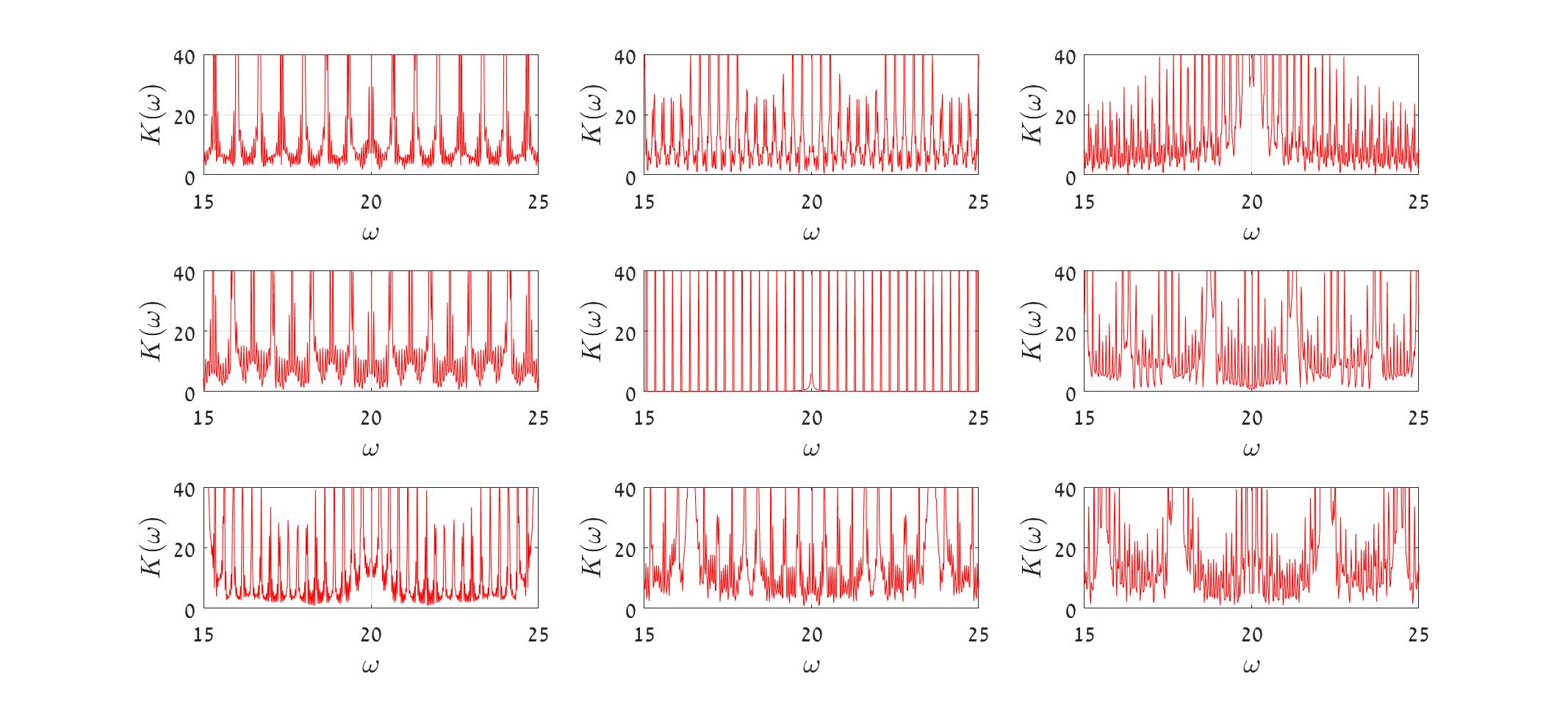}}}
    \caption{Iteration 1 for B = 4}
\end{figure}
\newpage
\hfill
\begin{figure} [!h]
    \centering
    \subfloat[Time Domain]{{\includegraphics[scale=0.20]{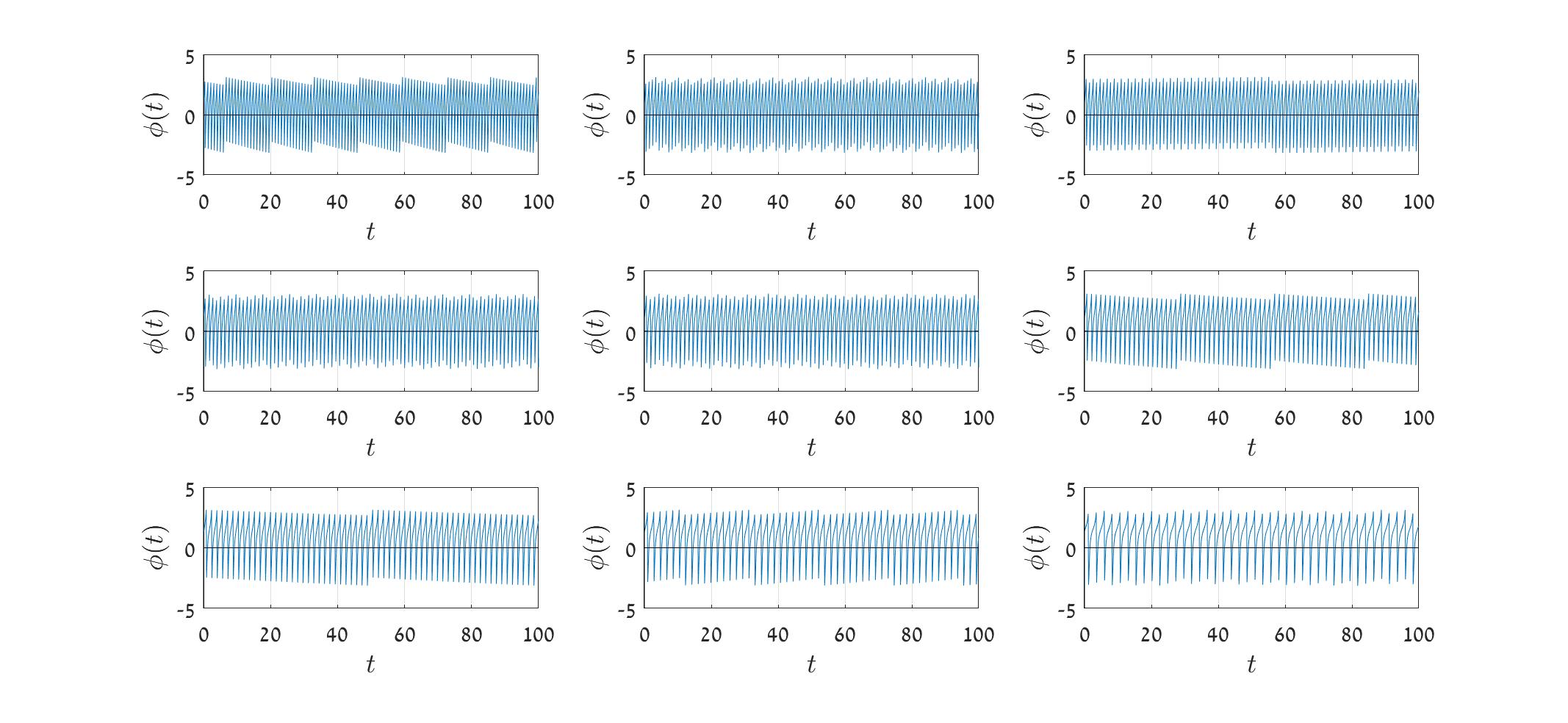}}}
    \qquad
    \subfloat[Frequency Domain]{{\includegraphics[scale=0.20]{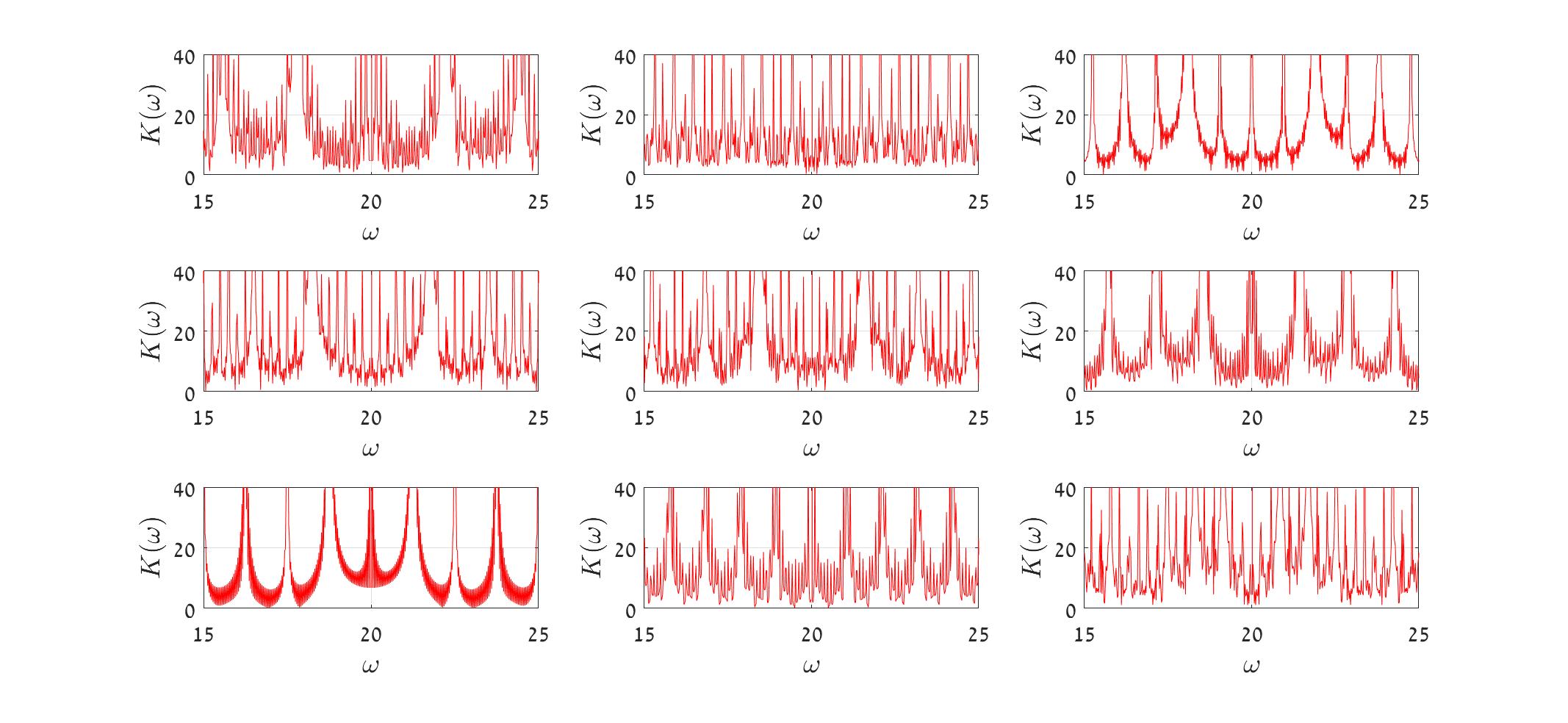}}}
    \caption{Iteration 2 for B = 4}
\end{figure}
\newpage
\hfill
\begin{figure} [!h]
    \centering
    \subfloat[Time Domain]{{\includegraphics[scale=0.20]{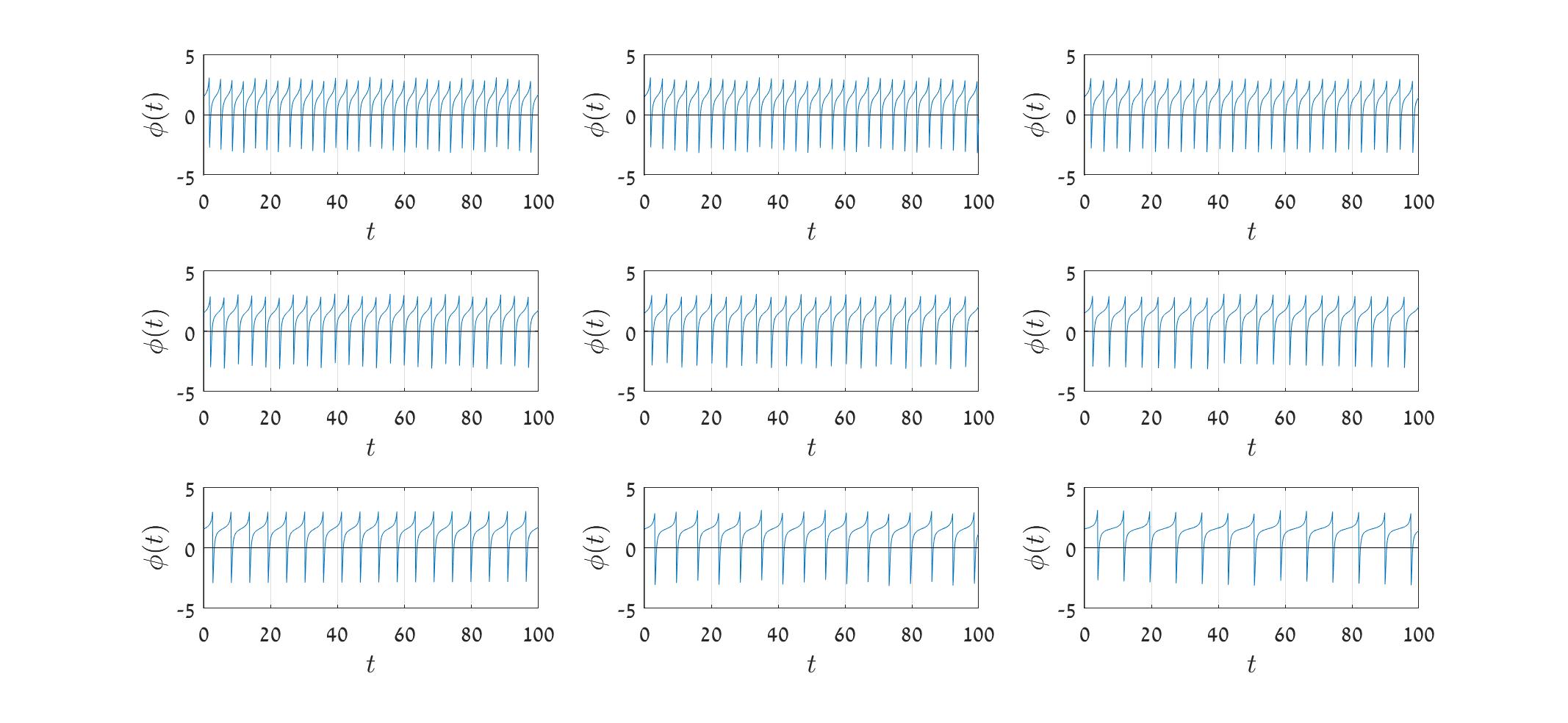}}}
    \qquad
    \subfloat[Frequency Domain]{{\includegraphics[scale=0.20]{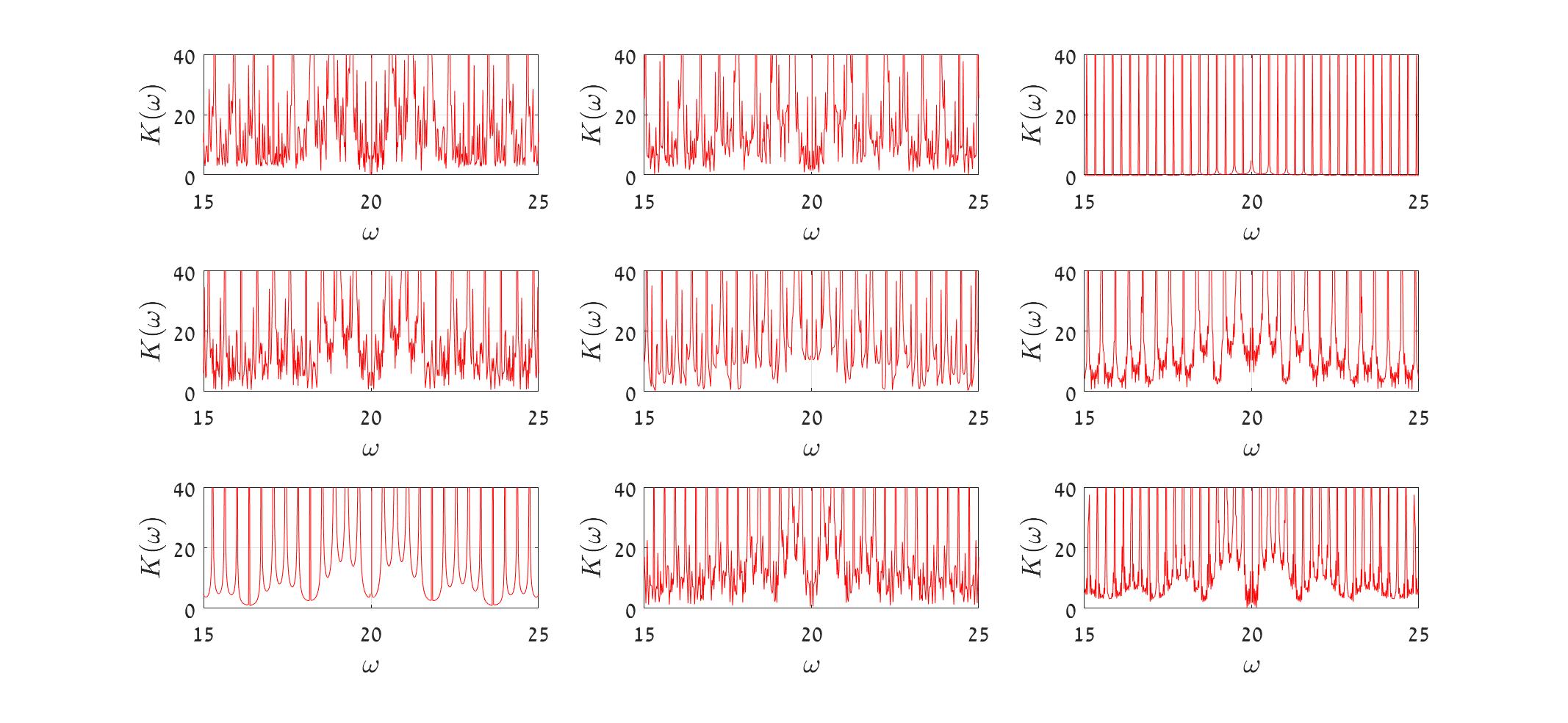}}}
    \caption{Iteration 3 for B = 4}
\end{figure}
\newpage
\hfill
\begin{figure} [!h]
    \centering
    \subfloat[Time Domain]{{\includegraphics[scale=0.20]{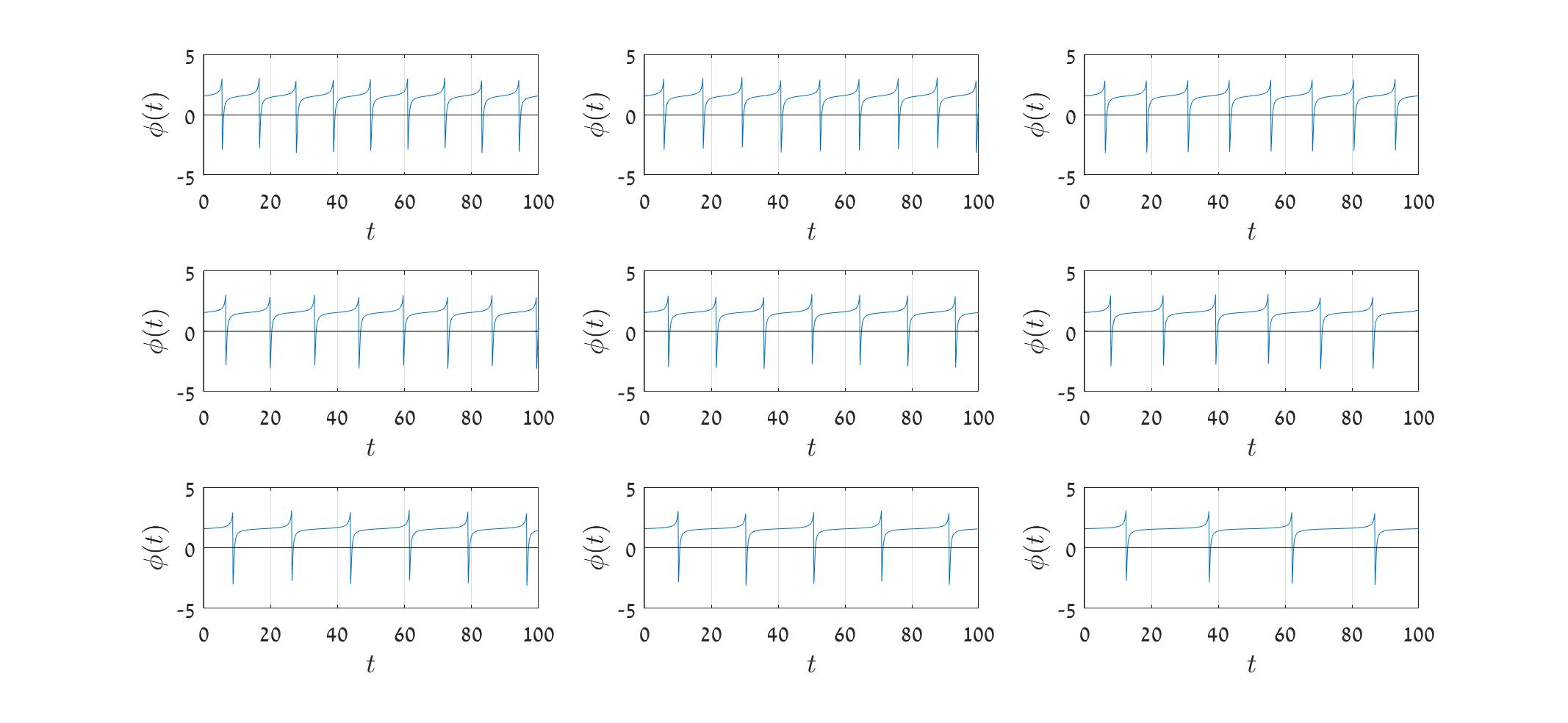}}}
    \qquad
    \subfloat[Frequency Domain]{{\includegraphics[scale=0.20]{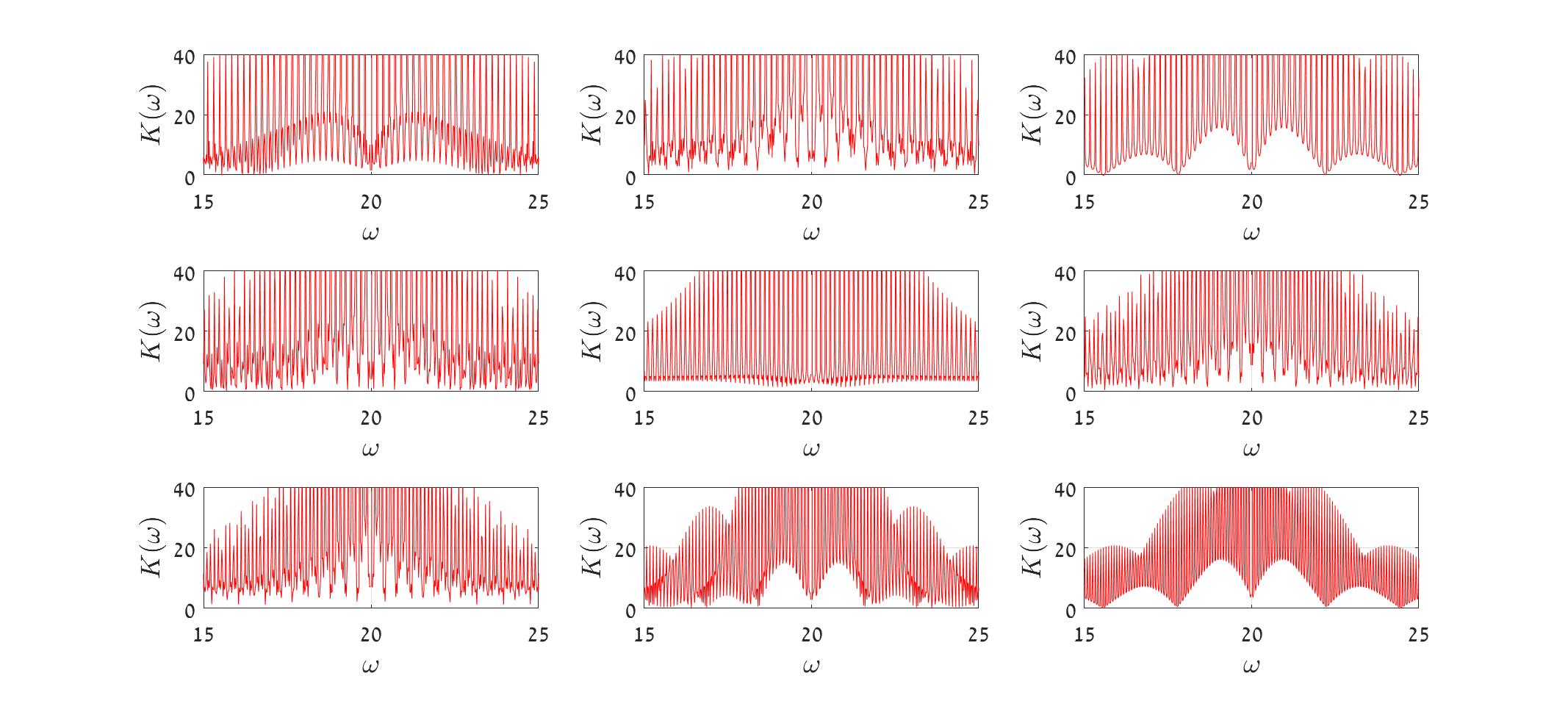}}}
    \caption{Iteration 4 for B = 4}
\end{figure}
\newpage
\hfill
\begin{figure} [!h]
    \centering
    \subfloat[Time Domain]{{\includegraphics[scale=0.20]{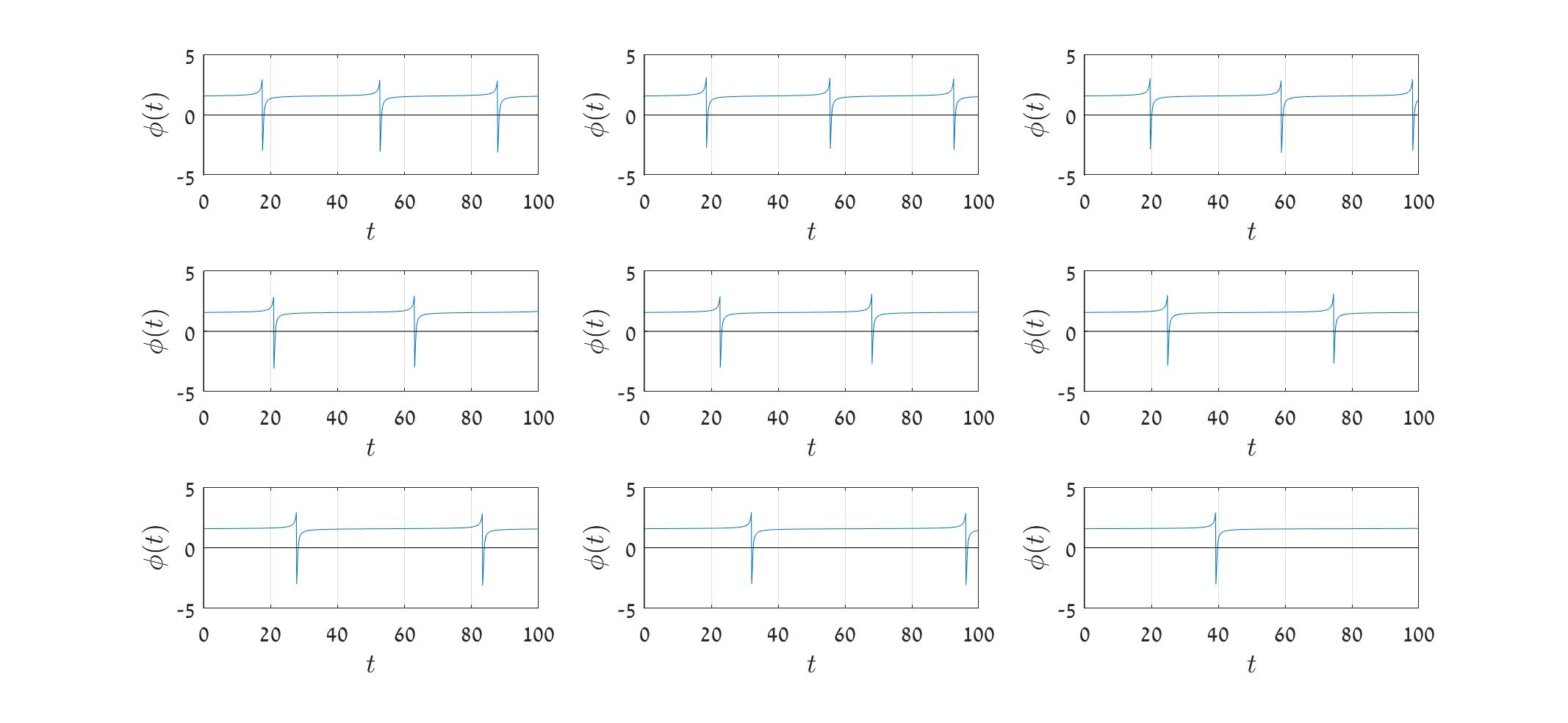}}}
    \qquad
    \subfloat[Frequency Domain]{{\includegraphics[scale=0.20]{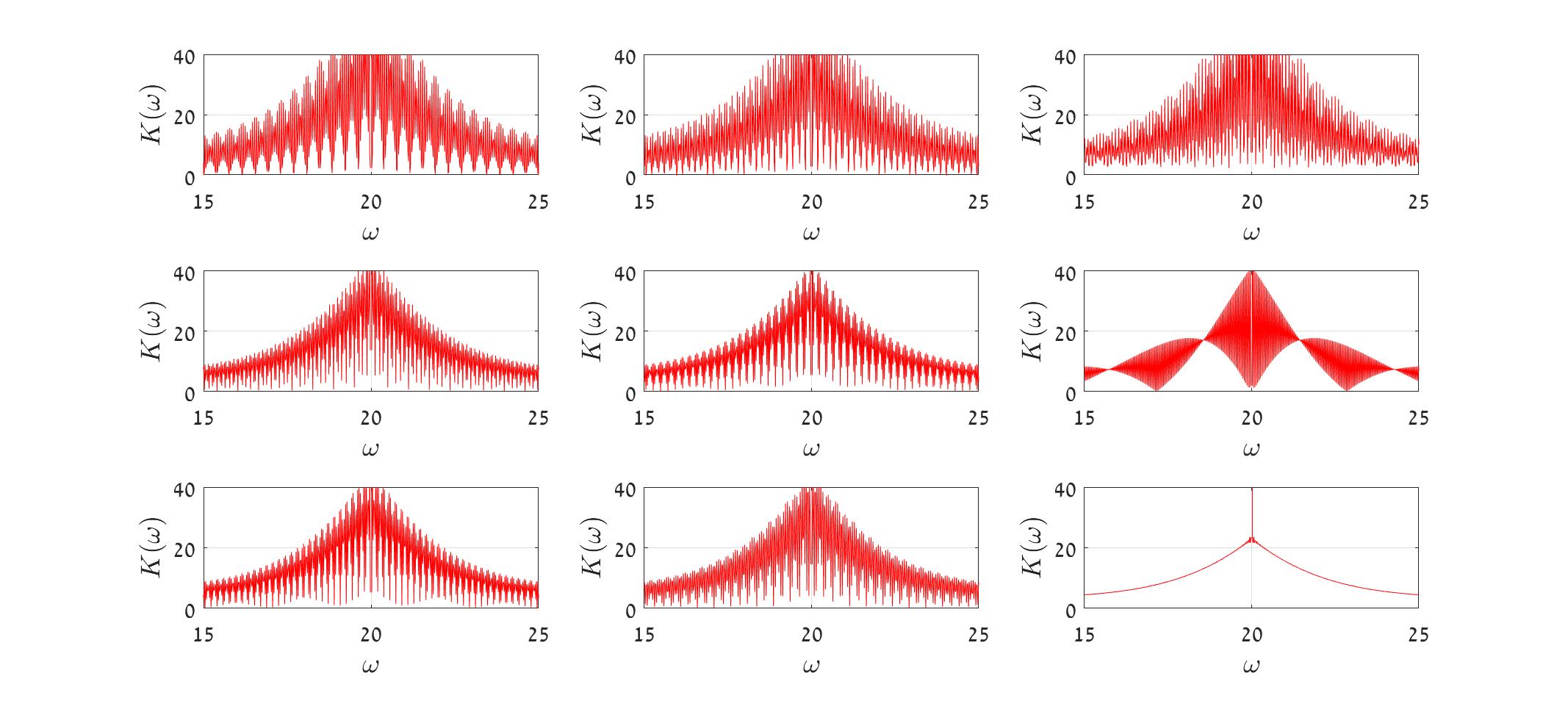}}}
    \caption{Iteration 5 for B = 4}
\end{figure}
\newpage
\hfill
\begin{figure} [!h]
    \centering
    \subfloat[Time Domain]{{\includegraphics[scale=0.20]{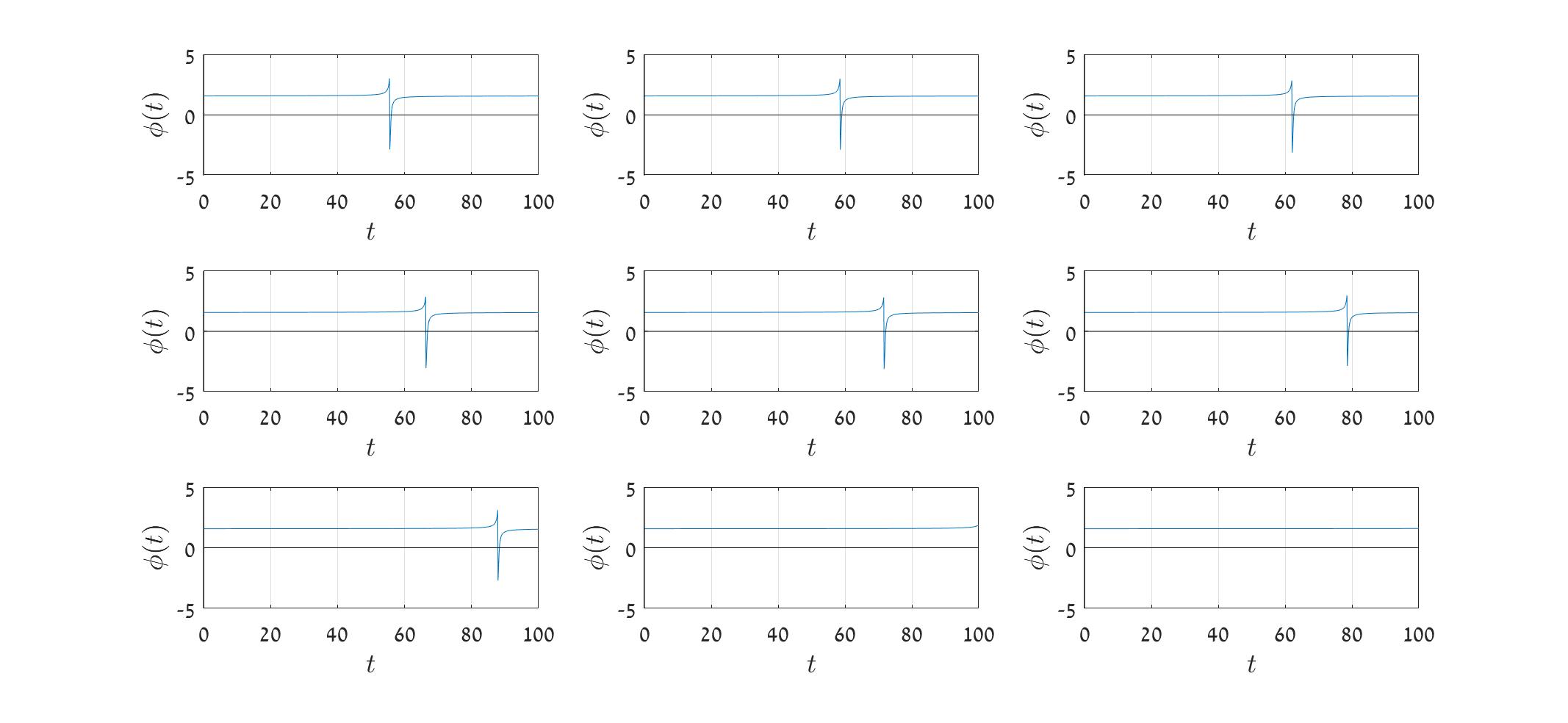}}}
    \qquad
    \subfloat[Frequency Domain]{{\includegraphics[scale=0.20]{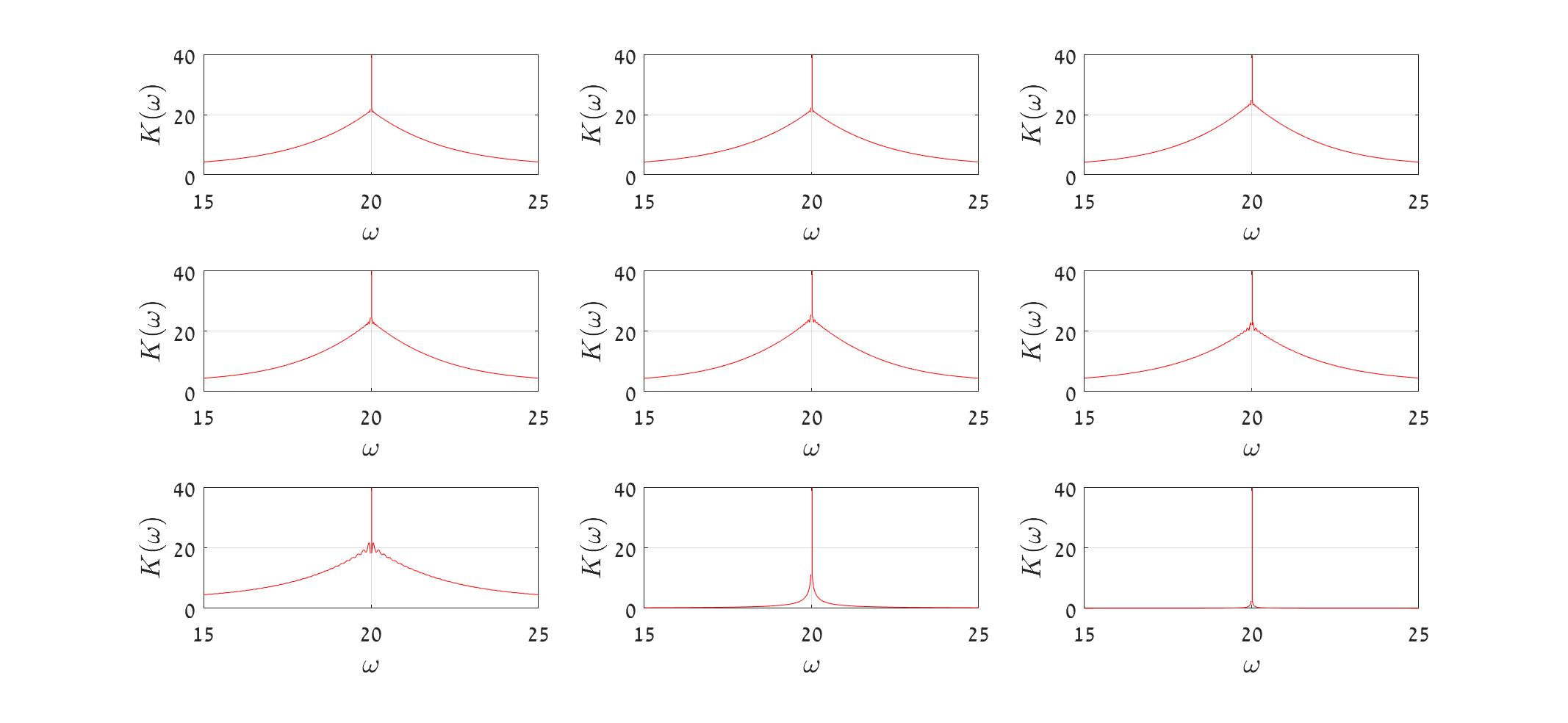}}}
    \caption{Iteration 6 for B = 4}
\end{figure}

\newpage
\subsubsection{Iterations for B = 5}
\hfill
\begin{figure} [!h]
    \centering
    \subfloat[Time Domain]{{\includegraphics[scale=0.20]{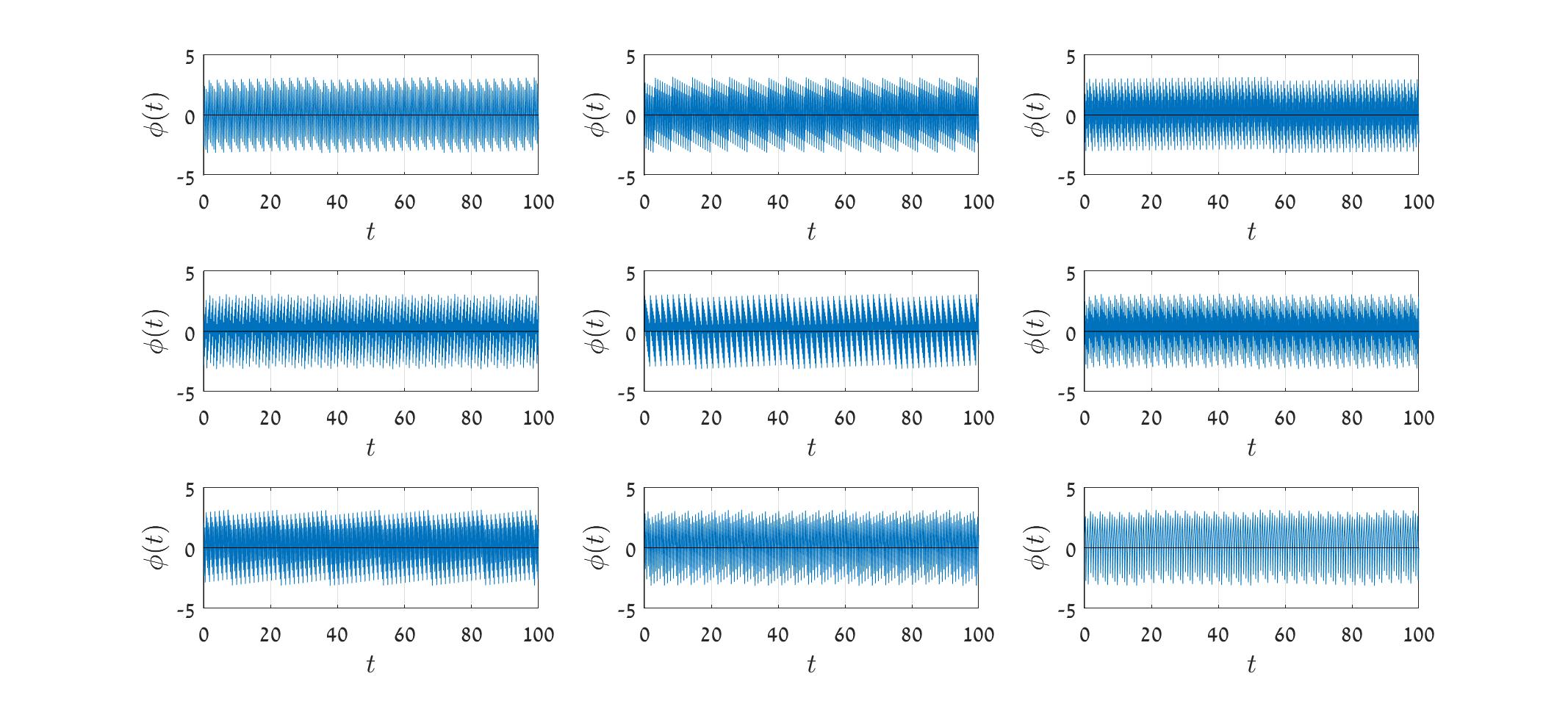}}}
    \qquad
    \subfloat[Frequency Domain]{{\includegraphics[scale=0.20]{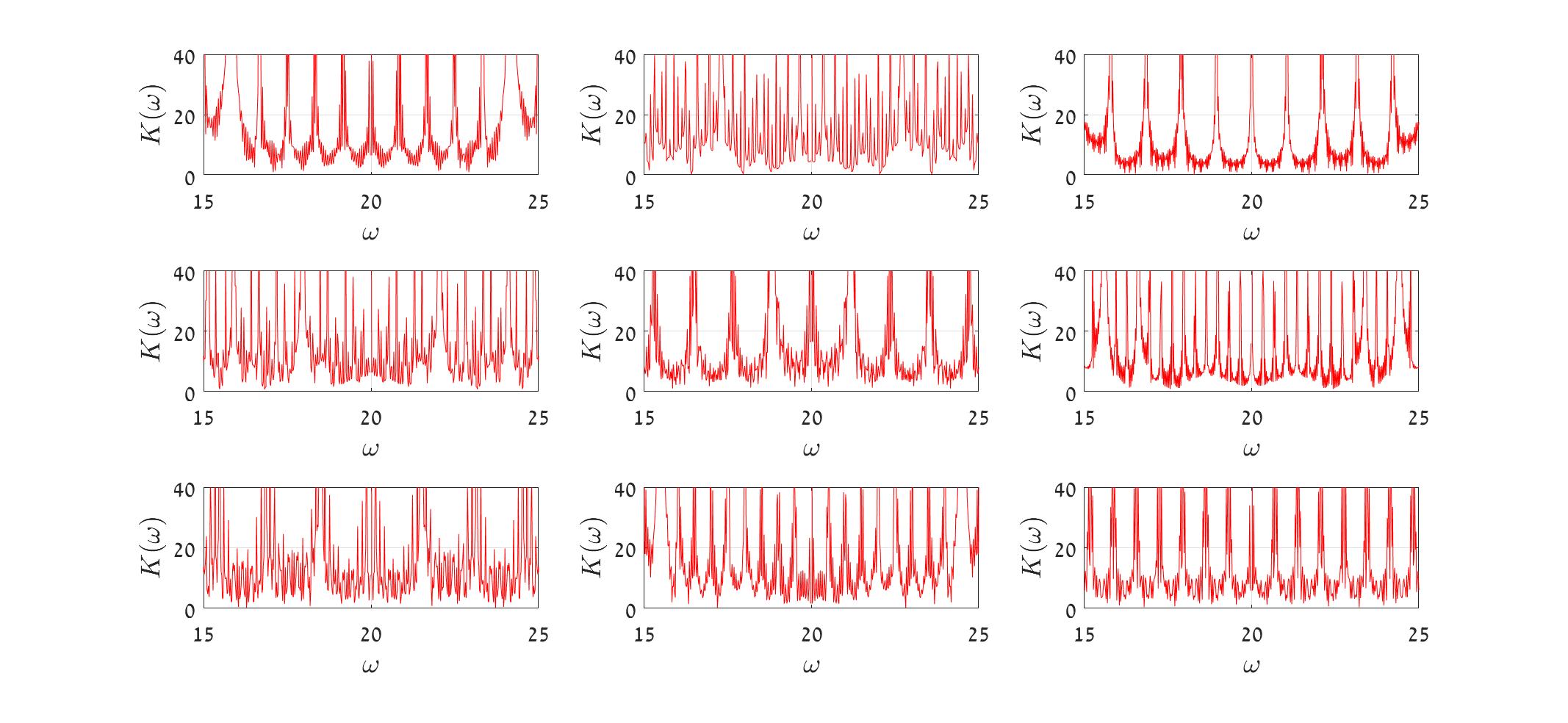}}}
    \caption{Iteration 1 for B = 5}
\end{figure}
\newpage
\hfill
\begin{figure} [!h]
    \centering
    \subfloat[Time Domain]{{\includegraphics[scale=0.20]{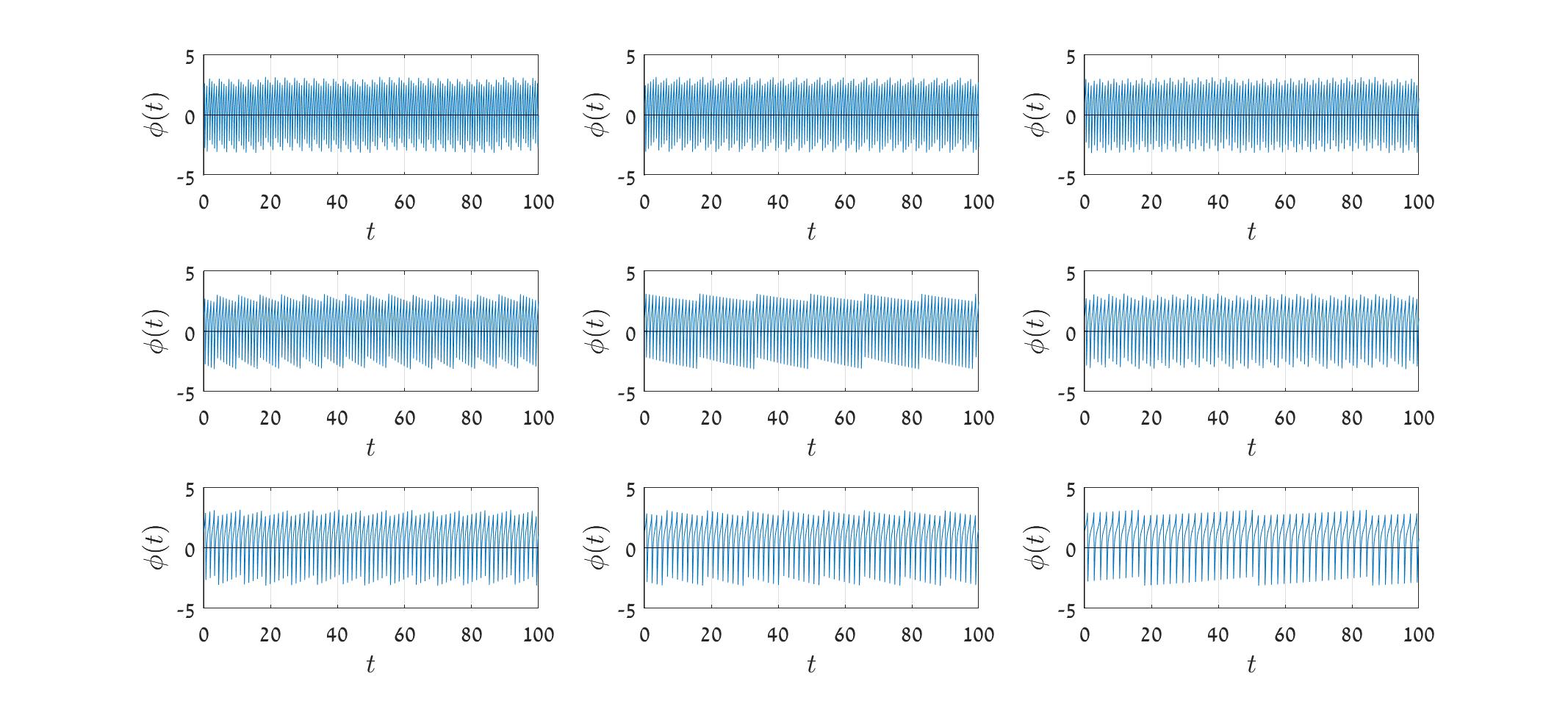}}}
    \qquad
    \subfloat[Frequency Domain]{{\includegraphics[scale=0.20]{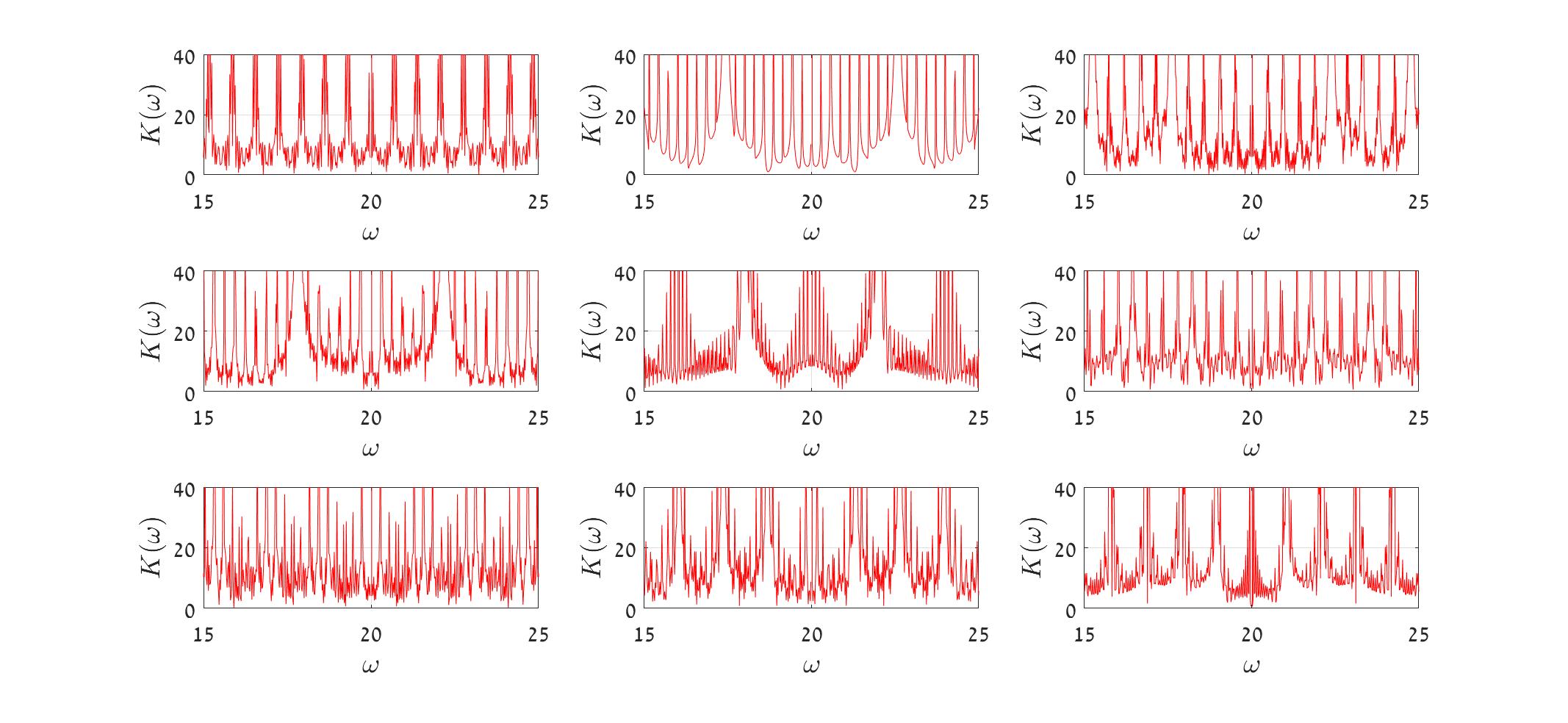}}}
    \caption{Iteration 2 for B = 5}
\end{figure}
\newpage
\hfill
\begin{figure} [!h]
    \centering
    \subfloat[Time Domain]{{\includegraphics[scale=0.20]{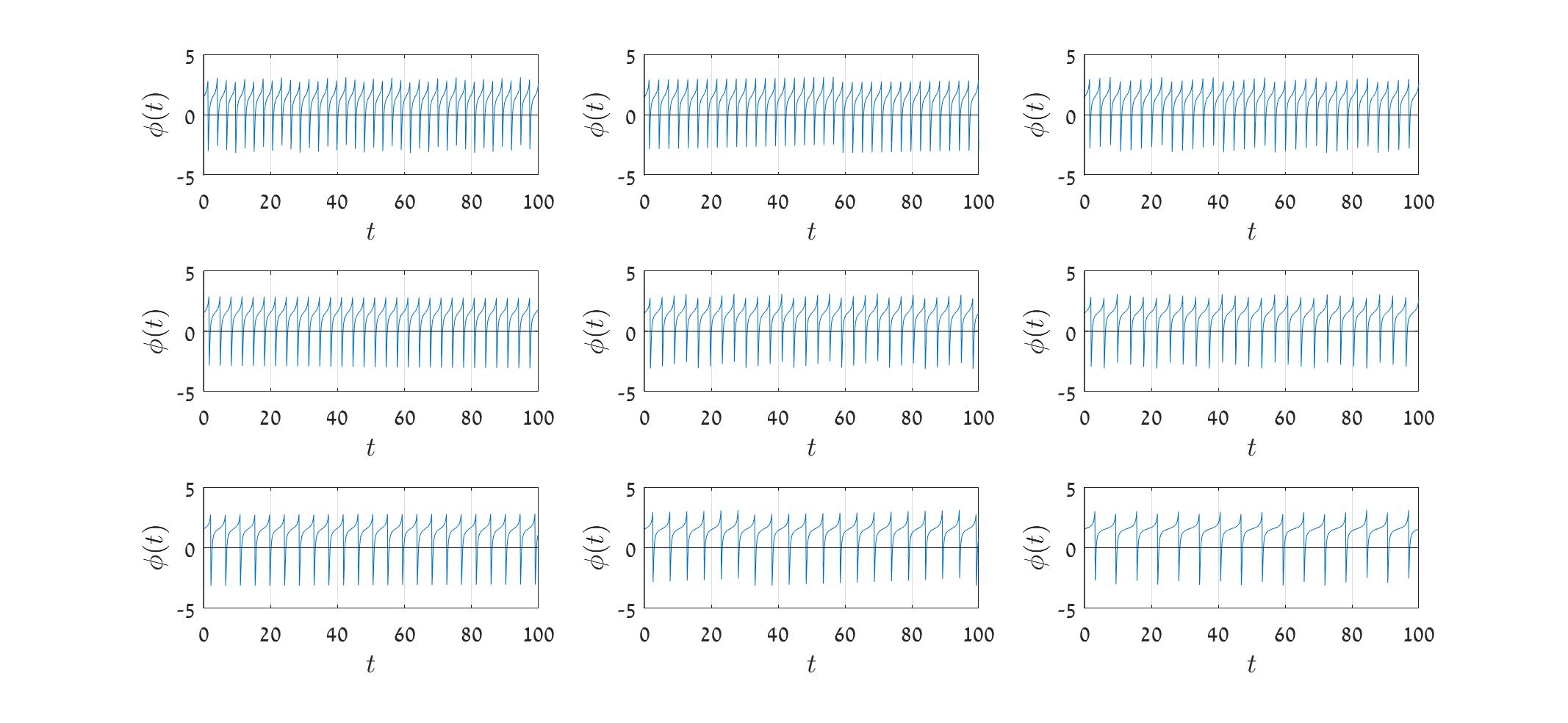}}}
    \qquad
    \subfloat[Frequency Domain]{{\includegraphics[scale=0.20]{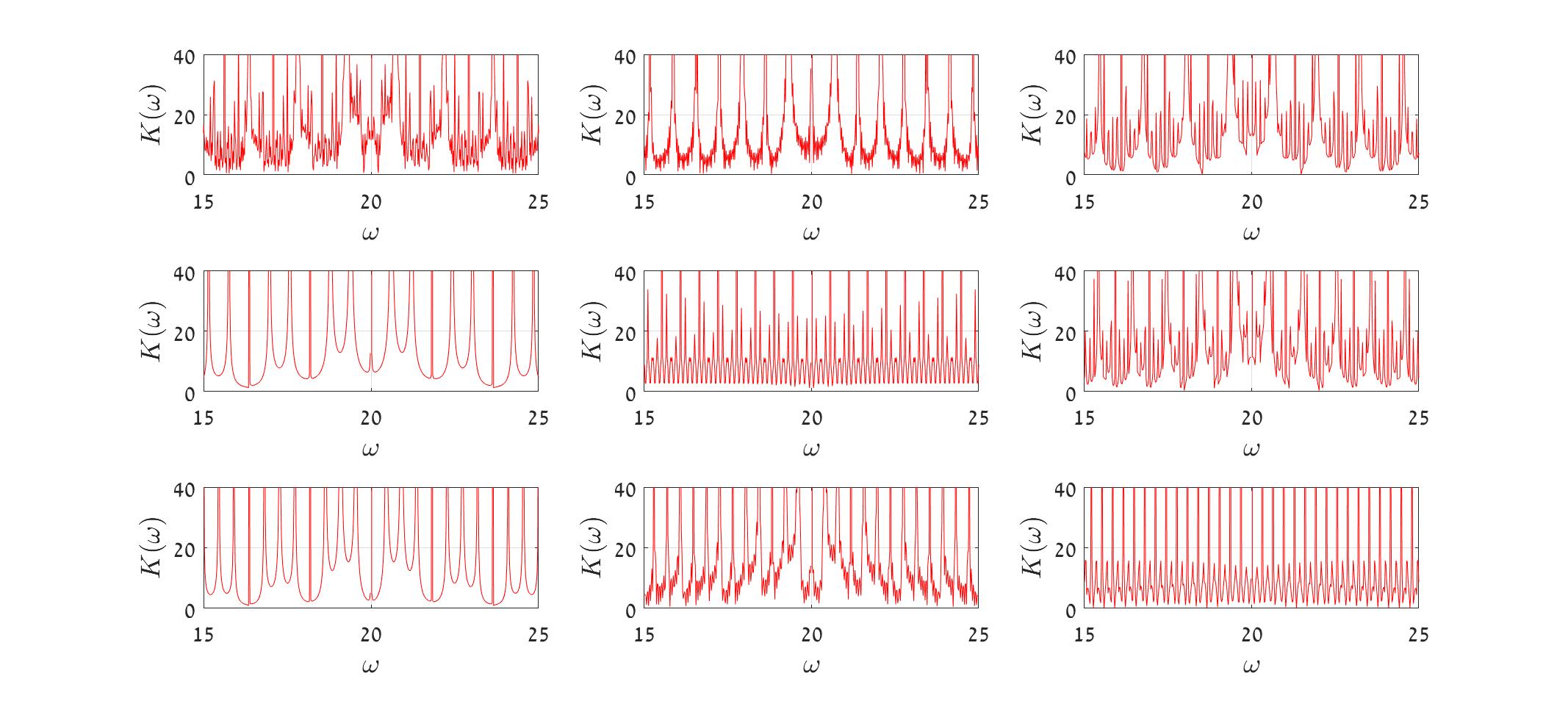}}}
    \caption{Iteration 3 for B = 5}
\end{figure}
\newpage
\hfill
\begin{figure} [!h]
    \centering
    \subfloat[Time Domain]{{\includegraphics[scale=0.20]{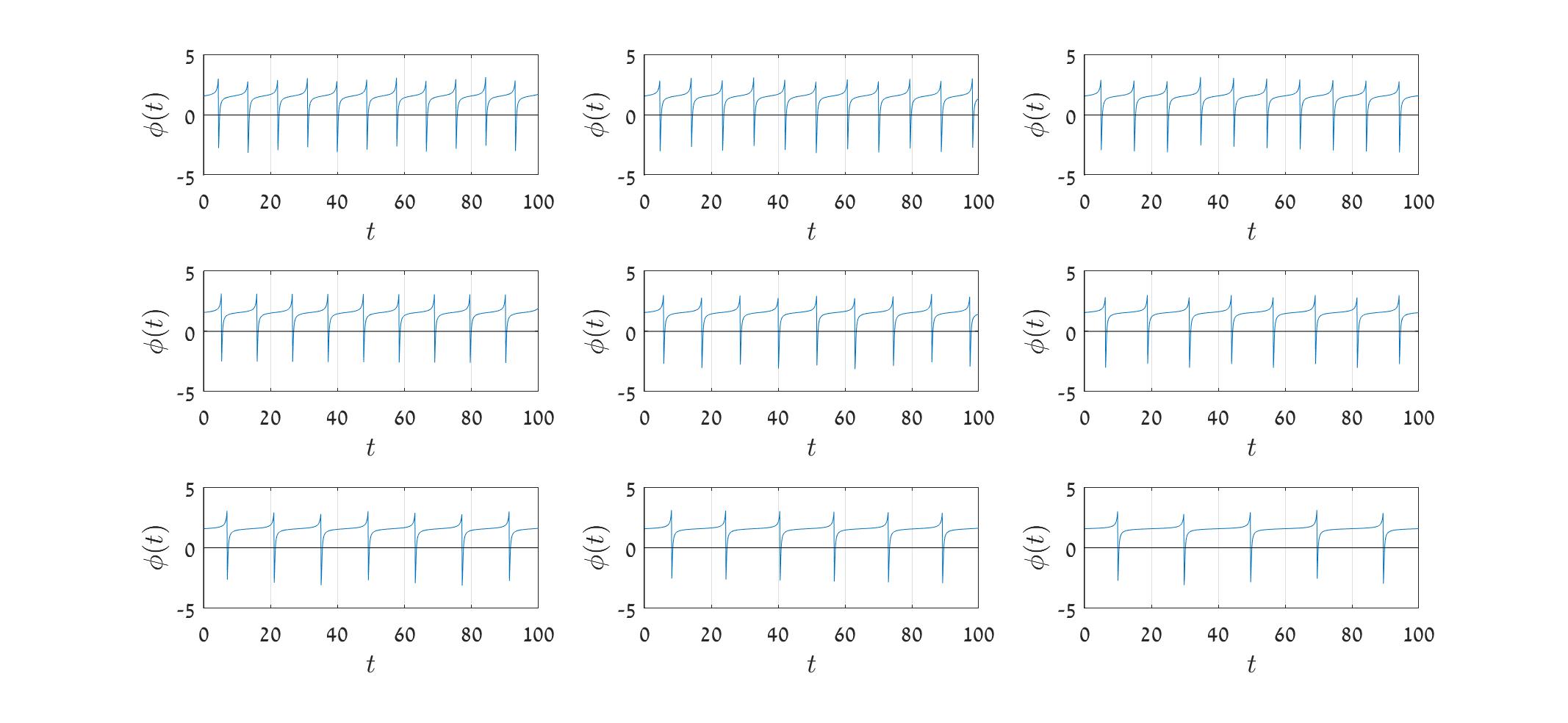}}}
    \qquad
    \subfloat[Frequency Domain]{{\includegraphics[scale=0.20]{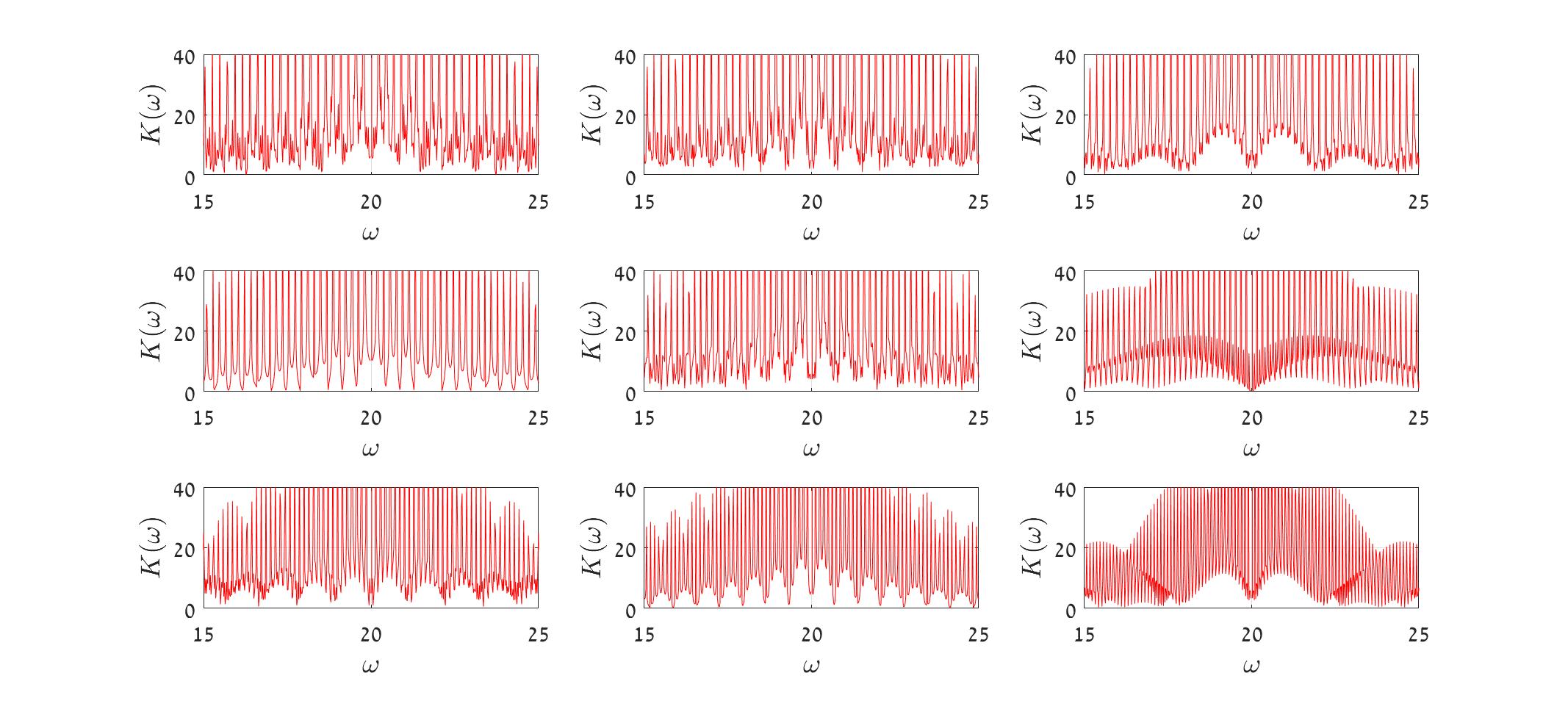}}}
    \caption{Iteration 4 for B = 5}
\end{figure}
\newpage
\hfill
\begin{figure} [!h]
    \centering
    \subfloat[Time Domain]{{\includegraphics[scale=0.20]{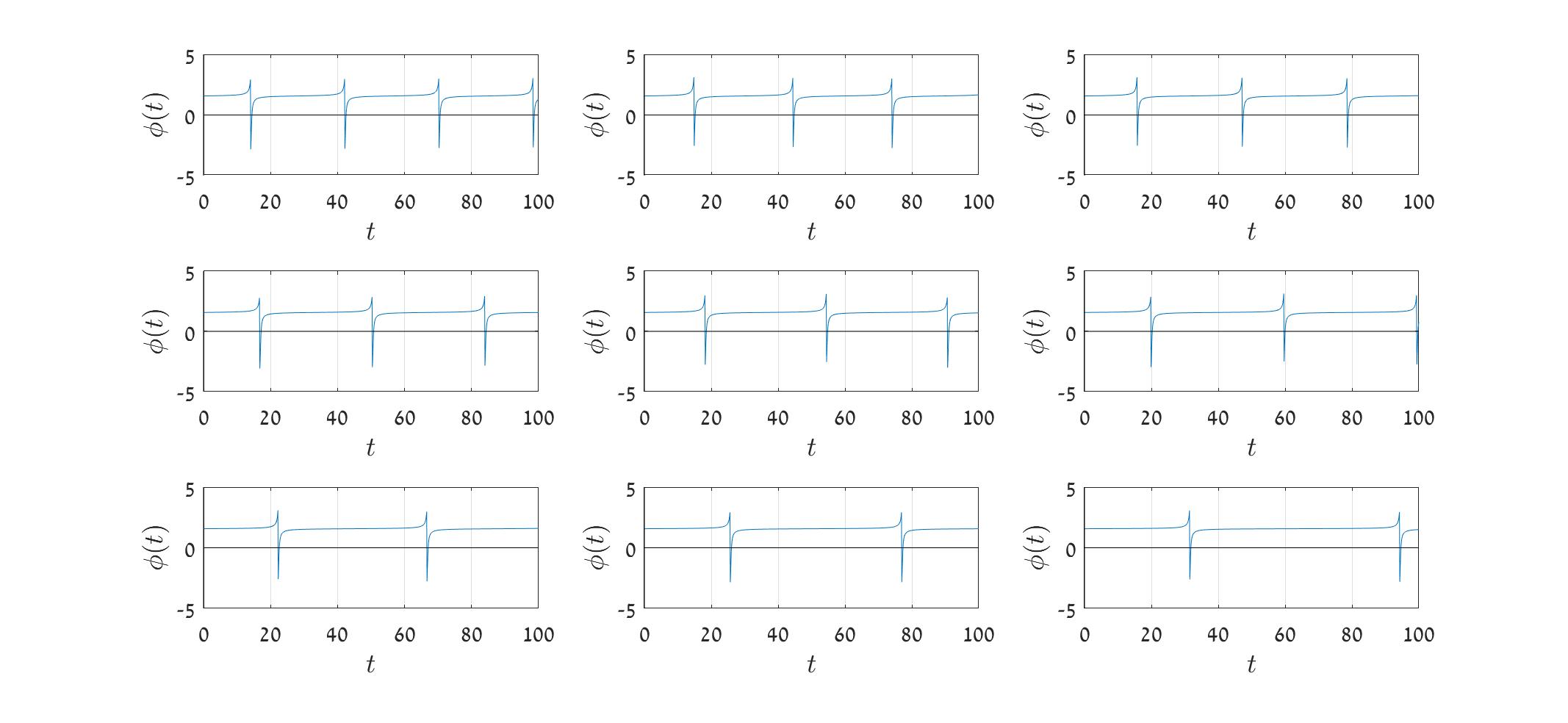}}}
    \qquad
    \subfloat[Frequency Domain]{{\includegraphics[scale=0.20]{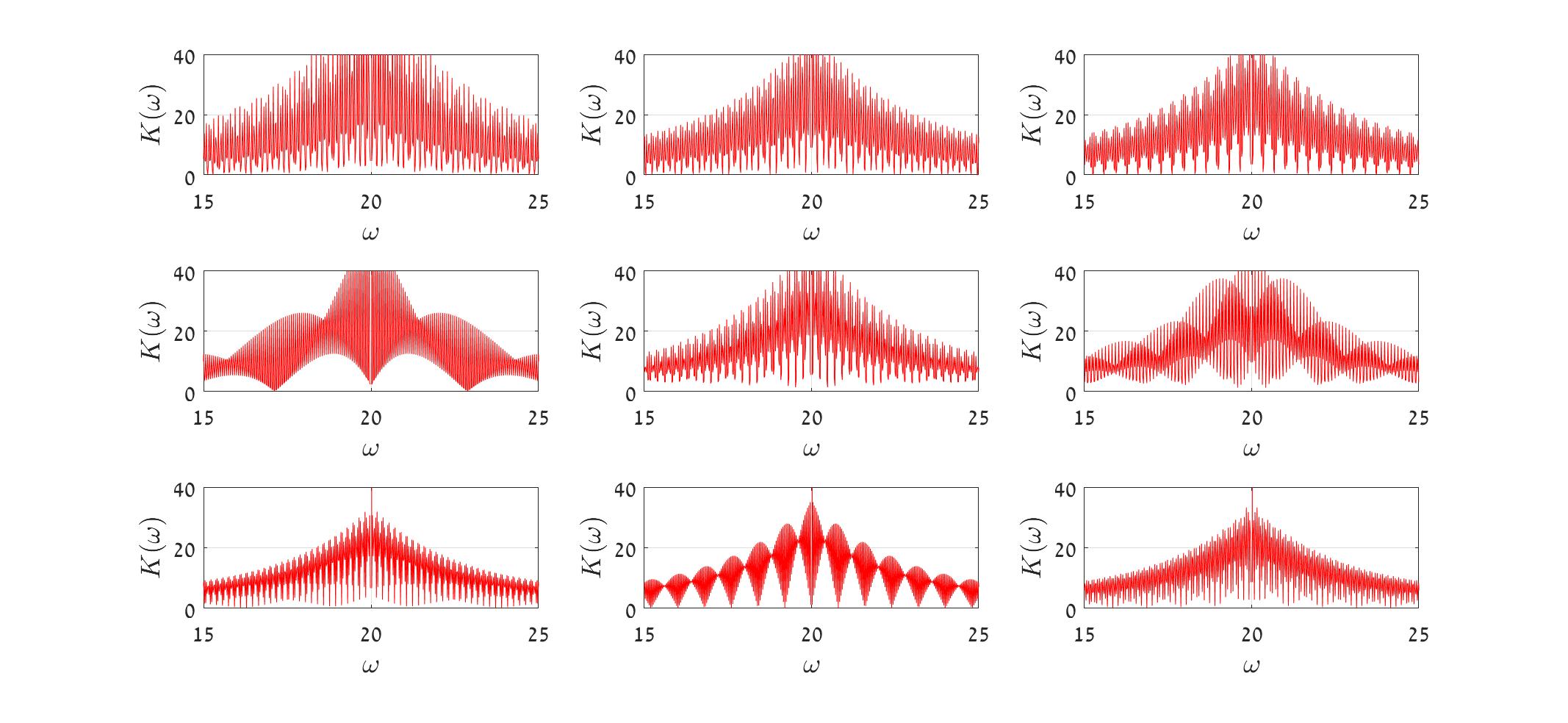}}}
    \caption{Iteration 5 for B = 5}
\end{figure}
\newpage
\hfill
\begin{figure} [!h]
    \centering
    \subfloat[Time Domain]{{\includegraphics[scale=0.20]{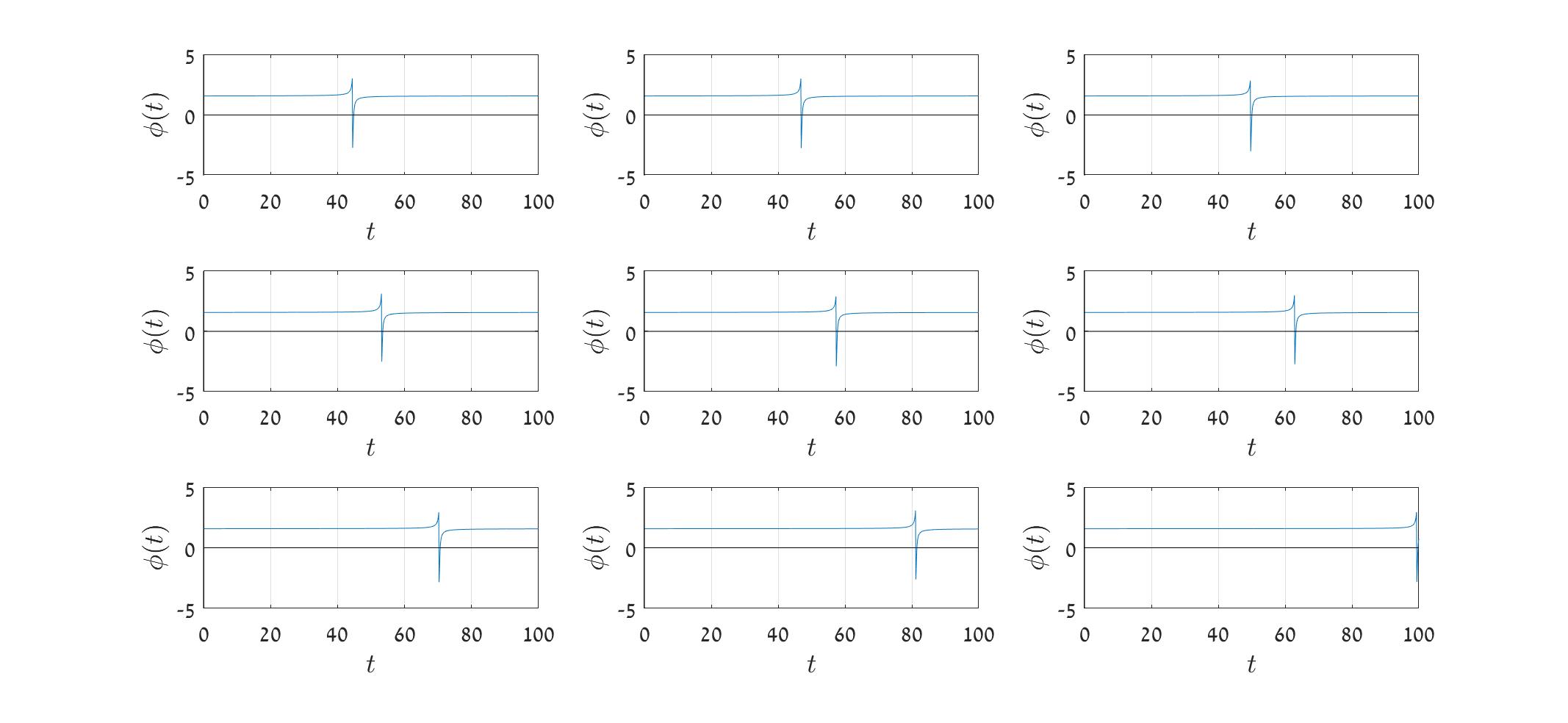}}}
    \qquad
    \subfloat[Frequency Domain]{{\includegraphics[scale=0.20]{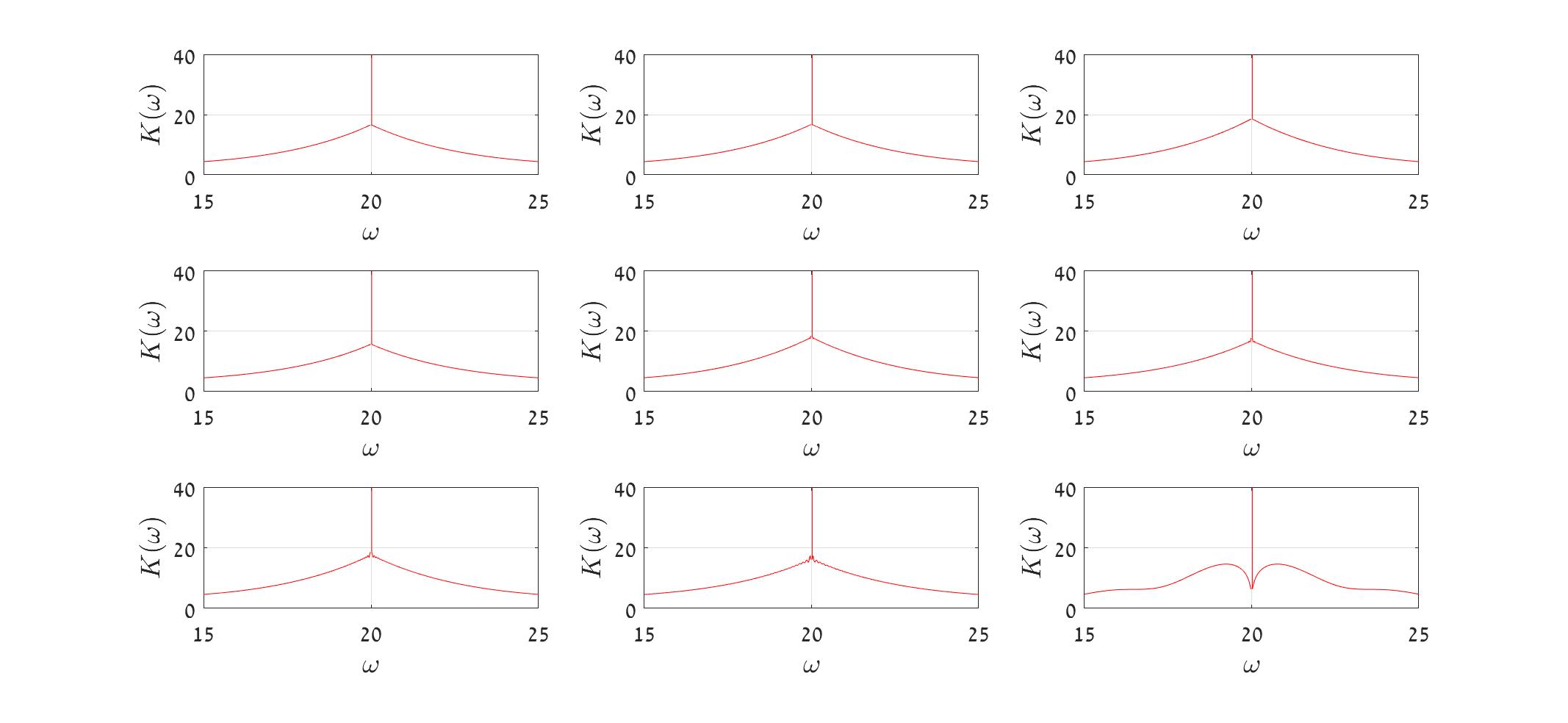}}}
    \caption{Iteration 6 for B = 5}
\end{figure}

\end{center}
\pagebreak 
\section{Observations and Conclusions}
This section lists the conclusions made based on the theoretical and numerical analyses made.
\begin{itemize}

\item{\makebox{} As we can see from the beam's spatial function simulation, the beam bends in the upward direction by a normalized amount of 1 - a result that matches our model, as discussed. We deduce that the function we got is then valid for this model and correlates well to the actual behaviour of the beam.}

\item{\makebox{} As we can see from the Beam’s temporal function simulation, the beam amplitude in time does converge to a value of $a_{ss} = \frac{1}{\pi}$ in the steady state and stays there over a large amount of time, as we predicted in the theory. We then conclude that the Simulink model does describe well the time dynamics of the beam.}

\item{\makebox{} As we can see from the Beam’s beam’s phase portrait simulation, the beam does converge to a steady limit cycle with an amplitude of $a_{ss} = \frac{1}{\pi}$ ,as predicted in the theory and illustrated in the temporal function. We then conclude that the control law is then valid and the derivative with respect to $\tau$ does also correlate to the predicted behaviour.}

\item{\makebox{} As it shows in the simulation results for the comb and specifically in the last ones, as the time signal becomes less dense in smaller set values of iteration, the gain in the frequency domain lowers and one can clearly see a frequency comb forming for the last iterations. These results match our theory, for the values of the penultimate and last iterations are very close to the point of the SNIC bifurcation, in which we predicted that the frequency comb will appear clearly. So, from that point of view, we conclude that in order to render a comb we would like our actual beam model to be as close as possible to the actual bifurcation values, as long as the beam can still withhold the situation.}

\item{\makebox{} The responsibility for capturing the SNIC bifurcation rests in the hands of the controller we added to the system, and the simulation results matched the theory even though we took a linear control model - meaning that linear control tools are applicable in the steady state for the parameters chosen.}

\item{\makebox{} Based on the simulation results, we can conclude that Adler's equation not only fits theoretically as an analytic equation, but is also a fitting model to incite the desired outcome.}

\item{\makebox{} The penultimate iteration displays a very dense comb, one that includes a very large spectrum of frequencies, that is favorable for the applications requiring the comb in the first place. The last iteration displayed a single curve that corrupted the comb, as in that iteration the values are too close to 1, and we assume that this behaviour is present due to the system's singularity in this point, as discussed in the theoretical background.}

\item{\makebox{} We see that increasing the parameter $B$ causes the spaces between the frequencies to increase and the whole comb to spread out An animation of this behaviour will be displayed in the presentation, and in that one can see that the tweaking $B$ causes the whole system to produce frequencies in different wave-like forms. We assume that this correlates to the fact that this parameter's value is a function of all of the physical sizes we discussed in this project, and thus a change by a normalized unit causes concrete visible changes. We conclude that the resulting comb we would want to produce with the beam would require specific parameters values and thus this has to be taken into consieration.}

\item{\makebox{} Looking at the simulation results, we also see that the time and frequency response of even such a simple model still demonstrate very rapid and sharp changes. Thus, we can conclude that the model is very sensitive to changes in phase, typically near values of bifurcation, and we predict that this will have to be taken into consideration in the future with the suggest model we would introduce.}

\item{\makebox{} Due to the properties of the SNIC bifurcation, meaning the fact that its period is infinite at the point of bifurcation, we cannot lock into a specific theoretical point in the phase space and call it our desired target, thus we also conclude that our aim is to get a close enough approximation to the desired output, one that will serve a suitable result for this study's problem.}

\item{\makebox{} As mentioned numerous times, the frequency comb is generated due to the behaviour of the phase, which meant that the phase is the cause for the bifurcation. We saw that this assumption holds as we predicted in the theory.}

\item{\makebox{} As the simulation results show, the frequency response is pretty hectic. We can predict that these results have to be improved in order for us to get a realistic result. Possible future plans for the continuation of this project can be to set up the simulation using different values, or alternatively use a analytical expression derived for this special matter by Steven W. Shaw \cite{shaw}, that will undoubtedly provide better results.}

\end{itemize}


\section{General Evaluation}

In this report, an elaborate discussion regarding a vibrating beam with non-linear dynamics was presented. The aim of this project was to excite the beam using external forcing and closed-loop feedback control in order to reach a state of SNIC bifurcation in the dynamical system that will hopefully produce a frequency comb from the vibrations of the beam. Starting with the stand-alone non-linear beam model, an initial mathematical formulation of the problem was investigated for the case in which the beam is externally perturbed and controlled - one that resulted in the non-linear Duffing equation. It is worth mentioning that the resulting equation contains constants that have not been determined for a specific beam with actual physical parameters. In that context, the project aimed to provide insight in the world of MEMS even though a general model was presented. To actually use this model, the system's parameters have to be set either by taking their value from literature tables or alternatively by calculating them using a Finite Elements Simulation software or other physical simulation tools. We saw that the forced, damped Duffing equation is impossible to solve analytically and therefore a more subtle model was used to obtain a set of equations that produced the solution for the vibrations of the beam - which is composed out of an amplitude and phase functions. This model takes the forcing and control of the beam into consideration. We saw that the phase function in our model can be presented in the form of Adler's equation and realized that this function is responsible for the formulation of the desired frequency comb. Simulations on the governing temporal and spatial equations of the beam were performed and all the theory presented in the background matched the results of the simulations, which was a good indication that the model is valid for analysis in the case presented. Ultimately, the main goal was achieved by conducting a simulation of Adler's equation to imitate the SNIC bifurcation behaviour to produce the desired entity - a frequency comb. The final simulation results also matched the theory, in that the SNIC bifurcation renders a frequency comb that was captured well at the penultimate iteration. To conclude, a firm theoretical layout of the problem was presented, all simulations of the desired outcomes matched our theory and a we can say that we have a model that is somewhat applicable for MEMS and other systems, noting that one have to define specification for specific desired frequency combs. Future plans can be to try this model experimentally and check the correlation of the experiment's results with the model shown.
\newpage
\section{Appendices}

\subsection{Derivation of the Adler Equation\protect\footnotemark} \label{adler_derivation}

\footnotetext{All figures in this appendix are from Adler's article}

The Adler equation describes the case of synchronization at its \textit{steady state} - in which phase difference between oscillator and external signal is constant, being when all transient behaviours vanish in time. Other than that, when no synchronization occurs, the equation gives frequency and wave form of the \textit{beat note} - periodic variations of frequency and amplitude. To do so, the equation contains a parameter which decides whether or not the transient term will vanish in time, thus producing an equivalent to the criteria for synchronization derived by other methods. The equation assumes that there are no aftereffects from different conditions which may have existed in the past. This simplifies the equation to a fair extant, but is applicable under specific conditions. Adler described those in detail in his article and went on to derive the equation using a model of two RCL circuits - a circuit that includes a resistor, capacitor and an inductor.

 \begin{figure}[h] \label{pics/Two RCL circuits}
\centering
\includegraphics[scale=0.3]{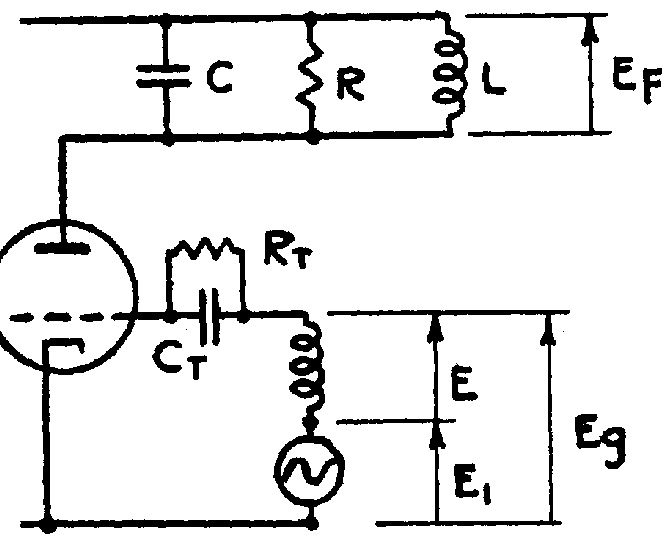}
\caption{Two RCL circuits}
\label{Two RCL circuits}
\end{figure}

\begin{itemize}
\item{\makebox[2cm]{$R$\hfill} is a resistor}
\item{\makebox[2cm]{$C$\hfill} is a capacitor}
\item{\makebox[2cm]{$L$\hfill} is an inductor}
\item{\makebox[2cm]{$E$\hfill} is the voltage induced in grid coil}
\item{\makebox[2cm]{$E_1$\hfill} is the voltage of impressed signal}
\item{\makebox[2cm]{$E_g$\hfill} is the resultant grid voltage}
\item{\makebox[2cm]{$E_F$\hfill} is the voltage across plate load}
\end{itemize}
 
 Using the RCL circuit - the Adler equation relates to \textit{voltages} in the circuit and their frequency of osculation in the branches of the circuit. This section will show the derivation of the equation as was shown by Adler in his article. Let the picture shown in fig. \ref{Two RCL circuits} be a vector representation of the voltages in the grid circuit as they are found at a given instant.

 \begin{figure}[h]
\centering
\includegraphics[scale=0.3]{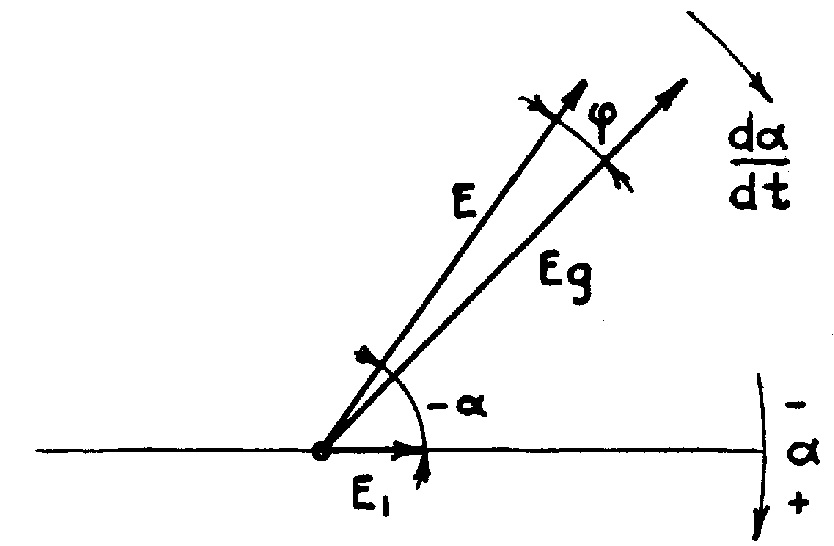}
\caption{Voltages in the grid circuit}
\end{figure}

Where

\begin{itemize}
\item{\makebox[2cm]{$\alpha$\hfill} is the angle between the induced voltage and the impressed signal}
\item{\makebox[2cm]{$\phi$\hfill} is the phases shift}
\end{itemize}

Let $E_1$ be a vector at rest with respect to our eyes. Any such vector will  symbolize an angular frequency $\omega_1$ - that of the external signal. A vector rotating clockwise with an angular velocity of $\frac{d\alpha}{dt}$ shall represent an angular frequency of $\omega = \omega_1 + \frac{d\alpha}{dt}$, or angular beat frequency of

\begin{equation}
    \Delta \omega = \omega - \omega_1 =  \frac{d\alpha}{dt}
\end{equation}

relative to the external signal. 
With no external signal impressed, $E_g$ and $E$ must coincide: the voltage $E$ returned through the feedback circuit must have the same amplitude and phase as the voltage $E$ applied to the grid. Those nonlinear elements which limit the oscillator amplitude will adjust the gain so that $\abs{E} = \abs{E_g}$ but the phase can only coincide at one frequency, the free-running frequency $\omega_o$. At any other frequency the plate load would introduce phase shift between $E_g$ and $E$. Fig. 8 shows a phase shift versus frequency graph for a single tuned circuit as assumed in fig. 6

 \begin{figure}[h]
\centering
\includegraphics[scale=0.3]{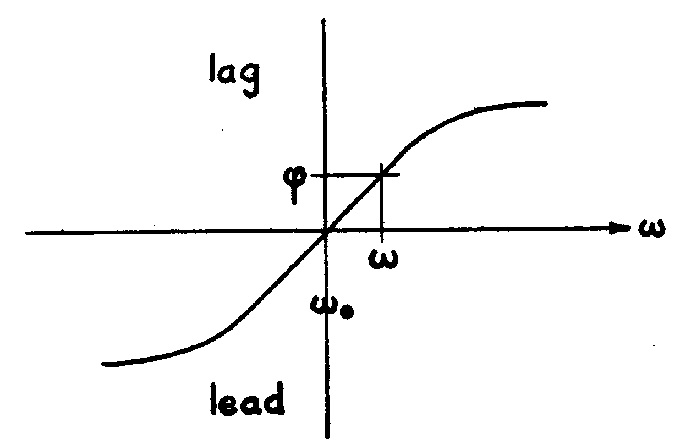}
\caption{Phase versus frequency for a simple tuned circuit}
\end{figure}

Let now an external voltage $E_1$ be introduced, and let fig. 7 represent the voltage vectors at a given instant during the beat cycle. Evidently, the voltage $E$ returned through the feedback circuit is now no longer in phase with the grid voltage $E_g$, the diagram shows $E$ lagging behind $E_g$ by a phase angle $\phi$. No such lag could be produced if the oscillator were still operating at its free frequency $\omega_o$, meaning that the frequency at this instant exceeds $\omega_o$ by an amount which will produce a lag equal to $\phi$ in the plate circuit. With $E_1 \ll E$ taken as an assumption by Adler for the validity of his argument - inspection by simple trigonometry of fig. 7 yields

\begin{equation}
    \phi(\alpha) = \frac{E_1 \sin{(-\alpha)}}{E} = - \frac{E_1}{E}\sin{(\alpha)}
\end{equation}

Another condition taken by Adler is that the pass band of the plate circuit is so wide that all frequencies are near its center, so only a small central part of the curve of $\phi$ vs. $\omega$ is taken into consideration, one that approaches a straight line with a slope of

\begin{equation}
    A = \frac{d\phi}{d\omega}
\end{equation}

Then, one can take a general linear curve function and apply it to an arbitrary phase angle as a function of another frequency [$\phi(\omega) = m\omega + n$]. Taking $m = A$, $\phi(\omega_0) = 0$, and setting $\Delta\omega_0 = \omega_0 - \omega_1 $ we get

\begin{equation}
    \phi(\omega) = A(\omega - \omega_0) = A[(\omega - \omega_1) - (\omega_0 - \omega_1)] = A(\Delta\omega - \Delta\omega_0)
\end{equation}

Plugging in eq. 18 to the RHS and eq. 19 to the LHS we get

\begin{equation}
    - \frac{E_1}{E}\sin{(\alpha)} = A\bigg(\frac{d\alpha}{dt} -  \Delta\omega_0\bigg)
\end{equation}

Dividing by A and setting $B = \frac{E_1}{E}\cdot\frac{1}{A}$ we finally obtain Adler's equation

\begin{equation}
    \frac{d\alpha}{dt} = -B \sin{(\alpha)} + \Delta\omega_0
\end{equation}

Setting $K = \frac{\Delta\omega_0}{B}$ we finally get

\begin{equation}
    \boxed{\frac{d\alpha}{dt} = B[K - \sin{(\alpha)} ]}
\end{equation}

\subsection{Complete Solution to the Beam PDE} \label{solving_PDE}

This appendix will show the complete solution to the beam's PDE for the temporal and spatial functions. After plugging $u_n(x,t)$ into equation \ref{beameq} we can solve the equation for the spatial and temporal functions

\begin{equation}
    \begin{gathered}
    \rho A \ddot{P_n(t)} X_n(x) +2 \zeta \dot{P_n(t)} X_n(x) -\\ -\frac{EA}{2l}\int_0^l \big[ P_n(t) X'_n(x) \big]^2 dx \cdot P_n(t) X''_n(x) + EI P_n(t) X''''_n(x) =\\
    = \big[ A\cos{(\omega_d t)} + \psi [P_n(t) X_n(x)] \big] \delta(x-x_0)
\end{gathered}
\end{equation}
Rearranging
\begin{equation} \label{rearraned_eq}
    \begin{gathered}
    \big[ \rho A \ddot{P_n(t)} +2 \zeta \dot{P_n(t)} \big] X_n(x) -\\ -\frac{EA}{2l}\int_0^l \big[ X'_n(x) \big] ^2 dx \cdot  X''_n(x) \cdot P^3_n(t) + EI P_n(t) X''''_n(x) =\\
    = \big[ A\cos{(\omega_d t)} + \psi [P_n(t) X_n(x)] \big] \delta(x-x_0)
\end{gathered}
\end{equation}
To solve for the spatial function $X_n(x)$, one takes the homogenous equation and set $\tau = 0$ to yields a linear equation that produces a solution for the eigenfunctions $X_n(x)$
\begin{equation}
\begin{gathered}
    \big[ \rho A \ddot{P_n(t)} +2 \zeta \dot{P_n(t)} \big] X_n(x) + EI P_n(t) X''''_n(x) = 0\\
    \rightarrow  \frac{\rho A \ddot{P_n(t)} +2 \zeta \dot{P_n(t)}}{P_n(t)} = -  EI \frac{X''''_n(x)}{X_n(x)}
    \end{gathered}
\end{equation}
assuming $P_n(t), X_n(x) \not\equiv 0 \ ; \forall n$. It can be seen that the RHS is a function of $t$ alone and the LHS is a function of $x$ alone. This is possible if and only if both sides are equal to a constant $\lambda _n$ for the $n$'th mode. So this bring us to a characteristic ordinary differential equation for the eigenfunctions $X_n(x)$
\begin{equation} \label{char_spatial}
X''''_n(x) + \frac{\lambda_n}{EI} X_n(x) = 0
\end{equation}
And also for $P_n(t)$
\begin{equation}
\ddot{P_n(t)} + \frac{2 \zeta}{\rho A} \dot{P_n(t)} - \frac{\lambda_n}{\rho A} P_n(t) = 0
\end{equation}
\subsubsection{The Separation Constant $\lambda _n$}
We now have to discuss of the value of $\lambda _n$, named the separation constant. This constant is assumed to be real, as in with a zero imaginary component, due to the problem being \textit{self adjoint}. This mathematical concept is relatively complex and is not in the scope of this project. So. if the constant is zero, then we'll get
\begin{equation}
\ddot{P_n(t)} + \frac{2 \zeta}{\rho A} \dot{P_n(t)}  = 0
\end{equation}
Taking $P_n(t) = e^{\mu_n t}$ yields an algebraic equation for the eigenvalues $\mu_n$ with respect to time
\begin{equation}
\mu_n ^2 + \frac{2 \zeta}{\rho A} \mu_n = 0 \rightarrow \boxed{{\mu_n}_1 = 0} \ ; \boxed{{\mu_n}_2 = - \frac{2 \zeta}{\rho A}}
\end{equation}
As known, these two values for $\alpha$ represent the eigenvalues that correlate to the appropriate eigenvectors that constitute a base for the vector space of the solutions for $P_n(t)$
\begin{equation}
P_n(t) = A_n e^{{\mu_n}_1} + B_n e^{{\mu_n}_2} = A_n + B_n e^{- \frac{2 \zeta}{\rho A}}
\end{equation}
This solution is asymptotically static (as $\lim_{t \to \infty} P_n(t) = A_n$) and therefore is not in our interest. Now, if $\lambda_n >0$ we'll get
\begin{equation}
\ddot{P_n(t)} + \frac{2 \zeta}{\rho A} \dot{P_n(t)} - \frac{ \abs{\lambda_n}}{\rho A} P_n(t) = 0
\end{equation}
Taking $P_n(t) = e^{\mu_n t}$ again
\begin{equation}
\mu_n ^2 + \frac{2 \zeta}{\rho A} \mu_n - \frac{ \abs{\lambda_n}}{\rho A} = 0 \rightarrow \boxed{{\mu_n}_{1,2} = \frac{-\frac{2 \zeta}{\rho A} \pm \sqrt{\big( \frac{2 \zeta}{\rho A} \big) ^2 +4 \frac{ \abs{\lambda_n}}{\rho A}}}{2} > 0 }
\end{equation}
It can be seen that the eigenvalues are positive and therefore the time solution diverges, which is physically unacceptable. Taking $\lambda_n <0$ we'll get
\begin{equation}
\mu_n ^2 + \frac{2 \zeta}{\rho A} \mu_n + \frac{ \abs{\lambda_n}}{\rho A} = 0 \rightarrow \boxed{{\mu_n}_{1,2} = \frac{-\frac{2 \zeta}{\rho A} \pm \sqrt{\big( \frac{2 \zeta}{\rho A} \big) ^2 -4 \frac{ \abs{\lambda_n}}{\rho A}}}{2} < 0 }
\end{equation}
These eigenvalues are both negative and therefore the solution converges in this case. This means that $\lambda _n$ has to be negative due to consideration of finite energy. 
\subsubsection{Eigenvalues for the Spatial Coordinate}
We'd like to find the eigenvalues for the spatial coordinate. Taking $X_n(x) = e^{\mu_n x}$ and plugging into equation \ref{char_spatial} yields an algebraic equation for the eigenvalues $\mu_n$ with respect to the $x$ coordinate
\begin{equation}
\mu_n^4 - \frac{\abs{\lambda_n}}{EI} = 0 \rightarrow \boxed{\left\{\mu_n\right\} = \left\{ \alpha _n , -\alpha _n , \alpha_n i, - \alpha _n i \right\}}
\end{equation}
where $\alpha _n = \sqrt[4]{\frac{\lambda_n}{EI}}$. The solution then will be
\begin{equation} \label{x1}
X_n(x) = A_n e^{\alpha _n x} + B_n e^{-\alpha _n x} + C_n e^{\alpha _n x i} + D_n e^{-\alpha _n xi}
\end{equation}
Recall Euler's formula
\begin{equation}
    e^{i \theta} = \cos(\theta) + i \sin (\theta)
\end{equation}
And the definition of the hyperbolic functions
\begin{equation}
    \sinh (x) = \frac{e^x - e^{-x}}{2} \ ; \ \cosh (x) = \frac{e^x + e^{-x}}{2}
\end{equation}
Using these two, we can rearrange equation \ref{x1} to the form
\begin{equation} \label{x2}
X_n(x) = \Tilde{A}_n \sinh (\alpha _n x) + \Tilde{B}_n \cosh (\alpha _n x) + \Tilde{C}_n \sin (\alpha _n x) + \Tilde{D}_n \cos (\alpha _n x)
\end{equation}
When the constants $\Tilde{A}_n, \Tilde{B}_n, \Tilde{C}_n, \Tilde{D}_n$ are determined using the boundary conditions formulated in equation \ref{boundary}. Plugging these, we get a set of $4$ linear equations for the constants.
\subsubsection{Boundary Conditions for the Suggested Solution}
The boundary conditions formulated in equation \ref{boundary} can be applied to the suggested solution of variable separation for the $n$'th mode
\begin{equation}
    u _n(0,t)  = 0 \rightarrow X_n(0)P_n(t) = 0 \rightarrow \boxed{X_n(0) = 0}
\end{equation}
as we assume that $\neg (P_n(t) = 0 \ ; \forall t) \equiv \exists t_i \ (P_n(t_i) \neq 0 )$ - meaning $P_n(t)$ is not zero for all $t$. As for the rest of the boundary conditions:
\begin{equation}
\begin{gathered}
u (l,t) = 0 \rightarrow X_n(l)P_n(t) = 0 \rightarrow \boxed{X_n(l) = 0}\\
\frac{\partial u}{\partial x} (0,t) = 0 \rightarrow X'_n(0)P_n(t) = 0 \rightarrow \boxed{X'_n(0) = 0}\\
\frac{\partial u}{\partial x} (l,t) = 0 \rightarrow X'_n(l)P_n(t) = 0 \rightarrow \boxed{X'_n(l) = 0}
\end{gathered}
\end{equation}
These conditions can be plugged into equation \ref{x2} to yield the following set
\begin{equation}
\begin{cases} 
      \Tilde{B}_n + \Tilde{D}_n = 0\\
     \alpha _n \Tilde{A}_n + \alpha _n \Tilde{C}_n = 0\\
      \Tilde{A}_n \sinh (\alpha _n l ) + \Tilde{B}_n \cosh (\alpha _n l ) + \Tilde{C}_n \sin (\alpha _n l ) + \Tilde{D}_n \cos (\alpha _n l ) = 0 \\
      \Tilde{A}_n \sinh (\alpha _n l ) + \Tilde{B}_n \cosh (\alpha _n l ) + \Tilde{C}_n \sin (\alpha _n l ) + \Tilde{D}_n \cos (\alpha _n l ) = 0 \\
      \alpha _n \Tilde{A}_n \cosh (\alpha _n l ) + \alpha _n \Tilde{B}_n \sinh (\alpha _n l ) + \alpha _n \Tilde{C}_n \cos (\alpha _n l ) - \alpha _n \Tilde{D}_n \sin (\alpha _n l ) = 0 \\
   \end{cases}
\end{equation}
Plugging the first two equations into the last ones we get a new set of two equations for $\Tilde{C}$ and $\Tilde{D}$ that can be expressed in matrix form
\begin{equation}
\begin{bmatrix}
\sin (\alpha _n l ) - \sinh (\alpha _n l ) & \cos (\alpha _n l ) - \cosh (\alpha _n l ) \\
\cos (\alpha _n l ) - \cosh (\alpha _n l ) & -\sin (\alpha _n l ) - \sinh (\alpha _n l )
\end{bmatrix} \begin{bmatrix}
\Tilde{C}_n \\
\Tilde{D}_n
\end{bmatrix}  = \begin{bmatrix}
0 \\
0
\end{bmatrix}
\end{equation}
This is a homogeneous set of linear equations that has a non-trivial solution if and only if the determinant of the coefficients matrix is zero (meaning actually that the equations are linearly dependant) and therefore we can calculate 
\begin{equation}
- \big[ \sin (\alpha _n l ) - \sinh (\alpha _n l ) \big] \big[\sin (\alpha _n l ) + \sinh (\alpha _n l ) \big] - \big[ \cos (\alpha _n l ) - \cosh (\alpha _n l ) \big] ^2 = 0
\end{equation}
Simplifying
\begin{equation}
\begin{gathered}
\sin ^2(\alpha _n l ) - \sinh ^2(\alpha _n l )  + \cos ^2(\alpha _n l ) - 2 \cos (\alpha _n l ) \cosh (\alpha _n l ) + \cosh^2 (\alpha _n l )  = 0 
\end{gathered}
\end{equation}
Let us look at two of the following hyperbolic identities
\begin{equation}
\sinh ^2(x) = \frac{\cosh (2x) -1}{2} \ ;  \cosh ^2(x) = \frac{\cosh (2x) +1}{2}
\end{equation}
Plugging these two and arranging we get the final characteristic equation of the beam
\begin{equation} \label{coscosh_eq}
    \boxed{\cos{(\alpha_n l)} \cosh{(\alpha_n l)} = 1}
\end{equation}

\subsection{Layout for Solving Adler's Equation} \label{solving_adler}
Adler's equation corresponds, using separation of variables, to solving the integral
\begin{equation}
    \int\frac{d\alpha}{K - \sin{(\alpha)}} = B\int_{t_0}^{t}  dt'
\end{equation}
The RHS is easy but the LHS is less so. It can be solved analytically using an integration technique called the tangent half-angle substitution - taking $s = \tan{(\frac{\alpha}{2})}$ and solving for $S$.\\
Let then
\begin{equation}
    ds = \frac{d\alpha}{2\cos^2{(\frac{\alpha}{2})}}
\end{equation}
Based on that substitution, one can draw a triangle of the form

\begin{center}
    
\begin{tikzpicture}[thick]
\coordinate (O) at (0,0);
\coordinate (A) at (4,0);
\coordinate (B) at (0,2);
\draw (O)--(A)--(B)--cycle;

\tkzLabelSegment[below=2pt](O,A){\textit{$s$}}
\tkzLabelSegment[left=2pt](O,B){\textit{$1$}}
\tkzLabelSegment[above right=2pt](A,B){\textit{$\sqrt{s^2+1}$}}

\tkzMarkRightAngle[fill=orange,size=0.5,opacity=.4](A,O,B)
\tkzLabelAngle[pos = 0.35](A,O,B)\

\tkzMarkAngle[fill= orange,size=0.7cm,%
opacity=.4](O,B,A)
\tkzLabelAngle[pos = 0.5](O,B,A){$\frac{\alpha}{2}$}

\end{tikzpicture}
\end{center}

So it shows that

\begin{equation}
    \sin{\Big(\frac{\alpha}{2}\Big)} = \frac{s}{\sqrt{s^2+1}}
\end{equation}
 
\begin{equation}
    \cos{\Big(\frac{\alpha}{2}\Big)} = \frac{1}{\sqrt{s^2+1}}
\end{equation}

Pluggint into eq. 29 one gets

\begin{equation}
    d\alpha = \frac{2}{s^2+1}ds
\end{equation}

Plugging into eq. 28

\begin{equation}
    \int\frac{\frac{2}{s^2+1}ds}{K - 2\frac{s}{s^2+1}} = 2\int\frac{ds}{K(s^2+1) - 2s}
\end{equation}

Which is now solved easily enough.

\subsection{MATLAB Simulation Codes}

In this appendix all of the codes of the simulations ran in this project are given according to their purpose. Each code segment represents an independent function that can be put in a single .m file and executed. Note that the codes were written in Matlab 2017a.

\subsubsection{Code for the Beam's Characteristic Equation Simulation} \label{char_eq_code}
\begin{verbatim}
% Beam's Characteristic Equation analysis
% Shay Kricheli, 2018

clc; clear;
fplot(@(x) cos(x)*cosh(x)-1);
grid;
axis([0 30 -5 5]);
xL = xlim; yL = ylim;
line(xL, [0 0], 'Color','black');  %y-axis
line([0 0], yL , 'Color','black');  %x-axis
h = ylabel('$cos(x)cosh(x)-1$'); set(h,'Interpreter','latex');
h = xlabel('$x$'); set(h,'Interpreter','latex');
set(gca,'fontsize',22);
\end{verbatim}
\subsubsection{Code for the Washboard Potential Simulation} \label{washboard_code}
\begin{verbatim}
% Washboard Potential analysis
% Shay Kricheli, 2018

clc; clear;
t = 0:0.1:50;
B = 80;
for delta = 70 : 80
    h = strcat('$\Delta\omega =$ ',num2str(delta));
    plot(t, -delta*t - B* cos(t),'DisplayName',h);
    hold on;
end
h = legend('show','Location','northeast'); set(h,'Interpreter','latex');
h = ylabel('$U (\phi)$'); set(h,'Interpreter','latex');
h = xlabel('$\phi$'); set(h,'Interpreter','latex');
grid;
axis([0 50 -3000 0]);
print('washboard','-dmeta','-r3000') % save photo
\end{verbatim}
\subsubsection{Code for the Beam's Spatial Function Simulation} \label{spatial_code}
\begin{verbatim}
% Beam Spatial Function analysis
% Shay Kricheli, 2018

clc; clear;
s = 0:0.001:4.75;
X_1 = @(s)(0.618*(sin(s)-sinh(s))-0.629*(cos(s)-cosh(s)));
plot(s, X_1(s));
grid;
axis tight;
h = ylabel('$X(s)$'); set(h,'Interpreter','latex');
h = xlabel('$s$'); set(h,'Interpreter','latex');
set(gca,'fontsize',22);
\end{verbatim}

\subsubsection{Code for the Temporal Function's Steady State Amplitude} \label{time_code_ss}

\begin{verbatim}
% Beam Time Function SS Amplitude analysis
% Shay Kricheli, 2018

clc; clear;
Q = 50; %quality factor
zeta = 1/(2*Q); %damping ratio
open_system('duffing_simulation.slx'); % open the simulink model
t = 1000; % run time
gamma = 1/(4*Q); %control effort
K = 0; % driving amplitude
%perform the simulation
sim = sim('duffing_simulation','StartTime','0','StopTime',num2str(t));
T = sim.get('T');
T_dot = sim.get('T_dot');
plot(linspace(0,t,size(T,1)),T(:,1));
grid;
axis([0 num2str(t) -1 1]);
h = ylabel('$T (\tau)$'); set(h,'Interpreter','latex');
h = xlabel('$\tau$'); set(h,'Interpreter','latex');

\end{verbatim}

\subsubsection{Code for the Phase Space Simulation} \label{phase_code}

\begin{verbatim}
% Beam Phase Space analysis
% Shay Kricheli, 2018

% Beam’s Complete Vibration analysis
% Shay Kricheli, 2018

clc; clear;
Q = 100; %quality factor
zeta = 1/(2*Q); %damping ratio
open_system('duffing_simulation.slx'); % open the simulink model
tau = 1000; % run time
gamma = 1/(4*Q); %control effort
K = 0; % driving amplitude
sim = sim('duffing_simulation','StartTime','0','StopTime',num2str(tau)); %perform the simulation
T = sim.get('T');
T_dot = sim.get('T_dot');

s = 0:0.001:4.75;
X = @(s)(0.618*(sin(s)-sinh(s))-0.629*(cos(s)-cosh(s)));
Y = X(s);

plot(T(:,1),T_dot(:,1));
grid;
axis tight;
h = ylabel('$\frac{dT(\tau)}{d\tau}$'); set(h,'Interpreter','latex');
h = xlabel('$T (\tau)$'); set(h,'Interpreter','latex');
set(gca,'fontsize',22);
pbaspect([1 1 1]);

\end{verbatim}

\subsubsection{Code for the Complete Vibrations Simulation} \label{complete_sim_code}

\begin{verbatim}
% Beam Phase Space analysis
% Shay Kricheli, 2018

% Beam’s Complete Vibration analysis
% Shay Kricheli, 2018

clc; clear;
Q = 100; %quality factor
zeta = 1/(2*Q); %damping ratio
open_system('duffing_simulation.slx'); % open the simulink model
tau = 1000; % run time
gamma = 1/(4*Q); %control effort
K = 0; % driving amplitude
sim = sim('duffing_simulation','StartTime','0','StopTime',num2str(tau)); %perform the simulation
T = sim.get('T');
T_dot = sim.get('T_dot');

s = 0:0.001:4.75;
X = @(s)(0.618*(sin(s)-sinh(s))-0.629*(cos(s)-cosh(s)));
Y = X(s);

for timeIndex=1:size(T,1)
     plot(s, Y*T(timeIndex));
     grid; axis([0 4.75 -0.5 0.5]);
     pause (0.01);  
 end

\end{verbatim}

\subsubsection{Code for the Adler Equation Simulation} \label{adler_code}

\begin{verbatim}
% Adelr Equation analysis
% Shay Kricheli

clc; clear;
t0 = 0;
time_series = 0:0.1:100; 

for B = 3:5
    RunAdler(B, time_series,t0)
end

function RunAdler (B, time_series, t0)
    AdlerPlotter (10, -1 , 2 , time_series , t0, B, 1);
    AdlerPlotter (2, -0.1 , 1.2 , time_series , t0, B, 2);
    AdlerPlotter (1.1, -0.01 , 1.02 , time_series , t0, B, 3);
    AdlerPlotter (1.01, -0.001 , 1.002 , time_series , t0, B, 4);
    AdlerPlotter (1.001, -0.0001 , 1.0002 , time_series , t0, B, 5);
    AdlerPlotter (1.0001, -0.00001 , 1.00002 , time_series , t0, B, 6);
end

function AdlerPlotter (start, jump , final, t, t0, B, iteration)

    l = 1;
    for K = start:jump: final
        eta = (K^2 - 1)^(0.5);
        AdlerSol{l} = 2*atan(1/K + eta/K*tan(eta*B*(t-t0)/2));
        subplot(3,3,l);
        plot(t,AdlerSol{l});
        axis([0 100 -5 5]);
        xL = xlim; yL = ylim;
        line(xL, [0 0], 'Color','black');
        line([0 0], yL , 'Color','black');
        grid;
        h = ylabel('$\phi (t)$'); set(h,'Interpreter','latex');
        h = xlabel('$t$'); set(h,'Interpreter','latex');
        set(gca,'fontsize',16);
        l = l + 1;
    end
    % save photo
    print(strcat ('TimeDomain',num2str(iteration),'B',num2str(B)),'-djpeg') 
    clf;
    for v = 1:l-1
        FourierAdler = fft(AdlerSol{v});
        omega = (0:length(FourierAdler)-1)*20/length(FourierAdler);
        subplot(3,3,v);
        plot(omega,abs(FourierAdler),'r',omega+20*ones(1,1001),
            abs(FourierAdler),'r')
        axis([15 25 0 40]);
        grid;
        h = ylabel('$K (\omega)$'); set(h,'Interpreter','latex');
        h = xlabel('$\omega$'); set(h,'Interpreter','latex');
        set(gca,'fontsize',16);
    end
    % save photo
    print(strcat ('FreqDomain',num2str(iteration),'B',num2str(B)),'-djpeg') 
end
\end{verbatim}

\end{document}